\documentclass[twocolumn]{openjournal}
\usepackage{xcolor}
\usepackage{textgreek}
\usepackage[utf8]{inputenc}
\usepackage[english]{babel}
\usepackage{hyperref}
\hypersetup{
    unicode, 
    colorlinks=true,
    linkcolor=linkcolor,
    citecolor=linkcolor,
    filecolor=linkcolor,
    urlcolor=linkcolor,
}
\usepackage{color,colortbl}
\definecolor{linkcolor}{rgb}{0.0,0.3,0.5}
\usepackage{tensind}
\tensordelimiter{?}
\DeclareGraphicsExtensions{.bmp,.png,.jpg,.pdf}
\usepackage{verbatim}
\usepackage[normalem]{ulem}
\usepackage{orcidlink}
\usepackage{soul}

\graphicspath{ {./figs/} }

\usepackage{newtxtext,newtxmath}
\usepackage[T1]{fontenc}
\usepackage{graphicx}	% Including figure files
\usepackage{amsmath, amssymb, graphicx, color, units, xspace}
\definecolor{orangemain}{HTML}{FFA630}
\definecolor{bluemain}{HTML}{00A7E1}
\newcommand{\withCE}{\textcolor{bluemain}{with CE}}
\newcommand{\withoutCE}{\textcolor{orangemain}{without CE}}

\usepackage{acro}
\DeclareAcronym{UFD}{
  short = UFD ,
  long  = ultra-faint dwarf
}
\DeclareAcronym{BHNS}{
  short = BHNS ,
  long  = black hole--neutron star
}

\DeclareAcronym{BNS}{
  short = BNS ,
  long  = binary neutron star
}

\DeclareAcronym{BBH}{
  short = BBH ,
  long  = binary black hole
}

\DeclareAcronym{NS}{
  short = NS ,
  long  = neutron star
}

\DeclareAcronym{BH}{
  short = BH ,
  long  = black hole
}

\DeclareAcronym{GW}{
  short = GW ,
  long  = gravitational wave
}

\DeclareAcronym{SN}{
  short = SN ,
  long  = supernovae
}

\DeclareAcronym{DWD}{
  short = DWD ,
  long  = double white dwarf
}

\DeclareAcronym{CE}{
  short = CE ,
  long  = common-envelope
}

\DeclareAcronym{SMT}{
  short = SMT ,
  long  = stable mass transfer
}

\DeclareAcronym{DCO}{
  short = DCO ,
  long  = double-compact object
}

\DeclareAcronym{ZAMS}{
  short = ZAMS ,
  long  = zero-age main sequence
}

\DeclareAcronym{CHE}{
  short = CHE ,
  long  = chemically homogeneous evolution
}

\newcommand{\Gpcyr}{\ensuremath{\,\rm{Gpc}^{-3}\,\rm{yr}^{-1}}\xspace}
\newcommand{\Rbhbh}{\ensuremath{\mathcal{R}_{\rm{BH-BH}}\xspace}}

\newcommand{\kms}{\ensuremath{\,\rm{km}\,\rm{s}^{-1}}\xspace}
\newcommand{\Msun}{\ensuremath{\,\rm{M}_{\odot}}\xspace}

\newcommand{\Zsun}{\ensuremath{\,\rm{Z}_{\odot}}\xspace}

\newcommand{\Gyr}{\ensuremath{\,\mathrm{Gyr}}\xspace}

\newcommand{\yearmin}{\ensuremath{\,\rm{yr}^{-1}}\xspace}

\newcommand{\GpcminThree}{\ensuremath{\,\rm{Gpc}^{-3}}\xspace}

\newcommand{\alphaCE}{\ensuremath{{\alpha}\xspace}}

\newcommand{\CfA}{Center for Astrophysics \textbar{} Harvard $\&$ Smithsonian,
60 Garden St., Cambridge, MA 02138, USA}

\newcommand{\Columbia}{Department of Astronomy, Columbia University, 538 West 120th Street, New York, NY 10027, USA}

\newcommand{\UCSD}{Department of Astronomy and Astrophysics, University of California, San Diego, La Jolla, CA 92093, USA}

\newcommand{\MPA}{Max Planck Institute for Astrophysics, Karl-Schwarzschild-Strasse 1, 85748 Garching, Germany}

\begin{document}

% \title{Common Envelopes Are Not Universally Required to Form Gravitational-Wave Sources}
%\\ \small{
% \title{Does the formation of Gravitational-Wave Sources Require Common Envelopes?}
% \title{The importance of Common Envelopes to form gravitational wave sources \\ in the isolated binary evolution paradigm}

\title{How Common Are Common Envelopes?\\
Quantifying Their Role in Forming Gravitational-Wave Sources}
%
% \vspace{-5cm}
\author{Floor S. Broekgaarden$^{1}$\orcidlink{0000-0002-4421-4962}}
\email{fbroekgaarden@ucsd.edu}
\author{Ana Lam$^{2}$\orcidlink{0009-0003-9165-9889}}
\author{Sasha Levina$^{1}$\orcidlink{0000-0003-1241-7615}}
\author{Jakub Klencki$^{3}$\orcidlink{0000-0002-7527-5741}}
\author{Kyle A. Rocha$^{1}$\orcidlink{0000-0003-4474-6528}}
\author{Lieke~van~Son$^{4}$\orcidlink{0000-0001-5484-4987}}
\author{Steffani M. Grondin$^{1}$\orcidlink{0000-0002-0444-8502}}
\author{Monica~Gallegos-Garcia$^{5,6}$\orcidlink{0000-0003-0648-2402}}
\author{Brian D.~Metzger$^{2}$\orcidlink{0000-0002-4670-7509}}
\author{Enrico Ramirez-Ruiz$^{7}$\orcidlink{0000-0003-2558-3102}}
\author{Angela Twum$^{7}$\orcidlink{0009-0006-4675-7596}}
\author{Melanie Santiago$^{1}$\orcidlink{0009-0001-1988-8383}}
\author{Julia Haynes$^{1}$\orcidlink{0009-0001-7549-2535}}
\author{Tyler B. Smith$^{8}$\orcidlink{0000-0002-1732-8040}}
\author{Amedeo Romagnolo$^{9,10}$\orcidlink{0000-0001-9583-4339}}
\author{Edo Berger$^{5}$\orcidlink{0000-0002-9392-9681}}
\author{Lucas~M.~de~S\'a$^{9}$\orcidlink{0000-0003-3109-9042}}

\affiliation{$^{1}$\UCSD}
\affiliation{$^{2}$\Columbia}
\affiliation{$^{3}$\MPA}
\affiliation{$^4$ Department of Astrophysics/IMAPP, Radboud University Nijmegen, PO Box 9010, 6500 GL Nijmegen, The Netherlands}
\affiliation{$^{5}$\CfA}
\affiliation{$^{6}$Harvard Society of Fellows, 78 Mount Auburn Street, Cambridge, MA 02138, USA}
\affiliation{$^{7}$Department of Astronomy and Astrophysics, University of California, Santa Cruz, CA 95064, USA}

\affiliation{$^{8}$Department of Physics and Astronomy,
University of California, Irvine, CA 92697, USA}
\affiliation{$^{9}$Universität Heidelberg, Zentrum für Astronomie (ZAH), Institut für Theoretische Astrophysik, Albert Ueberle Str. 2, 69120, Heidelberg, Germany}
\affiliation{$^{10}$Dipartimento di Fisica e Astronomia Galileo Galilei, Università di Padova, Vicolo dell’Osservatorio 3, I–35122 Padova, Italy}
% \affiliation{Department of Astrophysics, the Graduate Center, City University of New York, 365 Fifth Ave., New York, NY 10016, USA}

\begin{abstract}
A central goal of gravitational-wave astronomy is to use merging \ac{BBH}, \ac{BHNS}, and \ac{BNS} systems as fossils to reconstruct the formation and evolution of massive stars across cosmic time. 
In practice, this inference relies on population-synthesis models that map massive stellar binaries to merging compact objects. 
However, these models disagree on the dominant orbital-hardening mechanisms within isolated binary evolution, particularly on whether \ac{CE} evolution is required.
To address this, we compile and systematically compare formation-channel predictions from more than 200 isolated-binary population-synthesis simulation results, organized within a unified hierarchical taxonomy, with an online public interactive catalog.
We find that BBH and BHNS formation pathways span nearly the full allowed range from CE-dominated to without-CE-dominated evolution (0--100\%),  while often predicting similar merger rates, revealing a fundamental degeneracy: merger-rate measurements alone do not uniquely constrain the underlying evolutionary pathways. 
In contrast, BNS formation proceeds almost exclusively through channels involving at least one CE phase $(\gtrsim 90$--$100\%)$, suggesting CE evolution plays a qualitatively different role in BNS than in BBH and BHNS formation.
The relative contributions of with-CE and without-CE pathways are governed primarily by assumptions controlling mass-transfer stability, angular-momentum loss, CE efficiency, and supernova physics, which often act non-linearly and in correlated fashion, such that trends from one-at-a-time parameter variations do not generalize across simulation frameworks. 
Robust interpretation of gravitational-wave populations will therefore require transparent formation-channel definitions, reproducible analysis pipelines, systematic cross-code comparisons, and observational constraints that extend beyond merger rates alone.
%
%%%%
\end{abstract}

%

% Write your keywords here
\begin{keywords}
    {Massive Star Evolution, Gravitational Wave Astrophysics, Common Envelope}
\end{keywords}

\maketitle

%%%%%%%%%%%%%%%%%%%%%%%%%%%%%%%
%%%%%%%%%%%%%%%%%%%%%%%%%%%%%%
\section{Introduction}
\label{sec:intro}
%%%
The number of detected gravitational-wave sources from merging stellar-mass \ac{BBH}, \ac{BHNS}, and \ac{BNS} systems is rapidly increasing, revealing new insights into their masses, spins, and merger rates \citep[e.g.,][]{GWTC-4:2025, GWTC-4-populations:2025, GWTC-5:catalog, GWTC-5:populations}. 
These observations provide a unique opportunity to study massive stars across cosmic time where merging compact-object binaries detected through gravitational waves can be used as \emph{fossils} of massive binary evolution, encoding information about the formation, evolution, and explosive deaths of their progenitor stars and the environments in which they formed \citep[e.g.][]{MandelFarmer:2018, BHbook:2023, Chruslinska2024}.

One of the leading formation scenarios for these systems is the isolated binary evolution channel, in which two massive stars are born in a binary system that evolves isolated from dynamical interactions with other stars \citep[e.g.][]{mandel2018merging, Mapelli:2021review, MandelBroekgaarden:2021}. 
This channel is typically characterized by initially wide binaries that can evolve through episodes of mass transfer and/or other orbital-altering processes, such as supernova kicks, that tighten the binary and eventually produce compact-object systems that merge through gravitational-wave emission and can be observed today \citep[e.g.][]{MandelFarmer:2018, Marchant:2024}.
The isolated binary evolution channel has been shown to reproduce several observed properties of the BBH, BHNS, and BNS populations \citep[e.g.,][]{belczynski2016first, stevenson2017formation, Santoliquido:2020, Bavera:2021, Briel:2021, BroekgaardenBerger2021, vanSon:2022-nopeaks, Olejak:2022, DorozsmaiToonen:2022, Hendriks:2023, Xing:2024-BHNS-allZ, GWTC-5:populations}.

However, the physical details of the processes that tighten massive binaries within this scenario remain highly uncertain \citep[e.g.][]{Marchant:2021, Gallegos-Garcia2021, Ioro:2023sevn, Willcox:2023, Romagnolo:2025, Xu:2025, Klencki:2026-smt}. 
A key ingredient in this problem is the role of binary interactions, particularly mass transfer, in transforming wide binaries into compact systems that can merge within a Hubble time \citep[e.g.][]{Belczynski:2002, Marchant:2021, Li:2024}. 
Mass transfer governs orbital evolution and strongly affects the final properties of the compact-object binary, including masses, spins, and orbital separations \citep[e.g.][]{Dominik:2012, Neijssel:2019, Bavera:2021, vanSon:2022-nopeaks, Briel:2022, Olejak:2024, Klencki:2026-smt, Xu:2025}.
It also plays a central role in shaping a broad range of observable astrophysical phenomena  \citep[e.g.][]{Soker+19, Evans:2020, Pavlovskii:2017, Metzger22, Korb:2024, Wang:2024, KlenckiMetzger25, vanSon:2026}. 
Yet the stability and outcome of mass transfer in massive binaries remain poorly understood, and many key phases are difficult to observe directly \citep[e.g.][]{deMink:2007, Sana2012, schneider2015evolution, Agrawal:2023, Marchant:2024}. 
Gravitational-wave observations may therefore provide a powerful probe of these otherwise elusive evolutionary stages, provided we can robustly connect observed populations to their formation pathways \citep[e.g.][]{MandelFarmer:2017, Marchant:2024}.

Historically, it was widely assumed that dynamically unstable mass transfer leading to a \ac{CE} phase is required to produce merging compact-object binaries \citep[e.g.][]{SmarrBlandford1976, Paczynski:1976, Tutukov:1993, vanDenHeuvel:1994, taam2000common, TaurisvdH:2006,  Dominik:2012, ivanova2013common, Ivanova:2020book}. 
During a CE phase, unstable mass transfer leads to the engulfment of the companion in a shared envelope, driving rapid orbital shrinkage \citep[e.g.][]{Paczynski:1976}. 
If the envelope is successfully ejected, this can produce short-period binaries that merge within a Hubble time. 
As a result, CE evolution has long been considered a critical ingredient in forming gravitational-wave sources \citep[][and references therein]{Iaconi:2019, Ivanova:2020book, MandelBroekgaarden:2021, TaurisvandenHeuvel:2023}.

More recently, however, an alternative picture has emerged.  
A growing body of work suggests that CE onset and the successful  (complete) ejection of the shared envelope, particularly in massive binaries, may occur under more restrictive conditions than traditionally assumed \citep{Mennekens:2014, Kruckow:2016, EldridgeStanway:2016, Fragos:2019, Klencki:2020, Klencki:2021,Marchant:2021,Gallegos-Garcia2021}. 
At the same time, there is increasing evidence that mass transfer in massive binaries may be more stable than previously assumed \citep{vandenHeuvel:2017, Inayoshi:2017, Ge:2015, Ge:2020, Pavlovskii:2017, Olejak:2021CE, Bhattacharjee:2026}. 
 As a result, sequences of stable mass-transfer episodes may in some cases shrink binaries sufficiently without requiring a CE phase \citep[e.g.][]{vandenHeuvel:2017, Inayoshi:2017, Neijssel:2019, Bavera:2021, Marchant:2021, Gallegos-Garcia2021, Olejak:2024, Picco:2024, Xu:2025}. In these models, stable mass transfer can become a viable, and in some cases dominant, formation channel for BBH and BHNS mergers \citep[e.g.][]{Neijssel:2019, Gallegos-Garcia2021, vanSon:2022-nopeaks, Briel:2022, Ioro:2023sevn}, though many uncertainties remain \citep[e.g.][]{Klencki:2026-smt, Briel:2026}.
In contrast, \ac{BNS} systems appear more consistent across models, with most studies finding that their formation involves at least one CE phase, with only a small parameter space allowing for alternative pathways \citep[e.g.][]{Chruslinska:2018,  VignaGomez:2018, Wagg2022, Gallegos-Garcia:2023, Ioro:2023sevn, Tanaka:2023, vanSon:2024, Picco:2024, Chattaraj:2026}.
These results highlight a major question in massive binary evolution: whether \ac{CE} evolution is (truly) required to form the majority of gravitational-wave sources \citep[e.g.,][]{Klencki:2026-smt}.

More fundamentally, the diversity highlighted above reflects qualitatively different physical assumptions regarding how interacting binaries redistribute angular momentum and dissipate orbital energy. 
In \ac{CE} evolution, rapid inspiral, hydrodynamic drag, and envelope ejection dominate the orbital hardening \citep[e.g.,][]{Paczynski:1976, taam2000common, ivanova2013common, Ivanova:2020book}, whereas \ac{SMT} pathways rely on stable angular-momentum redistribution through Roche-lobe overflow without the onset of a dynamically unstable engulfment phase \citep[e.g.,][]{Soberman:1997, Pavlovskii:2017,  vandenHeuvel:2017, Marchant:2021, DorozsmaiToonen:2022}. As a result, the relative contributions of CE and without-CE formation channels are not merely a quantitative question of efficiency, but reflect fundamentally different physical pictures of how massive binaries respond to mass transfer and evolve toward compact-object mergers \citep[e.g.,][]{Gallegos-Garcia2021, Marchant2021,  Klencki:2026-smt}. 
Disentangling these physical pictures is therefore a prerequisite for robustly connecting gravitational-wave observations to the astrophysical processes shaping massive binary evolution.

% But generally it has become clear that isolated binary-evolution models can reproduce current gravitational-wave observations while invoking qualitatively different evolutionary histories \citep[e.g.][]{DorozsmaiToonen:2022, Boesky:2024popsynth}, raising the possibility that present observations do not uniquely constrain the underlying physics of massive binary evolution \citep[cf.][]{MandelBroekgaarden:2021}.
% These contrasting results highlight a major uncertainty in massive binary evolution: whether \ac{CE} evolution is (truly) required to form the majority of gravitational-wave sources \citep[e.g.,][]{Klencki:2026-smt}. 

Despite extensive work investigating the formation channels of BBH, BHNS, and BNS mergers within the isolated binary evolution scenario, most studies so far focus on a single compact-object class, a single population synthesis code, and/or a limited exploration of parameter variations. As a result, it remains difficult to disentangle differences arising from physical assumptions within a model from those driven by the underlying population-synthesis framework itself. A systematic comparison across studies is therefore needed to map the true diversity of the isolated binary evolution formation pathways.

In this work, we compile and compare formation-channel predictions from more than two hundred population-synthesis simulations in the isolated binary evolution paradigm. 
By bringing together results across different codes and mapping out their assumptions, we aim to disentangle parameter-driven variations from framework-dependent differences and to provide a global view of the current theoretical landscape.
Specifically, we address three key questions in the context of isolated binary evolution population synthesis models:
\begin{enumerate}
\item \emph{Is common-envelope evolution required to form BBH, BHNS, and BNS mergers?} 
\item \textit{What fraction of BBH, BHNS, and BNS mergers typically forms through pathways involving a CE phase?}
\item \emph{What drives the disagreement? Which physical assumptions and model choices are responsible for the large differences between studies?}
\end{enumerate}
By systematically organizing isolated binary evolution formation-channel predictions within a unified hierarchical taxonomy, analyzing their assumptions, and comparing them across the literature, this work aims to provide a foundation for interpreting gravitational-wave observations and identifying the key uncertainties that currently limit our understanding of massive binary evolution.
This is further accomplished by making all data and code to reproduce all results and figures in this work publicly available.

%%%%%%%%%%%%%%%%%%%%%%%%%%%%%%%%%%%%%%%%%%%%%%%%%%%%%%%%%%%%%%%%%%
%%%%%%%%%%%%%%%%%%%%%%%%%%%%%%%%%%%%%%%%%%%%%%%%%%%%%%%%%%%%%%%%%%
%%%%%%%%%%%%%%%%%%%%%%%%%%%%%%%%%%%%%%%%%%%%%%%%%%%%%%%%%%%%%%%%%%
%%%%%%%%%%%%%%%%%%%%%%%%%%%%%%%%%%%%%%%%%%%%%%%%%%%%%%%%%%%%%%%%%%
\section{Method: Data collection}
\label{sec:fc-method}
We compile intrinsic compact-object merger rates by formation pathway from the isolated binary-evolution literature. We restrict the sample to studies published within the past decade that (i) report \textbf{formation-channel--specific} merger rates and (ii) provide \textbf{intrinsic (astrophysical)} rates rather than detection-weighted (``detectable'') rates.
We exclude studies that only report detectable rates because such rates depend on assumed detector sensitivities and selection functions, which vary across analyses and make comparison challenging.
For transparency and reproducibility, the rate-extraction procedure for each study is documented in Appendix~\ref{app:fc_data-retrieval}.
We organize formation pathways into two hierarchical levels, illustrated in Figure~\ref{fig:formation-channel-cartoon}. An online interactive table displaying all the information can be found \href{https://floorbroekgaarden.github.io/Rates_of_Formation_Channels/interactive_figures_and_tables/formation_channel_rates_table.html}{online} and on \citet{Broekgaarden:Zenodo-Common-Common-Envelope}.

\subsection{Level~1 classification}
% At Level~1 we adopt three bins: (i) without CE, (ii) with CE, and (iii) other (unclassified residuals).
At Level~1, systems are classified into three broad categories according to whether they experience zero \acp{CE} (``without CE''), at least one \ac{CE} episode (``with CE''), or belong to a residual ``other'' category where the presence or absence of a \ac{CE} phase is not specified.
This distinction between with- and without-\ac{CE} evolution is chosen as a broad category to focus on the question whether the formation pathway involves a \ac{CE}.
Physically, this division approximately separates systems that undergo rapid hydrodynamic orbital shrinkage through envelope engulfment from those that avoid a dynamically unstable inspiral phase altogether. However, the boundary between these categories is not always sharply defined. In practice, binary interactions may span a continuum ranging from fully stable Roche-lobe overflow to delayed dynamical instability, grazing-envelope evolution, or partially unstable mass transfer. Consequently, the `with CE' and `without CE' terminology should be viewed as an effective phenomenological classification rather than a uniquely defined physical dichotomy.
The ``other'' category  is included for studies that aggregate a small residual fraction of systems into unspecified or unlabeled pathways, for example when many detailed channels are presented individually but a remainder is left without an explicit statement on \ac{CE} involvement \citep[e.g.][]{Olejak:2021CE, Neijssel:2019, Mestichelli:2025}.
In practice, this contribution is typically subdominant and only appears in our compilation for the formation-channel classification of \citet{Olejak:2021CE}.
When a study reports only finer sub-divisions (Level~2), we map each sub-channel to Level~1 and sum sub-channels to obtain the total ``without CE'' and ``with CE'' contributions. If a study reports rates for only a subset of channels, we treat unreported channels as negligible and set their rates to zero for that study. This convention is applied consistently across the compilation and is recorded alongside the extracted values.

\subsection{Without-CE versus stable mass-transfer channel terminology}
\label{sec:method-Without-CE-versus-SMT-terminology}
Importantly, throughout this work the term ``without CE'' refers strictly to the absence of a \ac{CE} episode and should not be interpreted as synonymous with a \ac{SMT} channel. Distinguishing binaries that avoid \ac{CE} evolution from those that specifically evolve through stable mass transfer has become increasingly relevant given growing interest in whether gravitational-wave sources can form entirely without a \ac{CE} phase \citep[e.g.][]{Bavera:2021, Briel:2022, Gallegos-Garcia2021, Gallegos-Garcia:2023}. Many studies therefore report results at the level of ``CE'' versus ``stable'', ``\ac{SMT}'', or ``only stable mass transfer'' channels \citep[e.g.][]{Bavera:2021, vanSon:2023sfrd, Hendriks:2023}.

However, these terms are not always used consistently across the literature. In several population-synthesis studies, the population labeled as SMT is identified operationally as all systems that avoid a \ac{CE} phase during their evolution \citep[e.g.][]{Bavera:2021, vanSon2023}. 
As a result, these selections can include not only binaries undergoing stable Roche-lobe overflow, but also systems that never experience mass transfer at all. 
Moreover, the term \ac{SMT} itself is also used inconsistently. In some studies it refers narrowly to only systems involving stable mass transfer both before \textit{and} after the first supernova \citep{Neijssel:2019, Broekgaarden2022, Ioro:2023sevn} (as we will see, this is sometimes referred to as the `classic SMT' or  `SMT+SMT' pathway), whereas in others it refers more broadly to any pathway involving stable mass transfer while avoiding \ac{CE} evolution \citep{vanSon:2022, Briel:2021, Bavera:2021}. 
Throughout this work we therefore reserve ``without CE'' for the broader Level~1 classification of any formation pathway that does not involve a CE, while using \ac{SMT} specifically for Level~2 channels involving stable mass transfer episodes.

An important example of such a non-interacting pathway is the `NON+NON' or ``lucky kick'' channel, in which binaries avoid both Roche-lobe overflow and \ac{CE} evolution entirely. These systems can nevertheless form merging compact-object binaries through favorable natal kicks that tighten the orbit sufficiently for the binary to remain bound and merge despite initially wide separations \citep[e.g.][]{VignaGomez:2018, Broekgaarden:2021iew, Chattaraj:2026, Mestichelli:2025}. Similar non-interacting pathways have also been proposed for the formation of ultra-wide pulsar systems \citep[e.g.][]{Stevenson:2022, vanderWateren:2024}.

Another pathway that typically avoids a \ac{CE} phase and is sometimes modeled within the same simulation is \ac{CHE} \citep[e.g.][]{deMinkMandel:2016, MandelDeMink:2016}. 
Although often assumed to be subdominant, several population synthesis studies find that \ac{CHE} may contribute a non-negligible fraction of the observable \ac{BBH} population \citep{Riley:2020, Stevenson:2022, vanSon:2024, Li:2025}. 
 The two studies included in our compilation that contain CHE \citet[][]{Li:2025, vanSon:2024} explicitly classify \ac{CHE} systems as avoiding Roche-lobe overflow and \ac{CE} evolution, and thus we group them within the broader level~1 ``without-CE'' category as a no mass transfer subchannel throughout this work.

{This CHE subchannel assignment carries two caveats, both arising from interaction that our taxonomy does not separately resolve. 
First, detailed stellar-evolution simulations suggest that tidally-induced \ac{CHE} binaries may evolve through prolonged contact phases \citep{Marchant2016}, which complicates their assignment to a strictly no-mass-transfer subchannel, even though such contact does not itself constitute a \ac{CE}. 
Second, most rapid population-synthesis codes model only tidally-induced \ac{CHE} in close binaries, whereas mass-transfer-induced quasi-homogeneous evolution, in which accretion-driven rejuvenation and rotational mixing render the accretor chemically homogeneous, is typically followed self-consistently only in detailed codes such as \textsc{BPASS} \citep[e.g.][]{Eldridge:2019}. 
Systems classified as forming without a \ac{CE} phase through stable mass transfer in \textsc{BPASS} (e.g. the SMT+NON channel) may therefore include a contribution from this mass-transfer-driven quasi-homogeneous pathway that is not separately resolved in our taxonomy.}

This diversity in terminology and classification criteria further motivates the need for a systematic framework to compare and categorize isolated binary evolution channels, a point we revisit in Section~\ref{sec:discussion-what-constitutes-a-formation-channel}.

\begin{figure*}
    \centering
    \includegraphics[width=0.91\textwidth]{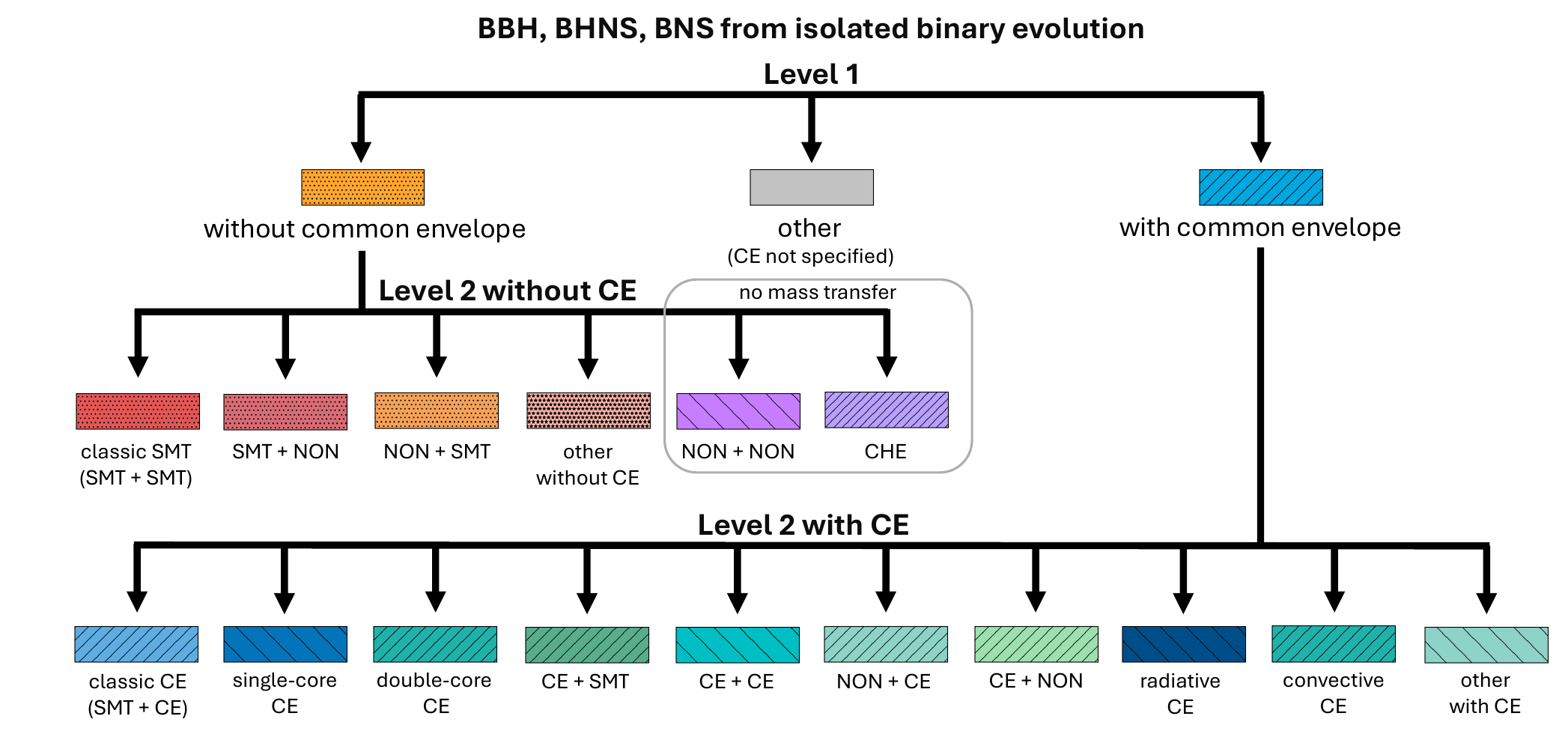} 
\caption{
Hierarchical formation-channel taxonomy used in this work to compile intrinsic merger rates from isolated binary evolution simulations of \ac{BBH}, \ac{BHNS}, and \ac{BNS} systems.
\textbf{Level~1} partitions mergers into systems evolving \emph{without CE} (no \ac{CE} episodes), \emph{with CE} (one or more \ac{CE} episodes), and an \emph{other} category for residual contributions where the original study does not further specify the \ac{CE} involvement. Throughout this work we primarily focus on the more consistently defined Level~1 classification.
\textbf{Level~2} illustrates the broader diversity of formation-channel definitions used across the literature. The Level~2 channels follow the original definitions adopted in each study and are therefore not always mutually exclusive. For the \emph{without-CE} branch, we distinguish the formation pathway involving stable mass transfer before and after formation of the first compact object (classic SMT; SMT$+$SMT), channels with stable mass transfer only before or after compact-object formation (SMT$+$NON and NON$+$SMT), systems without any mass-transfer episodes (NON$+$NON),  \ac{CHE}, and an ``other without CE'' residual category. In the studies considered here, the NON$+$NON and \ac{CHE} pathways correspond to channels without any mass-transfer episodes.
For the \emph{with-CE} branch, we include the classic CE pathway (SMT$+$CE), channels distinguished by the \ac{CE} type prior to compact-object formation (single-core CE and double-core CE), channels categorized by whether the \ac{CE} occurs before and/or after formation of the first compact object (CE$+$SMT, CE$+$CE, NON$+$CE, and CE$+$NON), envelope-structure subclasses (radiative and convective CE), and an ``other with CE'' residual category. Because different studies adopt different classification schemes, Level~2 channels can overlap conceptually. For example, systems classified as radiative or convective CE in one study may correspond to binaries grouped into single-core CE, double-core CE, classic CE, or other with-CE categories in another study. See text and Appendix~\ref{app:fc_data-retrieval} for further discussion. See \href{https://github.com/FloorBroekgaarden/Rates_of_Formation_Channels}{GitHub repository} to reproduce the figure.
}
    \label{fig:formation-channel-cartoon}
\end{figure*}
%

%%%%%%%
\subsection{Level~2 classification}\label{subsec:level-2-classification}
We further subdivide both the `without CE'' and `with CE'' Level 1 pathways using the Level 2 subchannel taxonomy illustrated in Figure~\ref{fig:formation-channel-cartoon}. In the text below, SN1 denotes the formation of the first compact object (via supernova or direct collapse). We use the shorthand SMT for stable mass transfer, CE for common-envelope evolution, and NON for the absence of mass transfer in the relevant evolutionary phase. The Level 2 taxonomy is intended as a flexible comparative framework to organize the heterogeneous channel definitions used throughout the literature. 
Level~2 channels follow the original definitions adopted in each study and are therefore not always mutually exclusive. Additional discussion of these study-dependent classifications is provided in Section~2.4.

\begin{itemize}
    \item \textbf{Level~2 without CE (no CE episodes):}
    For studies that provide interaction-sequence information around SN1, we distinguish the following subchannels:
    (i) SMT$+$SMT, i.e. stable mass transfer both before and after SN1 (often termed the ``only stable mass transfer'' or ``classic'' SMT pathway; e.g., \citealt{Olejak:2021CE,Briel:2022, Broekgaarden:2021iew, Sgalletta2024, Ioro:2023sevn, Boesky:2024gw});
    (ii) SMT$+$NON and NON$+$SMT, i.e.\ systems with stable mass transfer in only one of the two phases relative to SN1; and
    (iii) NON$+$NON, i.e.\ systems that do not undergo mass transfer in either phase around SN1.
    When a study reports CHE as a distinct formation pathway, we record it separately as CHE (in the studies considered here by definition involving no Roche-lobe overflow/SMT and no CE; \citealt{Li:2025}).
    Any residual or study-specific categories that do not map cleanly onto the above bins are recorded as ``other without CE''. This channel is only included in simulations such as \citet{Broekgaarden2022} where the authors split up the without-CE formation channel contribution into the classic stable mass transfer channel (SMT $+$ SMT) and everything else without a CE phase.
  \item \textbf{Level~2 with CE (one or more CE episodes):}
    When studies provide CE-subtype information, we retain it at Level~2. The most common refinement separates a ``classic CE'' channel from CE channels where at CE onset both stars are evolved or from different sequences of mass transfer:
    (i) \emph{classic CE}, commonly corresponding to an interaction sequence in which SMT occurs prior to SN1 and a CE episode occurs after SN1 (SMT $+$ CE in the pre-/post-SN1 shorthand; e.g., \citealt{Broekgaarden:2021iew,Olejak:2021CE,Boesky:2024gw,Sgalletta2024, Ioro:2023sevn});
    (ii) \emph{single-core CE} and \emph{double-core CE} (as the first mass transfer), where one or both stars, respectively, are evolved at the onset of the CE episode (e.g., \citealt{VignaGomez:2018, Broekgaarden:2021iew,Romagnolo:2023,Sgalletta:2025, Ioro:2023sevn}). 
    The double-core CE channel typically requires both stars to evolve on a similar timescale, and as a result is mostly effective for BNS progenitors with very equal mass ratio $q\gtrsim0.9$ and has historically been discussed in various contexts involving NSs \citep[e.g.][]{Brown:1995,  BetheBrown:1998, Dewi:2006bx, Belczynski:2010-pulsars, Hwang:2015, Ivanova:2020book, VignaGomez:2020}.
    (iii) Recent studies instead split CE episodes by envelope structure at onset; we follow the authors and record these as \emph{radiative CE} and \emph{convective CE} when reported  (\citealt{DorozsmaiToonen:2022}).
    (iv)  For studies that provide interaction-sequence splits relative to SN1, we follow these studies and map CE systems into the combinations SMT $+$ CE, CE $+$ SMT, CE $+$ CE, NON $+$ CE, and CE $+$ NON (e.g., \citealt{Briel:2022}).
    (v) Finally, we retain ``other with CE'' for residual CE-involving pathways, including accretion-induced collapse channels (with CE) when explicitly reported (\citealt{Pellouin:2025}). The  contributions of the other with CE channel are typically very small. 
\end{itemize}
%
%%%%%%%%%%%%

\subsection{Level 2 bookkeeping framework}
The Level 2 categories shown in Figure\ref{fig:formation-channel-cartoon} should not be interpreted as a single universally adopted or mutually exclusive classification scheme used throughout the literature. Instead, they represent a bookkeeping framework constructed to map heterogeneous author-defined formation channels onto a common taxonomy. Different studies typically classify systems according to different evolutionary criteria and often report only a limited subset of these channels (commonly 2–4 out of the broader set shown here). As a result, apparent overlaps between Level 2 categories usually do not occur within a single study, but instead arise because different works partition the evolutionary space using different definitions or levels of detail. For example, one study may distinguish systems by interaction sequence relative to SN1 (e.g., SMT+CE versus CE+SMT), while another separates CE systems according to envelope structure (radiative versus convective CE) or according to whether the CE is single-core or double-core. The taxonomy adopted here is therefore intended as a flexible comparative framework rather than a claim that all studies track the same mutually exclusive channel decomposition.

\subsection{Local merger rates}

All reported rates are converted to local intrinsic merger rates (redshift $z \approx 0$ or $z \approx 0.2$ in some studies) in units of \GpcminThree\yearmin, which facilitates comparison with gravitational-wave measurements and removes dependence on detector sensitivity. We denote these rates as $\mathcal{R}_{\rm{m}}$. While these local rates are most commonly published, we note that formation-channel contributions can evolve significantly with redshift \citep{Olejak:2022MNRAS.516.2252O, vanSon:2022,Boesky:2024popsynth}. Variations in assumed star-formation histories or in the precise definition of ``local'' redshift likely introduce minor differences in the inferred rates, as demonstrated in Section~\ref{sec:results-Simulation-Variations-and-Parameter-Dependencies}, and should be investigated in more detail in future work.

We compare the merger rates predicted by the population-synthesis simulations to the observationally inferred merger rates for BBH, BHNS, and BNS systems from \citet{GWTC-5:populations}, derived from the gravitational-wave detections reported in the GWTC-5 catalog \citep{GWTC-5:catalog}. 
Specifically, we adopt the joint merger-rate constraints inferred using the \textsc{FullPop} and \textsc{PixelPop} population models reported in Table~2 of \citet{GWTC-5:populations}. 
The BHNS and BNS rates in Table~2 are quoted at $z\approx0$, but the BBH rate of $27.5$--$49.4$~\Gpcyr is reported at $z\approx0.2$. 
Since our population-synthesis models report rates at redshifts varying between $z\sim0$--$0.2$, we extend the BBH constraint down to the $z\approx0$ range shown in Figure~5 of \citet{GWTC-5:populations}, giving a merged BBH observed rate of $13$--$49.4$~\Gpcyr that we use throughout this study.
These constraints already incorporate both statistical and population-model uncertainties.

For BBHs, \citet{GWTC-5:populations} report merger rates at $z \approx 0.2$, corresponding to the redshift range where the BBH merger rate is currently best constrained observationally. We therefore adopt a BBH merger-rate range of $27.5$--$49.4~\Gpcyr$. In contrast, the BNS and BHNS merger rates are reported at $z=0$, for which we adopt ranges of $5.1$--$154.7~\Gpcyr$ and $6.7$--$32.8~\Gpcyr$, respectively.

%%%%%%%%%%%%%%%%%%%%%%%%%%%%%%%%%%%%%%%%%%%%%%%%%%%%%%%%%%%%%%%%%%%%%%%%
%%%%%%%%%%%%%%%%%%%%%%%%%%%%%%%%%%%%%%%%%%%%%%%%%%%%%%%%%%%%%%%%%%%%%%%%
%%%%%%%%%%%%%%%%%%%%%%%%%%%%%%%%%%%%%%%%%%%%%%%%%%%%%%%%%%%%%%%%%%%%%%%%
%%%%%%%%%%%%%%%%%%%%%%%%%%%%%%%%%%%%%%%%%%%%%%%%%%%%%%%%%%%%%%%%%%%%%%%%

\section{Results}
\label{sec:results}

%%%
\begin{figure*}
    \centering
    \includegraphics[width=1\linewidth]{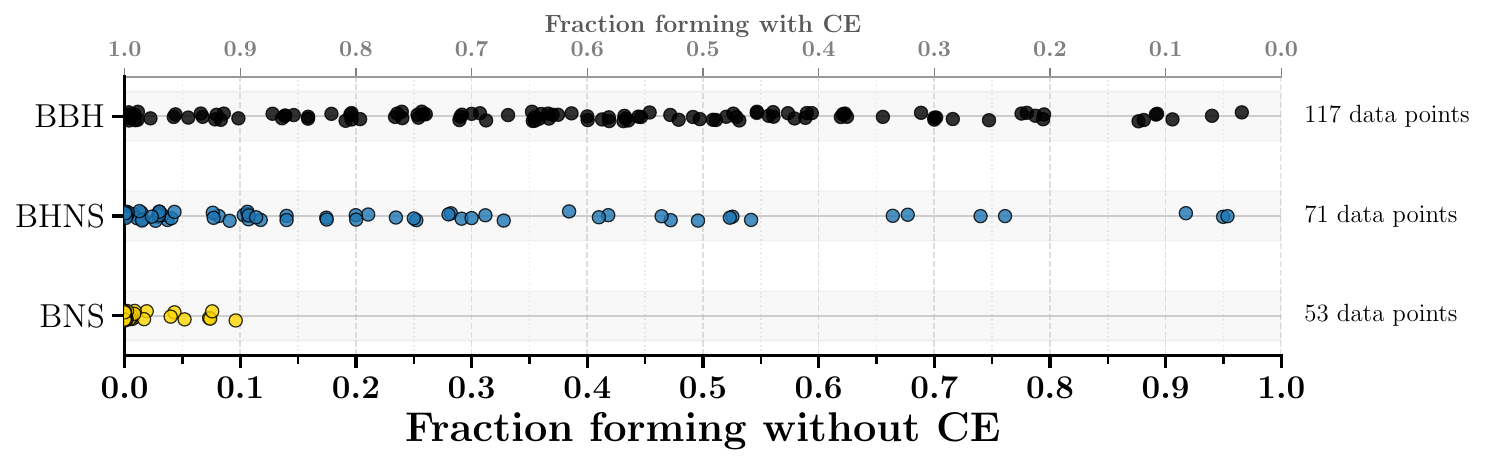}
\caption{Global overview of Level~1 formation-channel diversity across the compiled population-synthesis simulations. Each point shows the fraction of systems forming without a \ac{CE} phase for a given model, for \ac{BBH}, \ac{BHNS}, and \ac{BNS} mergers. Data points are given a small random vertical offset for visualization clarity.
\ac{BBH} and \ac{BHNS} populations span nearly the full range from CE-dominated to without-CE–dominated formation, whereas \ac{BNS} mergers show a strong preference ($\gtrsim90\%$ of all formed BNS) for channels involving \ac{CE} evolution.  
The \ac{BHNS} population additionally shows tentative clustering toward low without-\ac{CE} fractions, alongside substantial scatter between models. In Section~\ref{sec:results-Simulation-Variations-and-Parameter-Dependencies}, we explore the physical assumptions responsible for these differences, demonstrating that the diversity in outcomes cannot generally be reduced to a single parameter variation, population synthesis code, or stellar-evolution track framework.
The top axis indicates the corresponding \ac{CE} fraction (defined as $1 - $ fraction without CE). In a small number of cases ($5$ data points), this relation is approximate due to minor contributions from an ``other'' channel at Level~1; these contributions are negligible and do not affect the overall trends.
See  \href{https://floorbroekgaarden.github.io/Rates_of_Formation_Channels/interactive_figures_and_tables/formation_channel_rates_table.html}{interactive figure} and  \href{https://github.com/FloorBroekgaarden/Rates_of_Formation_Channels}{GitHub} for details and code.}
    \label{fig:summary-without-CE-channel}
\end{figure*}
%%%

%
\begin{figure*}
    \centering
\includegraphics[width=1\textwidth, trim={0 1.5cm 0 1.5cm}]{Fig3_CE_simple_lgd.pdf} 
% \vspace{-1cm}
\includegraphics[width=1\textwidth]{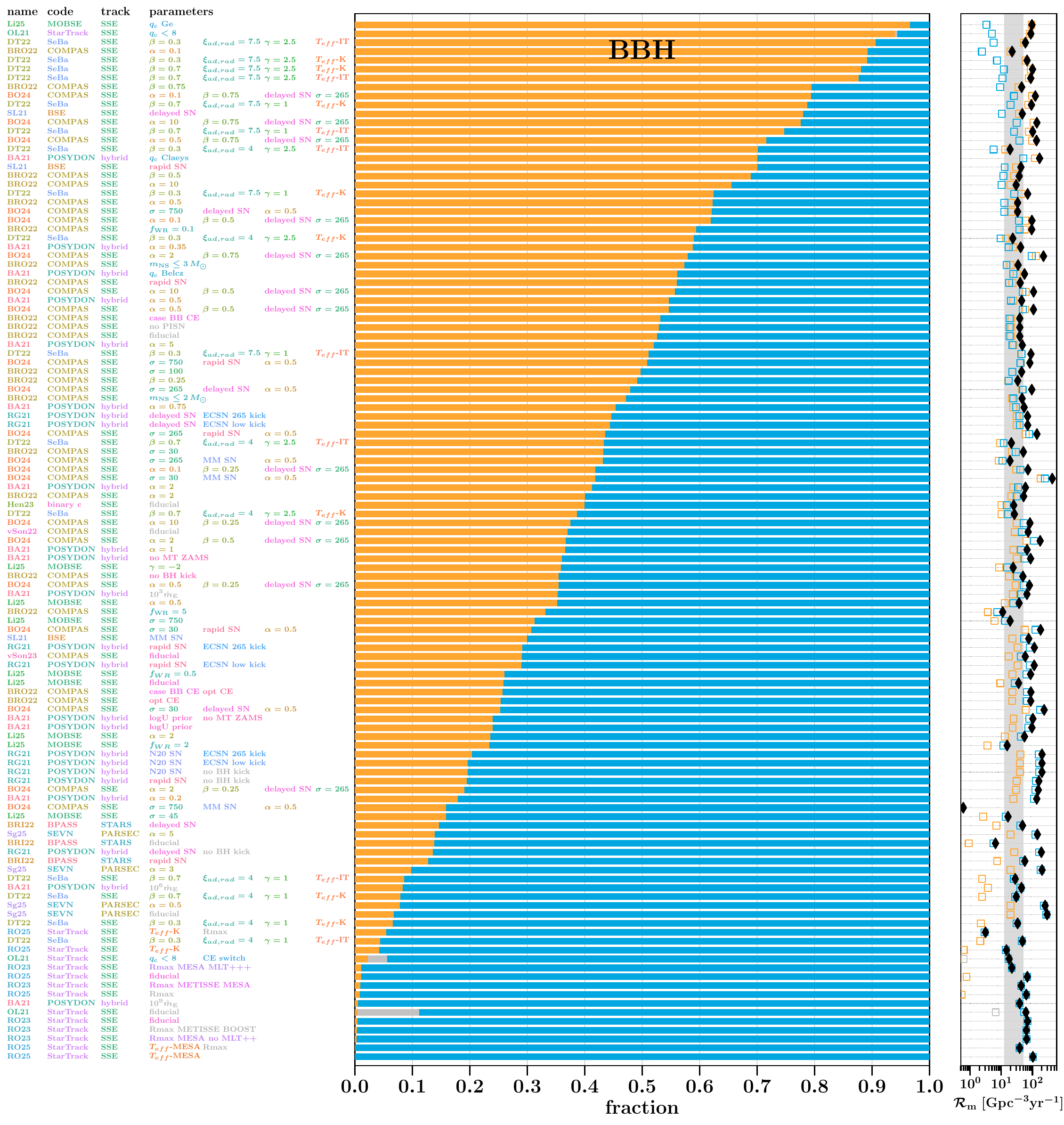} 
\caption{
\textbf{Middle panel:} Fractional contributions of Level~1 formation pathways (with and without CE) to the astrophysical BBH population across the compiled simulations. See Figure~\ref{fig:formation-channel-cartoon} for formation channels taxonomy.
\textbf{Right panel:} corresponding intrinsic local merger rates ($\mathcal{R}_{\rm{m}}$), together with the 90\% credible interval for the local BBH merger rate inferred from GWTC-5 at $z\approx0$--$0.2$ \citep[][gray shaded region]{GWTC-5:populations}. 
{The filled diamonds show the total intrinsic merger rate ($\mathcal{R}_{\rm m}$ at $z\approx0$--$0.2$), while the open squares show the corresponding rates split into the with-\ac{CE} (blue) and without-\ac{CE} (orange) contributions, colored as in the bar panel.}
\textbf{Left panel:} Labels in the far left encode the study, population synthesis code, stellar-evolution tracks, and key model parameters varied relative to the fiducial model (when available), including the CE efficiency  ($\alpha$), mass-transfer efficiency ($\beta$), supernova kick dispersion ($\sigma$), and critical mass ratio for mass transfer stability ($q_{\rm c}$). Tracks `hybrid' indicate a hybrid combination of MESA and COSMIC in an early version of POSYDON. 
Details of the data compilation and acronyms for the studies are provided in Appendix~\ref{app:fc_data-retrieval}, while a complete list of acronyms and parameter definitions is given in Appendix~\ref{app:fc_data-specs-acronyms}.
An online interactive table displaying all the values for each simulation and additional parameters and information can be found \href{https://floorbroekgaarden.github.io/Rates_of_Formation_Channels/interactive_figures_and_tables/formation_channel_rates_table.html}{online} and on \citet{Broekgaarden:Zenodo-Common-Common-Envelope}. 
See  \href{https://floorbroekgaarden.github.io/Rates_of_Formation_Channels/interactive_figures_and_tables/formation_channel_rates_table.html}{interactive figure} and  \href{https://github.com/FloorBroekgaarden/Rates_of_Formation_Channels}{GitHub} for details and code.}
 \label{fig:level-1-BBH}
\end{figure*}

\subsection{Overview of Formation Channel Diversity}
Figure~\ref{fig:summary-without-CE-channel} presents the compiled formation-channel contributions across more than one hundred population-synthesis simulations for BBH, BHNS, and BNS mergers. 
% The data retrieval is described in detail in Appendix~\ref{app:fc_data-retrieval}. 
We first summarize the Level~1 classification, distinguishing formation pathways into those with and without \ac{CE} evolution; shown in Figures~\ref{fig:level-1-BBH}--\ref{fig:level-1-bns}. 
We then follow with the more detailed Level~2 decomposition, for which the figures are presented in Figure~\ref{fig:level2-bbh}--\ref{fig:level2-bns}.

%%%%%%%%%%%%%%%%%%%%%%%%%%%%%%%%%%%%%
\subsubsection{Level 1: With CE versus without-CE pathways}
We begin by comparing the relative importance of systems that experience at least one \ac{CE} phase to those that avoid CE entirely across the compiled population-synthesis simulations.
To provide a global overview of the diversity in formation pathways across all simulations, Figure~\ref{fig:summary-without-CE-channel} shows the fraction of systems forming without a \ac{CE} phase for BBH, BHNS, and BNS mergers. 
Each point corresponds to a single population-synthesis model in our compilation.

This immediately illustrates the central result of this work: BBH and BHNS formation pathways span nearly the full allowed range from CE-dominated to without-CE–dominated formation pathways, whereas BNS systems cluster strongly toward CE-dominated pathways. The BHNS simulations also show a small clustering around small without-CE fractions.

For BBH mergers, the Level~1 classification exhibits the greatest diversity among the compact object merger classes as shown in detail in Figure~\ref{fig:level-1-BBH}. 
Both with-CE and without-CE dominated regimes are present, with many models occupying intermediate fractions, highlighting the strong sensitivity of BBH formation pathways to assumptions governing binary (interaction) physics. 
The associated local intrinsic merger rates span $\sim[1, 500]\Gpcyr$ and show no clear correlation with the CE channel contribution. 
This reveals a fundamental degeneracy: models with similar channel fractions can yield substantially different merger rates, while comparable total merger rates can arise from either CE- or without-CE–dominated pathways. 
This implies that merger rates alone do not uniquely constrain the underlying evolutionary history, as qualitatively different formation pathways can give rise to similar observed merger rates.

\begin{figure*}
    \centering
    \includegraphics[width=0.975\textwidth, trim={0 1.5cm 0 1.5cm}]{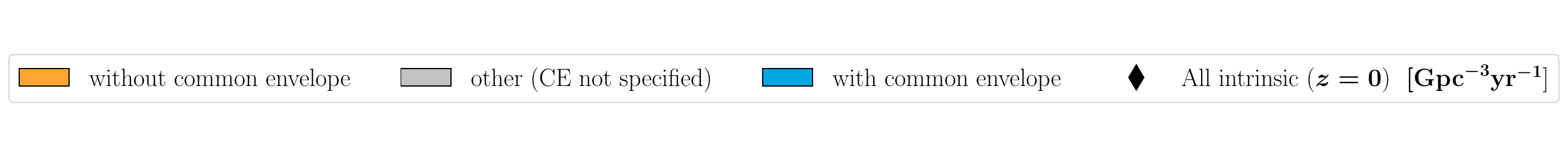} 
    \includegraphics[width=0.975\textwidth, trim={0 0.8cm 0 0cm}]{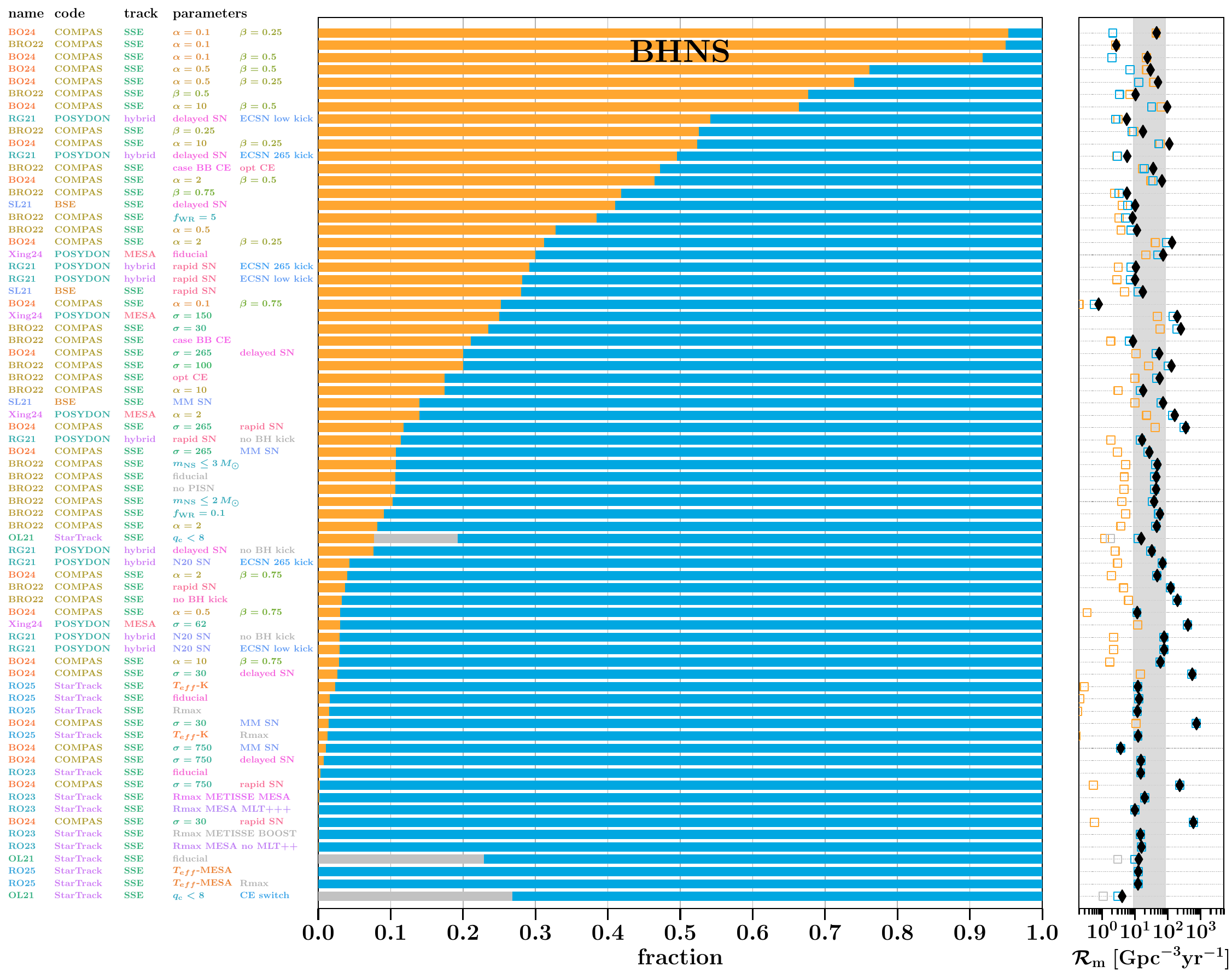} 
    \caption{Same as Figure~\ref{fig:level-1-BBH} for BHNS mergers. See  \href{https://floorbroekgaarden.github.io/Rates_of_Formation_Channels/interactive_figures_and_tables/formation_channel_rates_table.html}{interactive figure} and  \href{https://github.com/FloorBroekgaarden/Rates_of_Formation_Channels}{GitHub} for details and code.}
    \label{fig:level-1-bhns}
        %%%
        \vspace{0.7cm}
        %%%
    \includegraphics[width=0.975\textwidth, trim={0 0.8cm 0 0cm}]{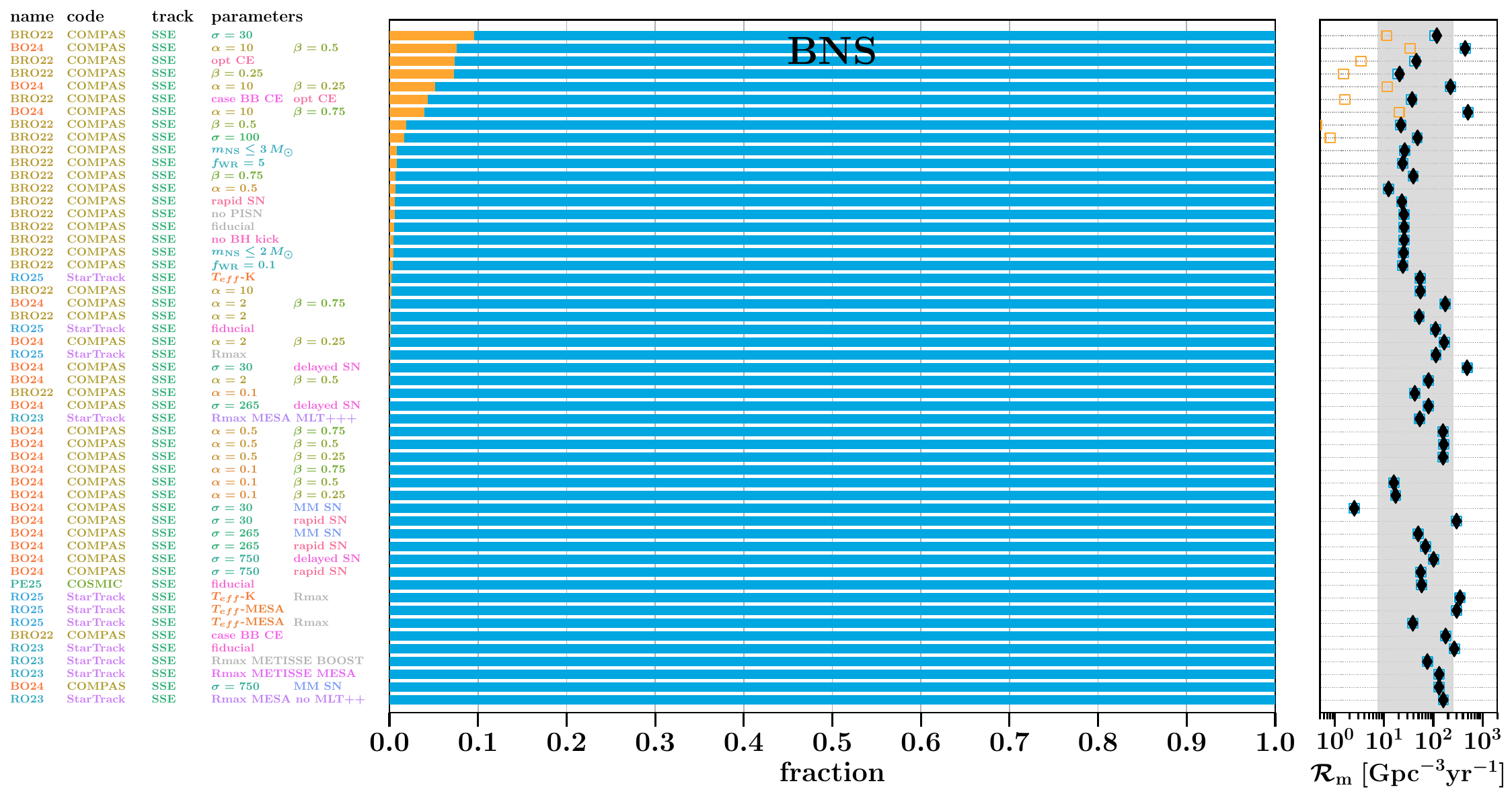} 
    \caption{Same as Figure~\ref{fig:level-1-BBH} for BNS mergers. See  \href{https://floorbroekgaarden.github.io/Rates_of_Formation_Channels/interactive_figures_and_tables/formation_channel_rates_table.html}{interactive figure} and  \href{https://github.com/FloorBroekgaarden/Rates_of_Formation_Channels}{GitHub} for details and code.}
    \label{fig:level-1-bns}
\end{figure*}
\begin{figure*}
    \centering
    \includegraphics[width=1\textwidth]{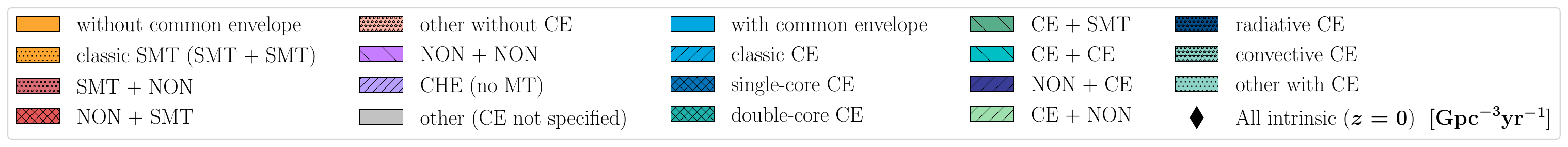} 
    \includegraphics[width=1\textwidth]{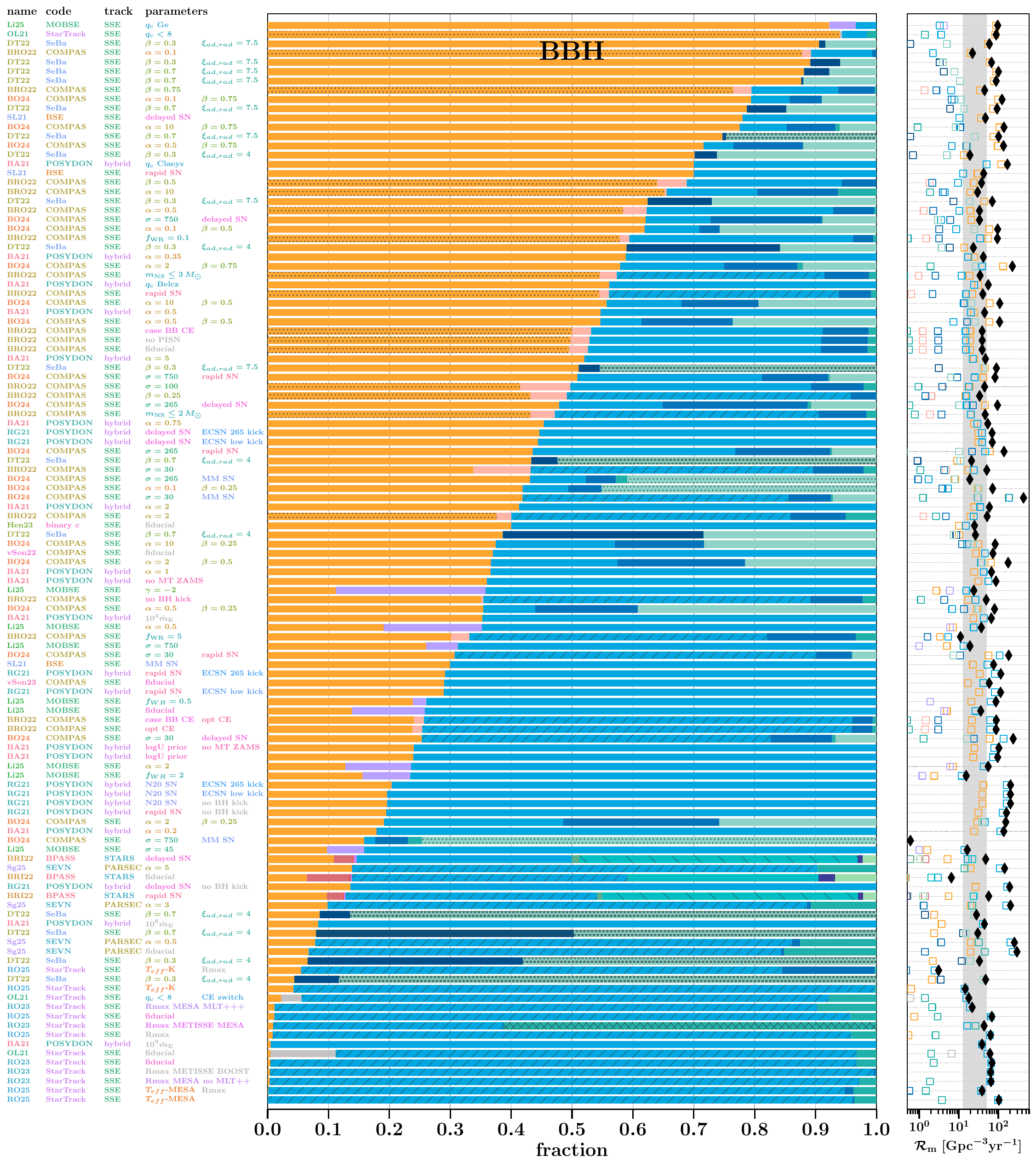} 
\caption{
Detailed Level~2 decomposition of BBH formation channels across the compiled population-synthesis simulations. Each horizontal bar represents a simulation, showing the fractional contribution of individual evolutionary subchannels within the isolated binary evolution paradigm. Subchannels are grouped into systems that evolve without a common-envelope (CE) phase (e.g., sequences involving only stable mass transfer or no interaction) and systems that experience at least one CE episode. 
The without-CE branch includes channels such as classic stable mass transfer (SMT+SMT), systems with stable mass transfer in only one evolutionary phase (SMT+NON or NON+SMT), and systems that avoid binary interactions altogether (NON+NON). The CE branch includes the classic CE pathway, as well as variations such as single-core and double-core CE events and sequence-based combinations of CE and stable mass transfer. The right panel shows the intrinsic BBH merger rate predicted by each simulation.
{The layout and study ordering follow those of Figure~\ref{fig:level-1-BBH}, with the addition of the Level~2 subchannel decomposition.}
The figure illustrates that the diversity observed at the Level~1 classification arises from a wide range of underlying evolutionary sequences rather than from a single dominant subchannel. Different population-synthesis models populate different combinations of these pathways, highlighting the complex interplay between binary interactions, stellar evolution, and supernova physics in shaping the BBH population.
See  \href{https://floorbroekgaarden.github.io/Rates_of_Formation_Channels/interactive_figures_and_tables/formation_channel_rates_table.html}{interactive figure} and  \href{https://github.com/FloorBroekgaarden/Rates_of_Formation_Channels}{GitHub} for details and code.}
    % \caption{
    % Detailed Level 2 decomposition of BBH formation channels showing the relative contributions of individual interaction sequences within the CE and no-CE branches. 
    % The figure illustrates the substantial diversity in subchannel contributions across simulations.
    % }
 \label{fig:level2-bbh}
\end{figure*}
\begin{figure*}
    \centering
    %%%%%%%%%%%%
    \includegraphics[width=0.97\textwidth, trim={0 0.1cm 0 1.5cm}]{figs/Fig3_CE_detailed_lgd.pdf}
    %%%%%%%%
    \includegraphics[width=0.97\textwidth, trim={0 0.8cm 0 0cm}]{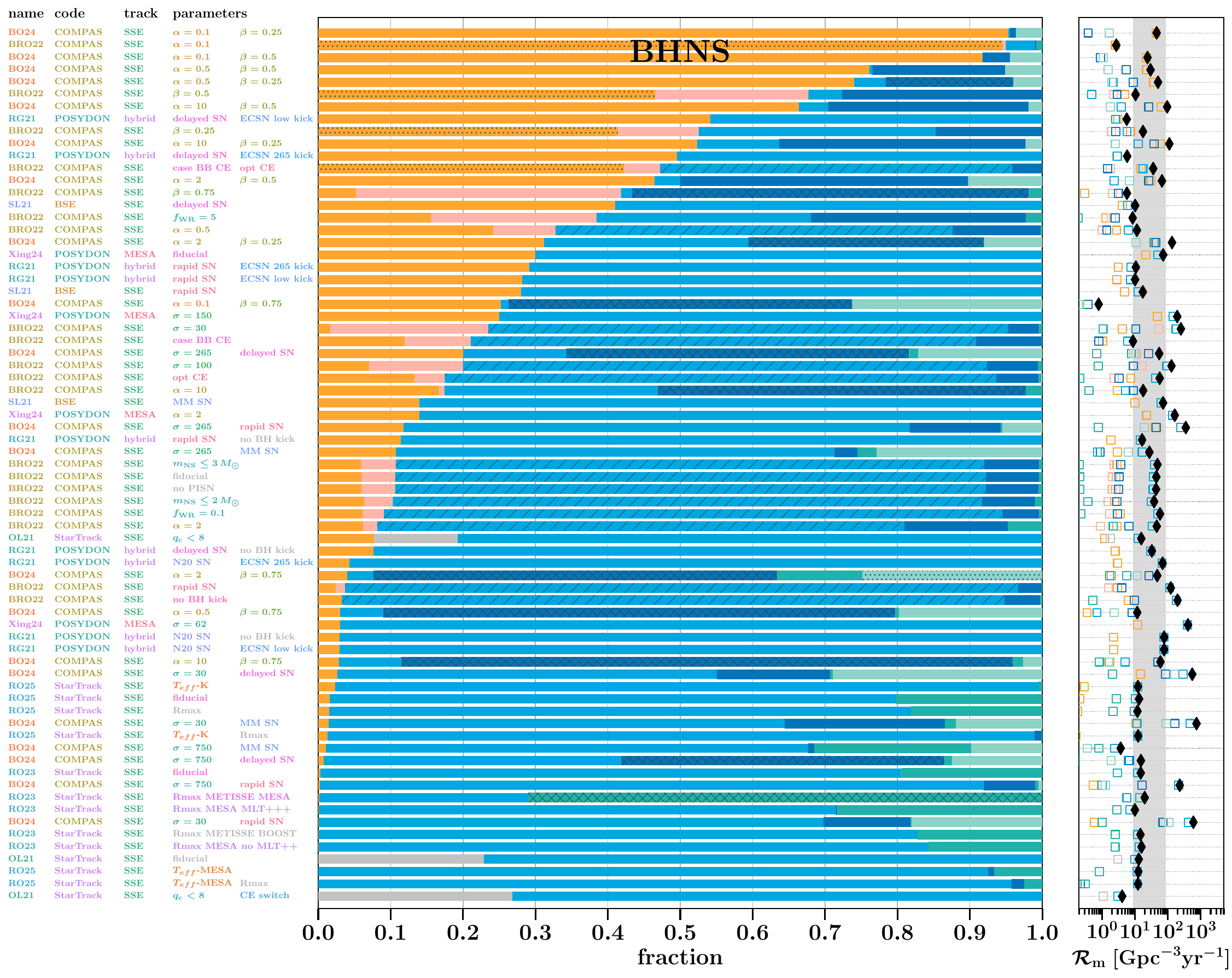} 
    \caption{Same as Figure~\ref{fig:level2-bbh} for BHNS mergers.  See  \href{https://floorbroekgaarden.github.io/Rates_of_Formation_Channels/interactive_figures_and_tables/formation_channel_rates_table.html}{interactive figure} and  \href{https://github.com/FloorBroekgaarden/Rates_of_Formation_Channels}{GitHub} for details and code.}
    \label{fig:level2-bhns}
    %%%%%%%%%%%%
    \includegraphics[width=0.97\textwidth, trim={0 0.8cm 0 0cm}]{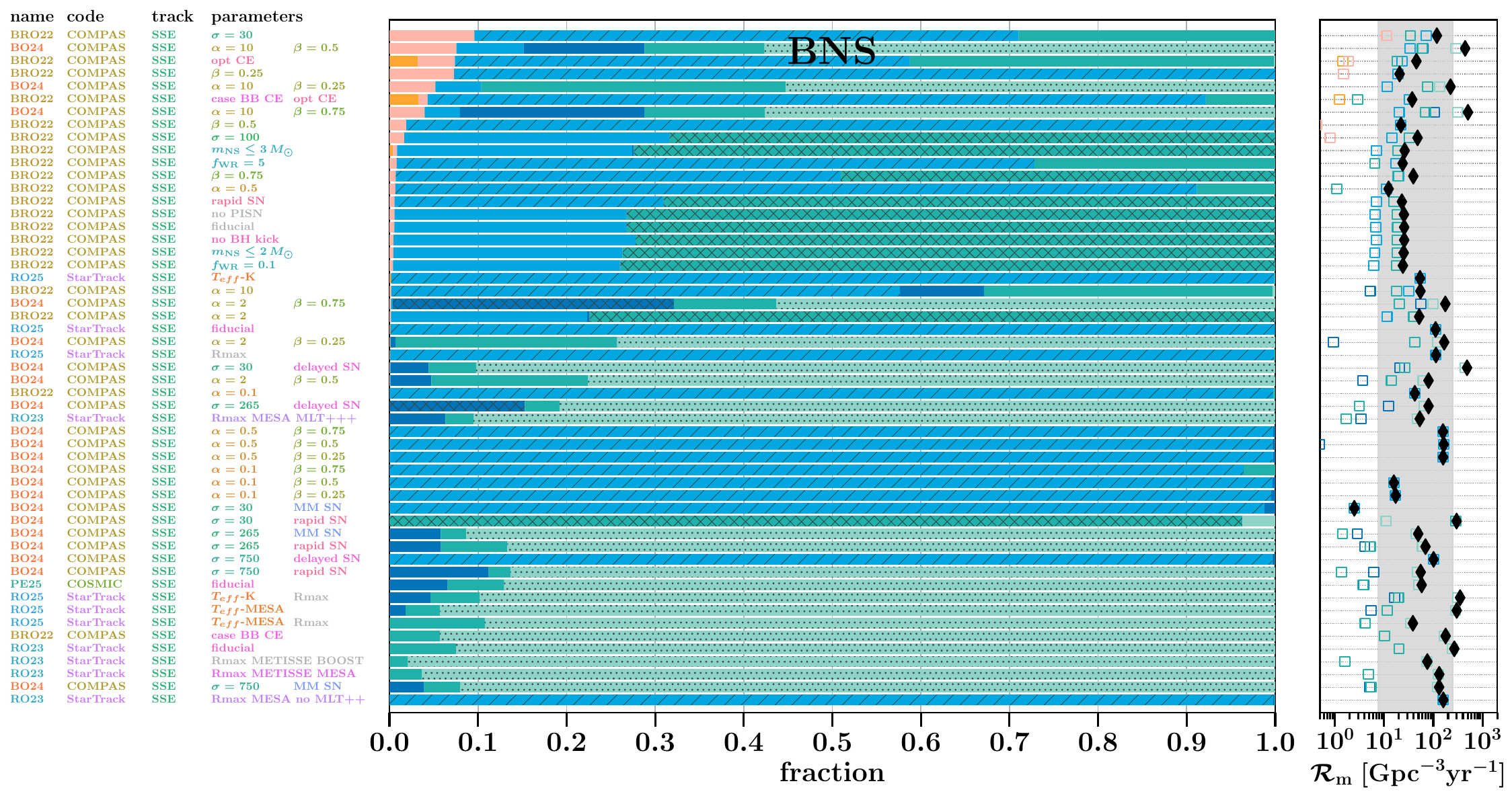} 
    \caption{Same as Figure~\ref{fig:level2-bbh} for BNS mergers. See  \href{https://floorbroekgaarden.github.io/Rates_of_Formation_Channels/interactive_figures_and_tables/formation_channel_rates_table.html}{interactive figure} and  \href{https://github.com/FloorBroekgaarden/Rates_of_Formation_Channels}{GitHub} for details and code.}
    %%%%%%%%%
    \label{fig:level2-bns}
\end{figure*}

The BHNS population displays somewhat less diversity than BBHs, as shown in Figure~\ref{fig:level-1-bhns}. 
In many simulations, CE pathways dominate, typically contributing $\gtrsim 70\%$ of BHNS mergers. 
However, a significant subset of models predicts substantial contributions from systems evolving without CE, and in some cases (particularly for COMPAS models with low CE efficiencies of $\alphaCE \sim 0.1$--$0.5$ and low mass transfer efficiencies of $\beta \sim 0.25$--$0.5$) BHNS formation is instead dominated ($\gtrsim 60\%$) by without-CE pathways. 
This demonstrates that no single evolutionary channel consistently dominates across studies. The predicted local intrinsic merger rates span $\sim[1, 400]\Gpcyr$, comparable to the BBH population.

In contrast, BNS mergers exhibit markedly more uniform behavior, as shown in detail in Figure~\ref{fig:level-1-bns}. 
Across the compiled simulations (and additional literature), all models consistently predict that the dominant formation pathway involves at least one CE phase, accounting for $\gtrsim 90\%$--$100\%$ of systems, with only a small minority ($\lesssim 10\%$) forming through channels without CE. 
Compared to the BBH and BHNS populations, the spread in channel fractions is significantly smaller, indicating strong agreement on the central role of CE evolution in BNS formation. The corresponding local intrinsic merger rates span $\sim[5, 1000]\Gpcyr$. 
We return to this comparatively robust prediction in Section~\ref{sec:discussion-bns-robust-result}, including a discussion of the rare without-\ac{CE} pathways.

%%%%%%%%%%%%%%%%%%%%%%%%%%%%%%%%%
\subsubsection{Level 2: Detailed Subchannel Decomposition}
Figures~\ref{fig:level2-bbh}--\ref{fig:level2-bns}  present the Level~2 decomposition of formation pathways when available, revealing additional complexity beyond the simple with CE versus without-CE distinction.  

Inside the without-CE branch, the classic only-stable-mass-transfer channel  (SMT + SMT; stable mass transfer before SN1 and stable mass transfer after SN1) frequently contributes significantly, although its importance varies widely across simulations. Systems undergoing stable mass transfer in only one evolutionary phase (SMT + NON or NON + SMT) and systems without any mass transfer (NON + NON) also contribute non-negligibly in several models, while \ac{CHE} appears as a minor but distinct channel where reported \citep{Li:2025}.

Within the CE branch, the classic CE pathway is typically the largest contributor, but significant diversity exists in the relative importance of single-core and double-core CE channels. Sequence-based combinations (e.g., CE+SMT or NON+CE) further illustrate the range of evolutionary histories captured across studies.

Overall, the Level~2 analysis demonstrates that the diversity observed at Level~1 is not driven by a clear single dominant subchannel, but instead reflects a broad distribution of evolutionary sequences across parameter space.
This indicates that the large spread in Level~1 channel contributions is not driven by a single dominant alternative pathway, but rather by a redistribution across multiple evolutionary sequences, reflecting the high-dimensional nature of binary evolution parameter space and the complex pathways to form gravitational-wave sources.

\begin{figure*}
    \centering
    \includegraphics[width=0.45\textwidth]{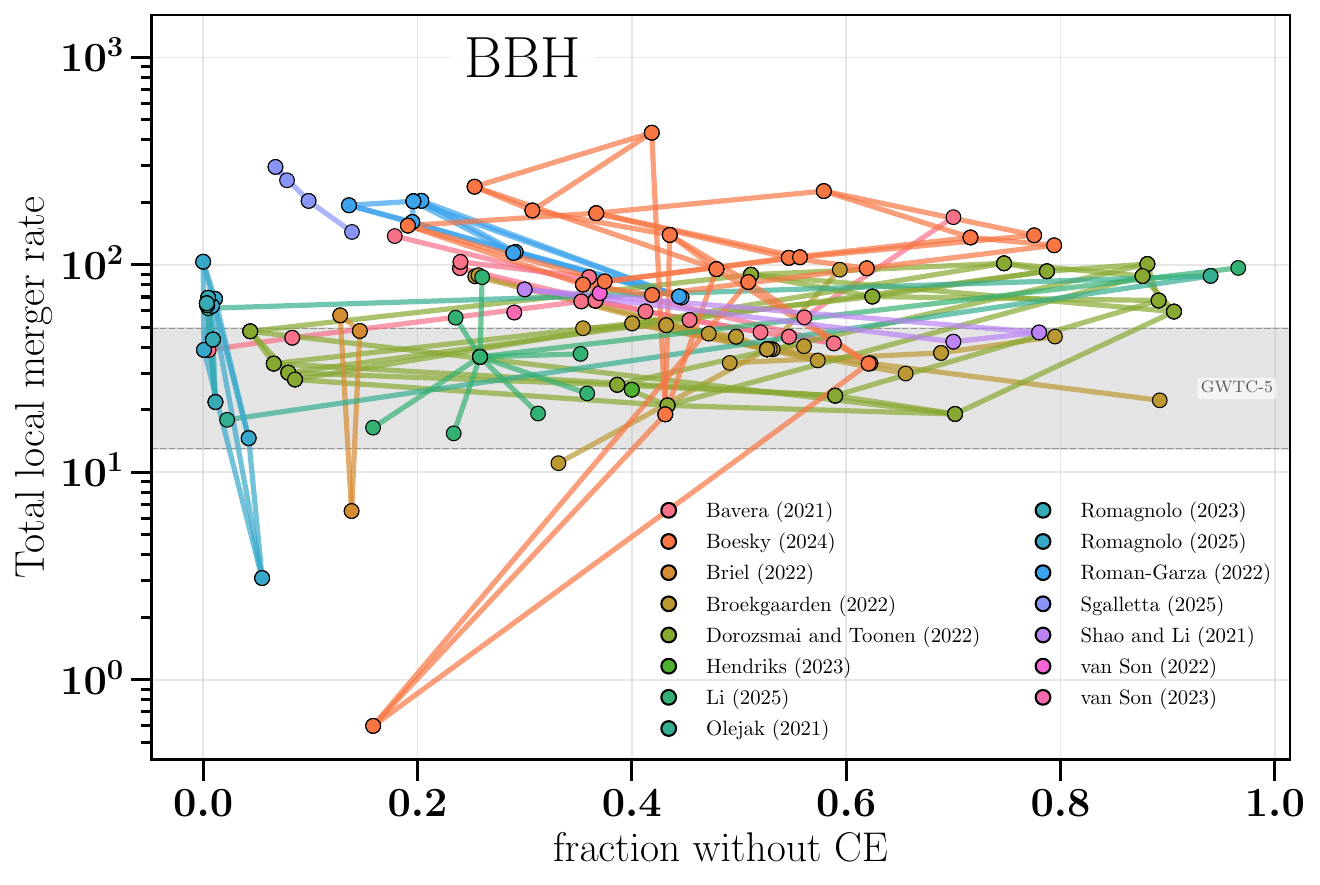} 
    %\\
    % \includegraphics[width=1\columnwidth]{figs/extra_Fig_parameter_family_delta_withoutCE_BHBH.pdf}
    % \vspace{0.5cm}
    \includegraphics[width=1\textwidth]{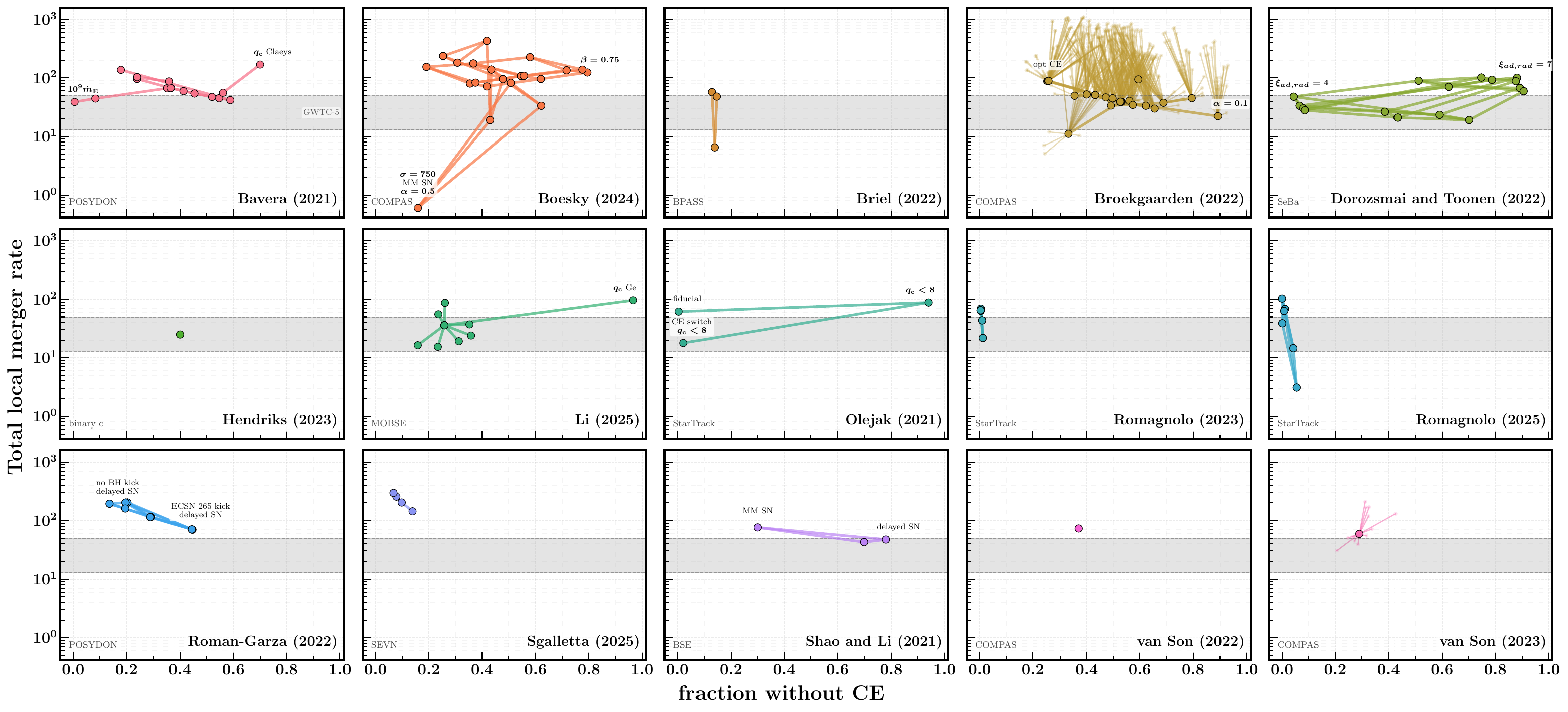}
    \caption{
    \textbf{Top:} Fraction of \ac{BBH} mergers forming without a \ac{CE} phase as a function of the total BBH intrinsic merger rate for the compiled population-synthesis simulations. Each point represents a single model, colored by study or stellar-evolution code. Lines connect models within the same study that vary a single parameter, illustrating the impact of controlled parameter changes. 
    \textbf{Bottom:} Same data grouped by study in individual subpanels. Each panel highlights how variations of specific binary-evolution parameters within a given population-synthesis framework affect the predicted fraction of BBH mergers forming without CE evolution.
    For \citet{Broekgaarden2022} and \citet{vanSon2023} additional lines are added for variations in the star formation history models.
    All panels show the 90\% credible interval for the local BBH merger rate inferred from GWTC-5 at $z\approx0$--$0.2$ \citep[][gray shaded region]{GWTC-5:populations}. 
    Some of the most extreme outlier models, corresponding to the lowest and/or highest without-\ac{CE} fractions, are annotated with their dominant model assumptions where appropriate. See the labels in Figure~\ref{fig:level-1-BBH} for the full parameter variations list.  See  \href{https://floorbroekgaarden.github.io/Rates_of_Formation_Channels/interactive_figures_and_tables/formation_channel_rates_table.html}{interactive figure} and  \href{https://github.com/FloorBroekgaarden/Rates_of_Formation_Channels}{GitHub} for details and code.}
    %
    % Within individual simulation grids, parameter variations often change the overall merger rate by orders of magnitude while leaving the relative contribution of CE and without-CE formation channels nearly unchanged. In contrast, the largest differences in formation-channel fractions occur between different population-synthesis codes. This behavior illustrates the presence of ``simulation silos,'' in which parameter studies within a single model framework explore only a limited subset of the broader uncertainty in binary-evolution predictions. The figure also highlights a degeneracy between merger rates and formation pathways: models producing similar BBH merger rates can nevertheless predict very different dominant evolutionary channels.
    % }
    \label{fig-comparison-bbh-simulation-silo}
\end{figure*}

%%%%%%%%%%%%%%%%%%%%%%%%%%%%%%%%%%%%%%%%%%%%%%%%%%%%%%
\subsection{Simulation Variations and Parameter Dependencies}
\label{sec:results-Simulation-Variations-and-Parameter-Dependencies}

The absence of clear clustering in the \ac{CE} formation-channel fractions by code, stellar tracks, or individual parameter choices in Figures~\ref{fig:level-1-BBH}--\ref{fig:level2-bns} already suggests that the origin of the variation is complex. Much of the diversity appears to arise either from differences between population-synthesis frameworks or from a small subset of influential parameters within a given code. Within individual studies, only a few parameter variations typically drive large changes in the relative importance of with-\ac{CE} and without-\ac{CE} pathways, although the identity of these key parameters differs between studies. Moreover, models adopting nominally similar parameter values (e.g., identical mass-transfer efficiency $\beta$) can still produce markedly different \ac{CE} fractions across different codes. This demonstrates that the effect of a single parameter depends on its coupling to other physical assumptions, such as stellar structure, angular-momentum loss, and mass-transfer stability. As a result, similar parameter choices do not lead to unique formation-channel outcomes, complicating attempts to isolate the impact of individual assumptions across population-synthesis frameworks.

These considerations motivate the more detailed within-code comparison of parameter variations presented below. Because all studies report results at the Level~1 classification, we focus on the relative contributions of systems evolving with and without a \ac{CE} phase, enabling a consistent comparison across models. We  focus on BBH and BHNS systems, which exhibit a broad range of \ac{CE} contributions, while BNS formation remains consistently dominated by channels involving \ac{CE} evolution (see Section~\ref{sec:discussion-bns-robust-result}).

\begin{figure}
    \centering
\includegraphics[width=\columnwidth]{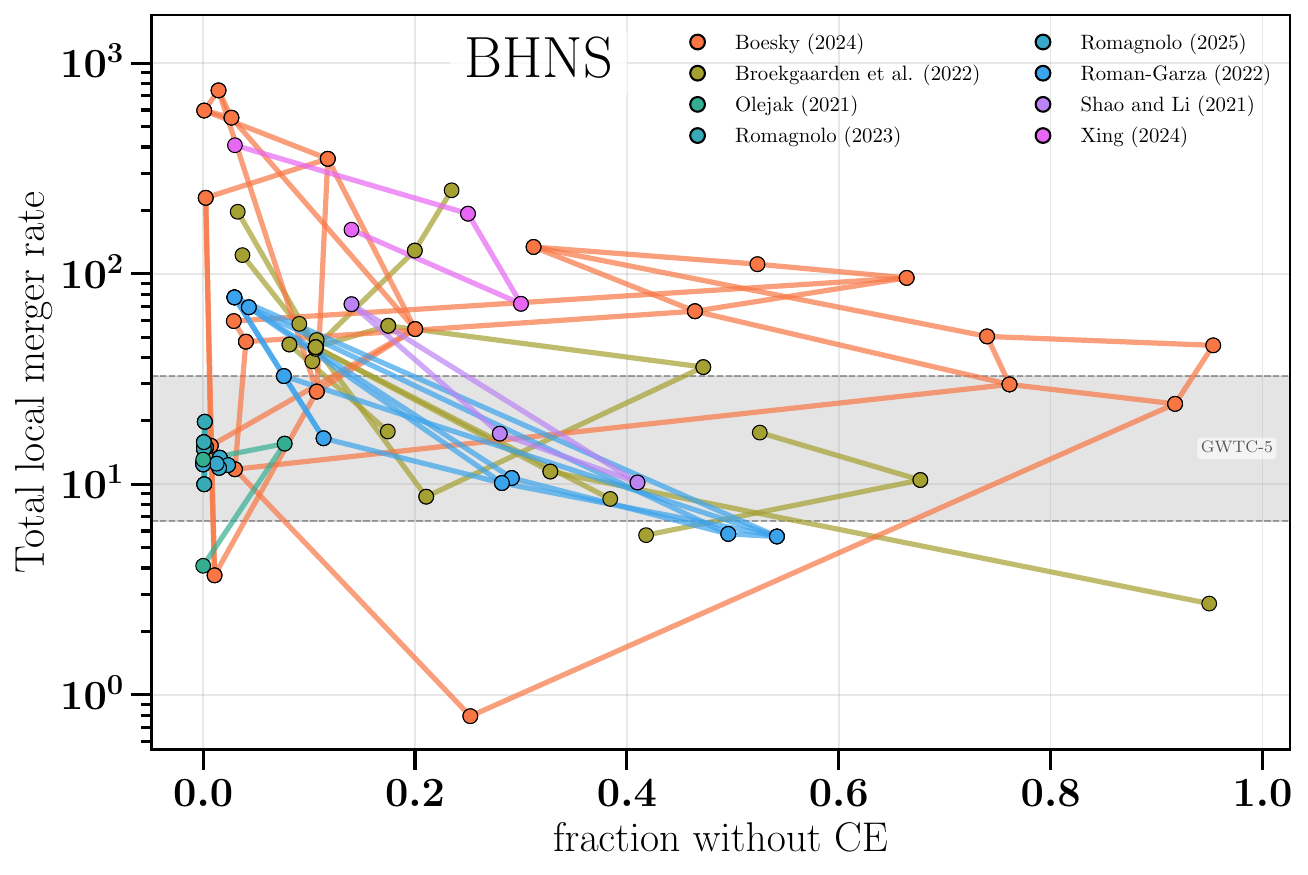} 
\includegraphics[width=\columnwidth]{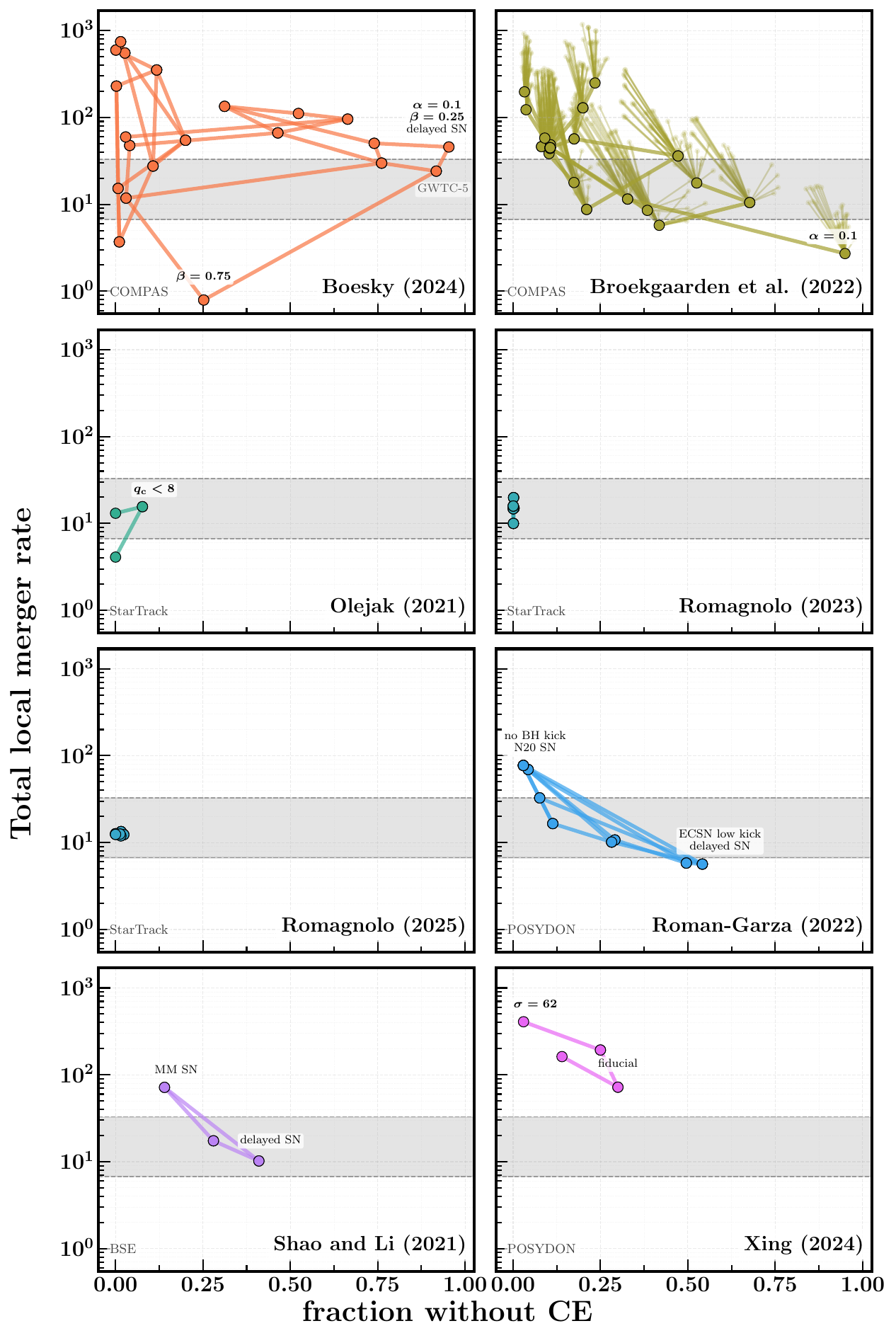} 
\caption{Same as Figure~\ref{fig-comparison-bbh-simulation-silo} for BHNS.
     For \citet{Broekgaarden2022} additional lines are added for variations in the star formation history models.
    See  \href{https://floorbroekgaarden.github.io/Rates_of_Formation_Channels/interactive_figures_and_tables/formation_channel_rates_table.html}{interactive figure} and  \href{https://github.com/FloorBroekgaarden/Rates_of_Formation_Channels}{GitHub} for details and code.}
    \label{fig-comparison-bhns-simulation-silo}
\end{figure}

Figures~\ref{fig-comparison-bbh-simulation-silo} (BBH) and \ref{fig-comparison-bhns-simulation-silo} (BHNS) illustrate the Level~1 channel contributions across the compiled simulations and parameter variations by connecting, within each study, models that differ by only a single parameter.  
This one-at-a-time variation strategy is standard practice in population-synthesis studies and is widely used to probe the sensitivity of predictions to uncertain aspects of massive binary evolution \citep[e.g.][]{Belczynski2010, Chruslinska:2018, Broekgaarden2022}. 
More recent work, however, has begun to explore simultaneous variations in multiple parameters \citep[e.g.,][]{DorozsmaiToonen:2022, Stevenson:2022, Boesky:2024gw, Boesky:2024popsynth}.

A perhaps counterintuitive result emerging from Figures~\ref{fig-comparison-bbh-simulation-silo} and ~\ref{fig-comparison-bhns-simulation-silo} is that a large increase in the relative contribution of the without-\ac{CE} formation channel does not necessarily correspond to a lower total merger rate (or a lower absolute merger rate from with-\ac{CE} channels; see Appendix~\ref{sec:appendix-absolute-merger-rates-parameter-variations}).
At first glance, one might expect the opposite behavior for two reasons. First, an increased contribution from without-\ac{CE} pathways could arise if binaries that would otherwise undergo a \ac{CE} phase instead evolve through stable (or no) interactions without entering a \ac{CE}. Second, avoiding \ac{CE} evolution often produces wider post-interaction binaries and therefore longer delay times, potentially reducing the fraction of systems that merge within a Hubble time (see Section~\ref{sec:discussion-observational-signatures}).

As shown in Figures~\ref{fig-comparison-bbh-simulation-silo} and \ref{fig-comparison-bhns-simulation-silo}, several simulations instead exhibit simultaneous increases in both the without-\ac{CE} fraction and the total merger rate. 
In particular for \acp{BBH}, some parameter variations produce a transition from predominantly with-\ac{CE} to predominantly without-\ac{CE} formation while maintaining comparable, or even higher, merger rates, indicating that the without-\ac{CE} contribution itself can increase substantially rather than merely replacing suppressed \ac{CE} systems. 
Overall, the lack of a clear monotonic relationship between formation-channel fractions and total merger rates highlights the difficulty of constraining formation pathways using merger rates alone. 
For \acp{BHNS}, some models do show lower merger rates at larger without-\ac{CE} fractions, although numerous exceptions remain.

\subsubsection{Key parameters driving formation-channel diversity}

A striking feature of Figures~\ref{fig-comparison-bbh-simulation-silo} and~\ref{fig-comparison-bhns-simulation-silo} is that some simulations tend to cluster within relatively narrow regions of parameter space when considered within a single code framework, while the largest differences emerge between different population-synthesis codes or a limited subset of specific parameter variations \citep[e.g.][]{Li:2025, Romagnolo:2023}. 
This behavior highlights what we refer to as population synthesis `simulation silos': variations within a given code, particularly single-parameter perturbations around a fiducial model, might probe only a limited region of the broader uncertainty space \citep[cf.][]{Broekgaarden:2026-lower-your-rates}.
In contrast, changes to the underlying physical framework (such as adopting a different population-synthesis code) or modifying a small number of particularly influential assumptions can lead to qualitatively different outcomes.

\begin{figure}
    \centering
    \includegraphics[width=1\columnwidth]{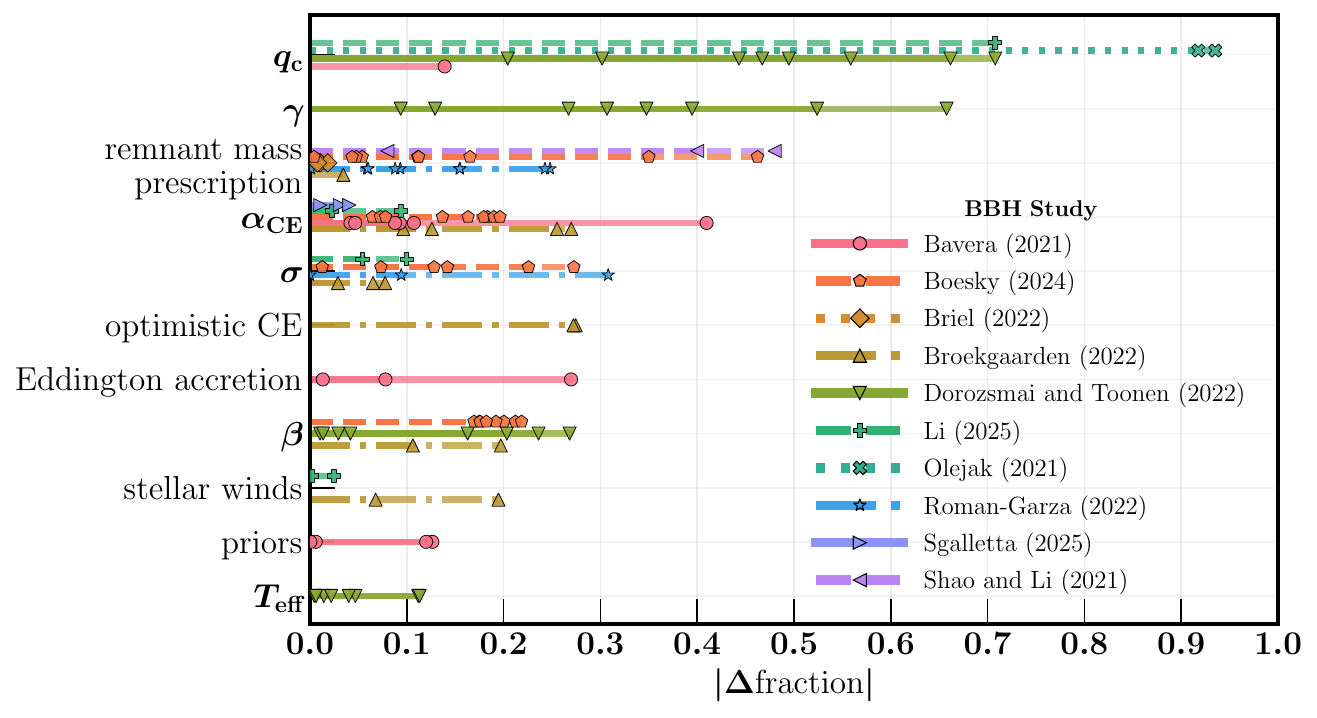}
    \includegraphics[width=1\columnwidth]{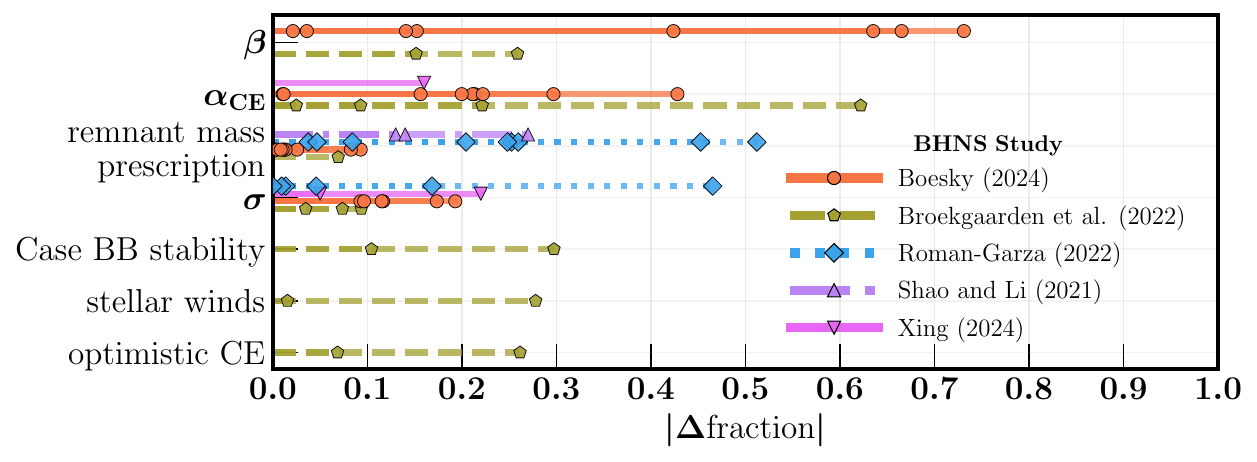}
\caption{
Change in the fraction of systems forming without a CE phase as a function of parameter-family variations for BBH (top) and BHNS (bottom) populations. Each scatter point shows the relative change in the without-CE fraction within a given study when varying a single model parameter, grouped by parameter family. Scatter points correspond to individual simulation variations within each family, while horizontal lines indicate the range of changes. Small vertical offsets (jitter) are applied to overlapping lines within the same parameter family for visual clarity and do not carry physical meaning. The figure only shows parameter variations that lead to changes of $|\Delta \mathrm{fraction}| \geq 0.1$.  See  \href{https://floorbroekgaarden.github.io/Rates_of_Formation_Channels/interactive_figures_and_tables/formation_channel_rates_table.html}{interactive figure} and  \href{https://github.com/FloorBroekgaarden/Rates_of_Formation_Channels}{GitHub} for details and code.%See text for symbols and acronyms.
}
\label{fig:BH-BH-without-CE-key-parameters-impact}
\end{figure}

Across the BBH and BHNS simulations, the diversity in predicted formation-channel contributions can be described primarily in the context of several key physical assumptions that regulate (i) the stability of mass transfer and (ii) the efficiency of orbital tightening. This is highlighted in  Figure~\ref{fig:BH-BH-without-CE-key-parameters-impact} that quantifies the parameter impacts on the formation channel contributions across BBH and BHNS for parameters that affect it by at least an absolute change in the without-CE formation channel contribution of $\Delta \rm{fraction} \gtrsim 0.1$ within a study, and Figures~\ref{fig:BH-BH-without-CE-parameter-axis} (BBH) and~\ref{fig:BH-NS-without-CE-parameter-axis} (BHNS) that quantify the CE contribution as a function of the parameter for a range of population synthesis parameters. 
Below we summarize the dominant parameters and their impact on BBH and BHNS populations across the simulations, noting that some aspects of this discussion are necessarily limited by the set of models included in our comparison. % (see Section~\ref{sec:fc-discussion} for further detail).

\begin{itemize}
%%%%%%%%%%%%%%%%%%%%%%%%%%%%%%%%%%%%%%%%%%%%%%%%%%%%%%%%%%%%%%%%%%%%%%%%%%%%%%%%%%%%%%%%%%
\item \textit{\textbf{Mass-transfer stability criteria regulate whether systems avoid CE, with models adopting higher $q_{\rm c}$ or $\xi_{\rm ad, rad}$ values leading to significantly higher without-CE contributions without necessarily lowering the total merger rate.}}
The assumed critical mass ratio for stability ($q_{\rm c}$), or equivalently the donor response parameter $\xi_{\rm ad, rad}$\footnote{$\xi_{\rm ad, rad} = (d \log R_{\rm{d}} / d \log M )_{\rm{ad}}$  is the mass radius exponent of giants with radiative envelopes \citep[with $R_{\rm{d}}$ the donor radius,][]{Soberman:1997}.}, sets whether Roche-lobe overflow proceeds stably or leads to CE evolution\footnote{There are notable exceptions to the $q_\mathrm{c}$ formalism which explicitly model the donor's response to mass loss with detailed stellar models in population synthesis codes such as \textsc{BPASS} \citep{Eldridge:2017} and \textsc{POSYDON} \citep{Fragos+2023ApJS,Andrews+2025ApJS}, see Appendix~\ref{app:fc_data-specs-acronyms} for details.}. Models that adopt higher values for $q_{\rm c}$ or $\xi_{\rm ad, rad}$ significantly expand the region of parameter space in which binaries avoid CE evolution.

For BBHs, this leads to some of the largest (outlier) variations in the without-CE fraction in Figure~\ref{fig-comparison-bbh-simulation-silo}, with models adopting high stability thresholds (corresponding to $q_{\rm c} \gtrsim 3$--$8$) finding that the majority of systems evolve through stable mass transfer in the simulations from \citet{Olejak:2021CE, DorozsmaiToonen:2022, Li:2025}, and overall, stability criteria impact the without CE BBH fraction by up to unity (Figure~\ref{fig:BH-BH-without-CE-key-parameters-impact}).
Our overview shows that such changes in the stability criterion do not necessarily reduce the total BBH rate, but often yield comparable, or even higher, BBH merger rates. This likely reflects that increased mass-transfer stability shifts parts of the parameter space that would previously have merged during a \ac{CE} episode into systems that can instead survive and form merging BBHs through stable mass transfer \citep[cf.][]{Gallegos-Garcia2021}.
 For BHNS systems, \citet{Olejak:2021CE} has explored this, finding a modest rise in the fraction forming without CE.  The simulations from \citet{Broekgaarden2022} show that by changing case BB mass transfer to always be dynamically unstable (instead of always stable) mass transfer, can reduce the with-CE BHNS fraction by inducing mergers during this phase.

Another way to modify the effective stability boundary is to vary the assumptions for when stars develop convective envelopes, which directly affects the adopted critical mass ratio. 
For example, \citet{DorozsmaiToonen:2022, Romagnolo:2025} vary the effective temperature $(T_{\rm{eff}})$ as a proxy for the evolutionary stage at which a convective envelope forms. However, this variation has only a modest impact, changing the total BBH fraction forming without CE by at most $\sim 10\%$ (Figure~\ref{fig:BH-BH-without-CE-key-parameters-impact}).
This relatively weak sensitivity likely reflects that a large fraction of BBH progenitors in the without-CE channels undergo Case A or early Case B mass transfer, before the donor develops a deep convective envelope. As a result, their mass-transfer stability is largely unaffected by changes to the assumed convective-envelope boundary \citep[cf.][]{DorozsmaiToonen:2022, Briel:2026}.

%%%%%%%%%%%%%%%%%%%%%%%%%%%%%%%%%%%%%%%%%%%%%%%%%%%%%%%%%%%%%%%%%%%%%%%%%%%%%%%%%%%%%%%%%%
\item \textit{\textbf{Angular-momentum loss ($\gamma$) is a key parameter regulating the dominant BBH formation channel.}}
The efficiency with which mass loss removes angular momentum from the binary, often parameterized through $\gamma$-like prescriptions, has a first-order impact on the relative contribution of with- and without-CE BBH formation channels \citet[e.g.][]{Marchant:2021, Willcox:2023, DorozsmaiToonen:2022}.
As shown in Figure~\ref{fig:BH-BH-without-CE-key-parameters-impact}, the $\gamma$ variations of \citet{DorozsmaiToonen:2022} produce some of the largest shifts among all parameter studies: increasing $\gamma$ from 1 to 2.5 substantially increases the without-CE contribution and, in some models, changes the population from predominantly CE-driven to without-CE dominated (Figure~\ref{fig:BH-BH-without-CE-parameter-axis}).
In these models, $\gamma$ represents the specific angular momentum carried away by mass lost during non-compact-object mass transfer, normalized to the binary's specific angular momentum \citep[cf.][]{PortegiesZwart:1996}.
Physically, higher specific angular-momentum loss (higher $\gamma$) leads to more efficient orbital shrinkage during stable mass transfer, allowing binaries to merge within a Hubble time without a CE phase, whereas lower specific angular-momentum loss (lower $\gamma$) causes orbital widening and suppresses merging BBHs through the without-CE channel \citep{DorozsmaiToonen:2022}.
This connects directly to the ongoing debate over whether stable mass transfer can simultaneously remain stable and shrink the orbit sufficiently to produce merging BBHs \citep{vanSon:2022, Willcox:2023, Klencki:2026-smt}.

%%%%%%%%%%%%%%%%%%%%%%%%%%%%%%%%%%%%%%%%%%%%%%%%%%%%%%%%%%%%%%%%%%%%%%%%%%%%%%%%%%%%%%%%%%
\item \textit{\textbf{Common-envelope efficiency ($\alpha_{\rm CE}$) values in the range $\sim0.5$--$2$ typically produce the lowest without-CE contribution for BBH and BHNS mergers.}}
The CE efficiency parameter $\alphaCE$ determines how efficiently orbital energy is used to eject the envelope during CE evolution and therefore whether binaries survive the CE phase and emerge in sufficiently tight orbits to merge within a Hubble time.
Figure~\ref{fig:BH-BH-without-CE-key-parameters-impact} shows that, within the simulations included in our compilation, variations in $\alphaCE$ can change the without-CE fraction by up to $\sim0.4$ for BBHs and $\sim0.6$ for BHNSs, contributing to some of the outliers in Figures~\ref{fig-comparison-bbh-simulation-silo} and~\ref{fig-comparison-bhns-simulation-silo}.

Across most studies, both low and high $\alphaCE$ values reduce the with-CE BBH and BHNS rates and thereby increase the relative importance of without-CE pathways, with enhanced without-CE fractions consistently found outside the canonical range of $\sim0.5$--$2$ (Figures~\ref{fig:BH-BH-without-CE-parameter-axis} and~\ref{fig:BH-NS-without-CE-parameter-axis}; based on \citealt{Bavera:2021, Broekgaarden2022, Boesky:2024gw, Sgalletta2024, Li:2025, Xing:2024-BHNS-allZ}).
This  ``sweet spot'' arises because very low $\alphaCE$ values lead to more binaries merging during the CE phase as more orbital energy is required, while very high $\alphaCE$ values produce post-CE separations that are too wide for the systems to merge within a Hubble time \citep[e.g.][]{Bavera:2021, Broekgaarden2022, Kruckow:2018}.
In other words, inefficient CE ejection (small $\alphaCE$) requires binaries to deposit more orbital energy into the envelope, increasing the likelihood of stellar mergers during CE, whereas highly efficient CE ejection (large $\alphaCE$) leaves binaries at wider separations and reduces the merger rate because fewer systems inspiral within a Hubble time.

A small exception to this trend is the $\alphaCE=0.2$ model of \citet{Bavera:2021}, which exhibits an additional dip in the without-CE fraction due to an enhanced with-CE BBH merger rate at this specific CE efficiency.
The authors find that, for this model, the post-CE orbital separations of BH--star systems peak in a regime that efficiently produces BBHs with short merger times, while still avoiding mergers during the CE phase.
This corresponds to a second ``sweet spot,'' distinct from the broader $\alphaCE \sim 0.5$--$2$ regime, which instead reflects the tails of the post-CE separation distribution (see their Figure~3).
Overall, this highlights that the dependence on $\alphaCE$ is complex and not strictly monotonic: although global trends favor reduced CE contributions at extreme $\alphaCE$ values, specific combinations of initial conditions and orbital evolution (likely correlated with the physical model assumptions for stellar expansion and mass transfer stability) can produce localized enhancements in the with-CE channel. 

Given the substantial uncertainties in CE evolution and the growing recognition that the standard $\alphaCE\lambda$ formalism provides only a highly simplified description of the underlying physics \citep[e.g.,][]{Fragos:2019, IaconiDeMarco:2019, Ivanova:2020book, HiraiMandel:2022, Lau2022}, it is important for future studies to explore a broad range of CE efficiencies, including potentially correlated variations in $\alphaCE$, when assessing their impact on BBH, BHNS, and BNS merger populations.
We further note that comparisons between studies are complicated by differences in the adopted envelope-binding-energy parameter $\lambda$, which can vary substantially between population-synthesis frameworks depending on whether fixed values or stellar-structure-dependent fitting formulae are used \citep[e.g.][]{XuLi:2010, Marchant:2021, IaconiDeMarco:2019, Ivanova:2020book, Sgalletta:2026}.
Since binary evolution modeling such as the mapping between pre- and post-CE orbital separations depend primarily on the combination $\alphaCE\lambda$, variations in $\lambda$ can significantly muddy direct comparisons between nominal $\alphaCE$ assumptions across different studies and codes. 
In this context, it is perhaps notable that many studies nevertheless consistently find a minimum in the without-CE contribution specifically around the same range of  $\alphaCE \sim 0.5$--$2$, despite substantial differences in the underlying CE implementations across population-synthesis frameworks.

%%%%%%%%%%%%%%%%%%%%%%%%%%%%%%%%%%%%%%%%%%%%%%%%%%%%%%%%%%%%%%%%%%%%%%%%%%%%%%%%%%%%%%%%%%
\item \textit{\textbf{Mass-transfer efficiency ($\beta$) can strongly alter the relative importance of with- and without-CE formation channels, particularly for BHNS mergers.}}
The mass-transfer efficiency parameter $\beta$, which describes the fraction of transferred mass accreted during mass transfer between two non-compact stars, affects both the stability of Roche-lobe overflow and the orbital evolution of the binary. 
In the population-synthesis simulations considered here, the mass-transfer efficiency $\beta$ is typically either modeled as a function of the thermal timescale of the accretor star \citep[e.g.][]{COMPAS:2022}, or assumed to take a fixed value throughout the evolution \citep[e.g.][]{Belczynski:2008}. In Figure~\ref{fig:BH-BH-without-CE-key-parameters-impact} and ~\ref{fig:BH-BH-without-CE-parameter-axis} we consider for $\beta$ the (subset of) model variations where $\beta$ is varied between fixed values.
The impact of changing $\beta$ is intrinsically complex because increasing $\beta$ can influence binary evolution in several competing ways \citep{DorozsmaiToonen:2022, Willcox:2023}.

On the one hand, more efficient accretion increases the mass of the companion accretor star, which can help binaries remain bound following the first supernova  \citep[e.g.][]{Renzo:2019, Renzo:2021, Broekgaarden2022}.
Efficient accretion during the first mass-transfer phase can also generate highly unequal BH--star binaries (where the star can now be much more massive than the BH because it accreted efficiently earlier). Such systems might be particularly effective at shrinking their orbits during subsequent reverse stable mass transfer, thereby enhancing the formation of merging BBH and BHNS systems through stable pathways \citep[e.g.][]{Olejak:2024, Klencki:2026-smt, Schurmann:2025}.

On the other hand, more conservative mass transfer can also destabilize Roche-lobe overflow. Increased $\beta$ causes the accretor to respond more strongly to the incoming mass and can lead to stronger orbital contraction during the transfer phase, effectively lowering the critical mass ratio for stable mass transfer and making runaway mass transfer more likely for a wider range of donor-accretor mass ratios \citep[e.g.][]{DorozsmaiToonen:2022, vanSon:2022-nopeaks}.
Moreover, higher $\beta$ reduces the amount of mass and angular momentum lost from the system, which can limit orbital shrinkage in some evolutionary phases and prevent binaries from reaching sufficiently tight separations to merge within a Hubble time \citep[e.g.][]{Willcox:2023, vanSon:2022-nopeaks}.
The relative importance of these effects depends sensitively on the evolutionary stage, donor structure, and the modeling assumptions for mass transfer and angular momentum loss.

As a result, variations in $\beta$ produce markedly different trends in the without-CE contribution across studies, as illustrated in Figure~\ref{fig:BH-BH-without-CE-parameter-axis}. For BBHs, the COMPAS-based models of \citet{Broekgaarden2022} and \citet{Boesky:2024gw} show that increasing $\beta$ generally increases the fraction of mergers formed without CE. In contrast, the SeBa-based models of \citet{DorozsmaiToonen:2022} exhibit little change, or even a decrease, in the without-CE contribution with increasing $\beta$. These contrasting trends demonstrate that the impact of $\beta$ cannot be understood solely in terms of its effect on the immediate stability of a mass-transfer episode. Rather, $\beta$ influences the subsequent evolution of both stars and the orbit, with the net effect depending on the interplay of multiple mass-transfer phases, stellar rejuvenation, orbital evolution, and compact-object formation.

For BHNS systems, the models in our compilation that vary $\beta$ \citep{Boesky:2024gw, Broekgaarden2022} exhibit the opposite trend from their BBH populations: increasing $\beta$ decreases the without-CE contribution (Figure~\ref{fig:BH-NS-without-CE-parameter-axis}), changing the without-CE BHNS fraction by up to $\sim0.75$ (Figure~\ref{fig:BH-BH-without-CE-key-parameters-impact}). In these models, the trend is driven by a decrease in the absolute without-CE formation rate, accompanied by a modest increase in the with-CE rate\footnote{See \href{https://floorbroekgaarden.github.io/Rates_of_Formation_Channels/interactive_figures_and_tables/formation_channel_rates_table.html}{interactive figures}.}. Notably, these are the same models in which increasing $\beta$ leads to a larger without-CE contribution for BBH mergers. This contrast demonstrates that identical parameter variations can shift the relative importance of formation pathways in opposite directions for different compact-object populations. The differing response likely reflects the distinct binary configurations that produce BBH and BHNS mergers, including differences in typical mass ratios, compact-object formation histories, and the sequence of subsequent binary-interaction phases.

\begin{figure}
    \centering
    \includegraphics[width=1\columnwidth]{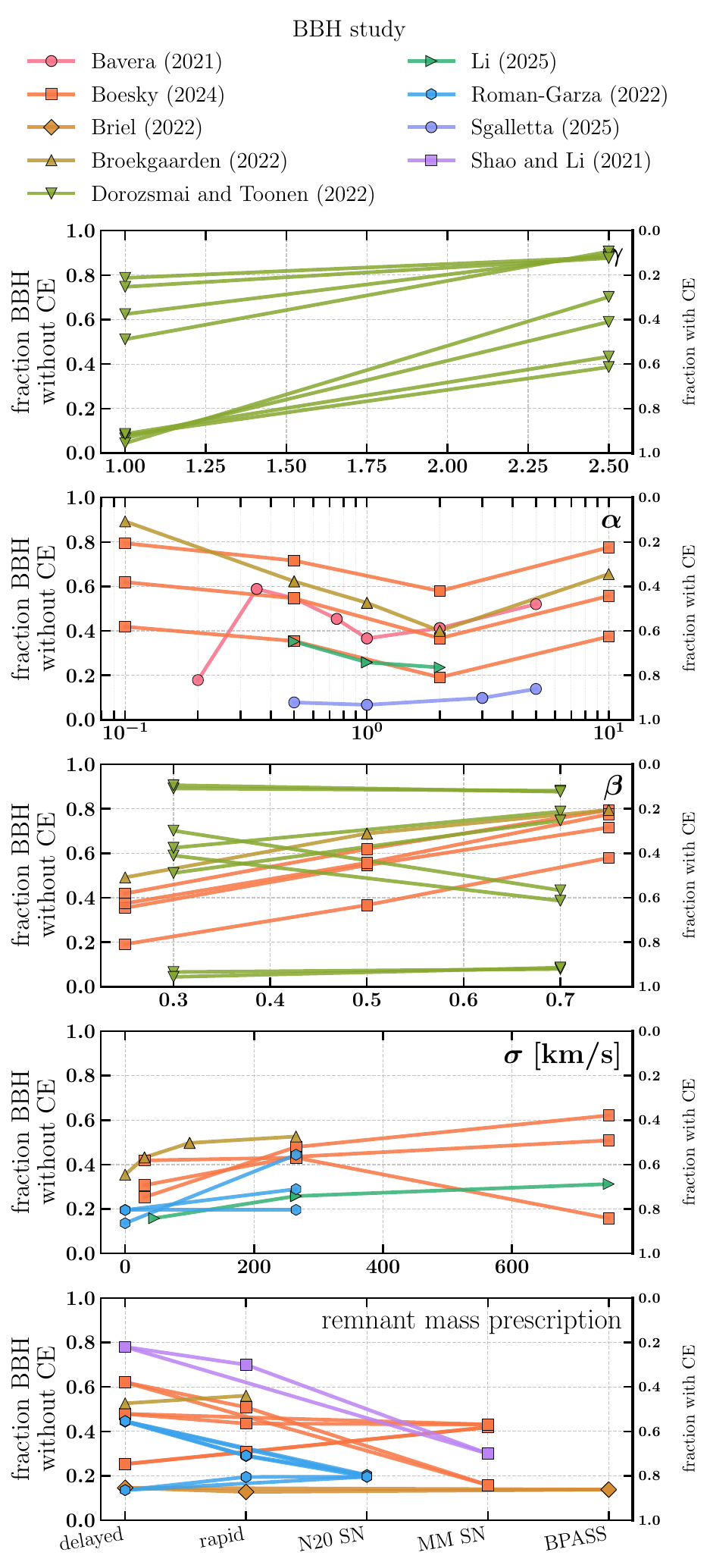}
\caption{
Dependence of the fraction of BBH mergers forming without a CE phase on commonly varied binary-evolution parameters. Each panel shows simulations in which a single parameter is varied while other assumptions are kept fixed within the same population-synthesis study. Shown parameters include the angular momentum loss parameter $\gamma$, CE efficiency parameter $\alpha_{\rm CE}$, the mass-transfer efficiency $\beta$, supernova natal kick magnitude parameter $\sigma$, and the adopted supernova remnant prescription. Each point represents a simulation, with lines connecting models belonging to the same simulation grid to highlight controlled (single) parameter variations. 
For the remnant-mass prescription panel, the x-axis represents discrete model assumptions rather than a continuous parameter range. Consequently, individual parameter variations may connect non-adjacent x-axis positions, leading to triangular connection patterns within a given study. 
 Multiple lines for a given study indicate simultaneous variation of additional parameters (i.e., multi-dimensional grids).
 See \href{https://floorbroekgaarden.github.io/Rates_of_Formation_Channels/interactive_figures_and_tables/formation_channel_rates_table.html}{interactive figure} and  \href{https://github.com/FloorBroekgaarden/Rates_of_Formation_Channels}{GitHub} for details and code.}
\label{fig:BH-BH-without-CE-parameter-axis}
\end{figure}

\begin{figure}
    \centering
    \includegraphics[width=1\columnwidth]{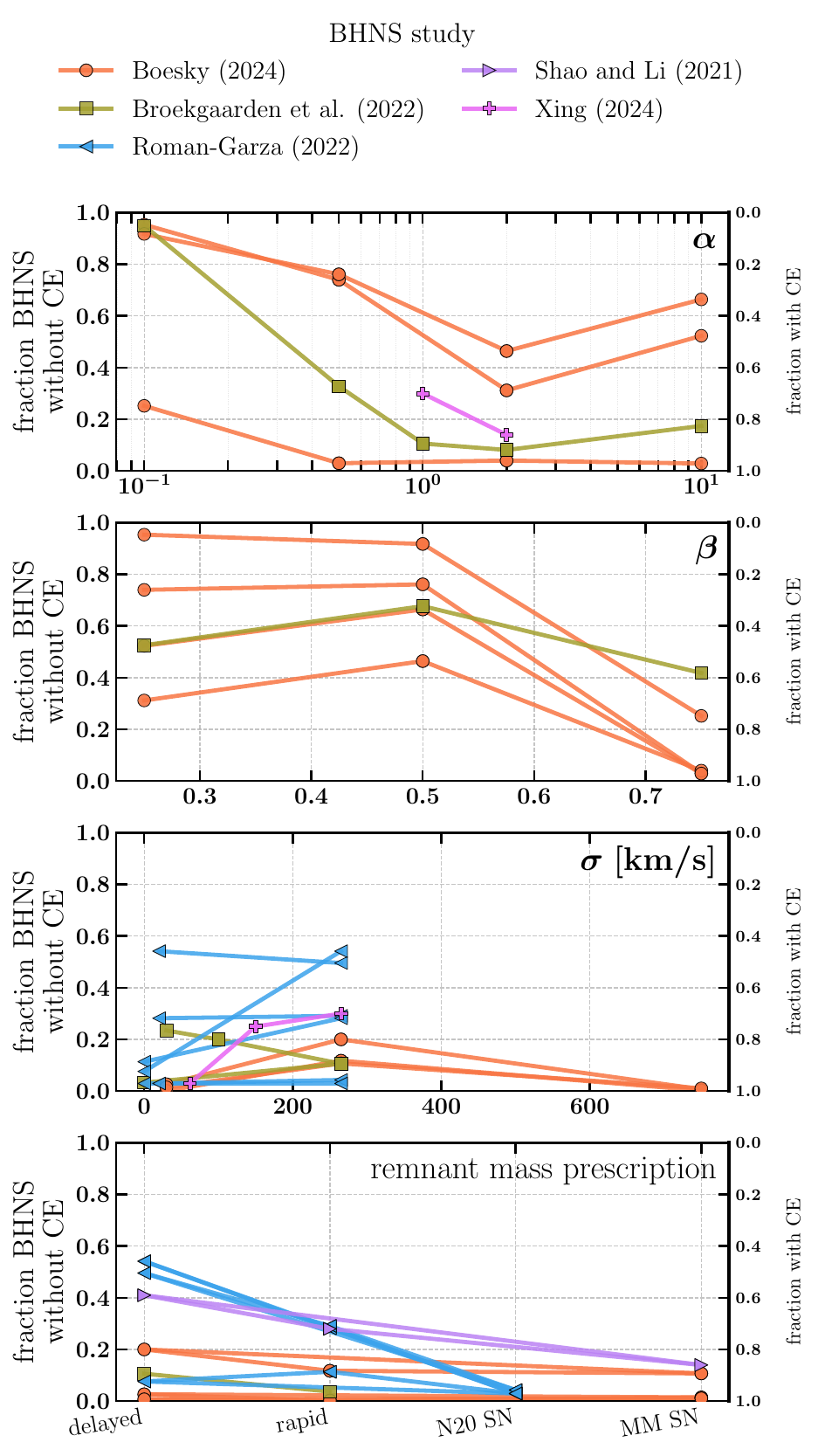}
\caption{
Same as Figure~\ref{fig:BH-BH-without-CE-parameter-axis} for BHNS mergers. The panel with $\gamma$ is not shown as no BHNS simulations include variations in $\gamma$. 
For the $\sigma$ panel, the quoted $\sigma$ values are not always defined identically across studies. In some models the natal-kick dispersion is applied to both NSs and BHs, while in others it is applied only to NS kicks, only a subset of NS kicks (e.g. only electron-capture supernovae) or only to BH kicks.  See  \href{https://floorbroekgaarden.github.io/Rates_of_Formation_Channels/interactive_figures_and_tables/formation_channel_rates_table.html}{interactive figure} and  \href{https://github.com/FloorBroekgaarden/Rates_of_Formation_Channels}{GitHub} for details and code.
}
\label{fig:BH-NS-without-CE-parameter-axis}
\end{figure}

%%%%%%%%%%%%%%%%%%%%%%%%%%%%%%%%%%%%%%%%%%%%%%%%%%%%%%%%%%%%%%%%%%%%%%%%%%%%%%%%%%%%%%%%%%

\item \textbf{\textit{The magnitude of natal kicks imparted to compact objects, parameterized by $\sigma$, plays an important role in setting the relative contribution of with- and without-CE formation channels for BBH and BHNS mergers.}}
We find that the assumed distribution of compact-object natal kick magnitudes, parameterized with $\sigma$, can alter the without-CE contribution by up to $\sim0.3$ for BBHs and $\sim0.5$ for BHNSs (Figure~\ref{fig:BH-BH-without-CE-key-parameters-impact}).
Natal kicks affect both binary survival and the post-supernova orbital evolution of the system.
Higher kick magnitudes (larger $\sigma$ values) increase the probability that binaries are disrupted during compact-object formation and therefore reduce the number of systems that remain bound.
At the same time, natal kicks can significantly alter the post-supernova orbital properties by changing the orbital separation and inducing eccentricity, which can either promote or suppress mergers within a Hubble time depending on the kick magnitude, direction, and pre-supernova binary configuration.

For BBH mergers, models with low or zero BH kicks generally produce higher absolute merger rates for both with- and without-CE channels because fewer binaries are disrupted during compact-object formation.
Since the natal-kick prescriptions in population synthesis models primarily affect low-mass BHs, this trend also reflects that the BBH merger-rate distribution peaks at relatively low BBH masses. 
Moderate kicks can preferentially enhance the without-CE contribution by tightening binaries that would otherwise remain too wide to merge within a Hubble time.
As a result, several studies find somewhat larger without-CE BBH fractions for moderate or high kick dispersions ($\sigma \gtrsim 100,\kms$), with increases of $\sim0.2$--$0.4$ in the relative without-CE contribution (Figure~\ref{fig:BH-BH-without-CE-parameter-axis}).
This behavior reflects that many of the without-CE channels often lack the strong orbital tightening associated with a CE phase, making natal kicks particularly important for shrinking the orbit sufficiently to produce merging BBHs \citep[e.g.,][]{Klencki:2026-smt, Chattaraj:2026}.

A related trend is seen for BHNS systems in Figure~\ref{fig:BH-NS-without-CE-parameter-axis}, where moderate natal kicks ($\sigma \sim 100$--$300,\kms$) can enhance the without-CE contribution by increasing the absolute without-CE BHNS merger rate in studies such as \citet{RomanGarza:2021, Xing:2025} and \citet{Boesky:2024gw}.
However, other simulations instead find little effect or even a decrease in the without-CE contribution for similar kick assumptions \citep[e.g.][]{Broekgaarden2022, RomanGarza:2020}.
These differences highlight the complex interplay between natal kicks and other binary-evolution assumptions, including mass transfer, orbital separations, and compact-object formation physics, in determining the dominant BBH and BHNS formation pathways.

%%%%%%%%%%%%%%%%%%%%%%%%%%%%%%%%%%%%%
%%%%%%%%%%%%%%%%%%%%%%%%%%%%%%%%%%%%%
\item \textit{\textbf{Remnant-mass prescriptions strongly affect the CE contribution to BBH and BHNS formation.}}
The adopted compact object remnant mass prescription can play a major role in determining the dominant formation pathway for BBH and BHNS mergers, changing the without-CE contribution by up to $\sim 0.5$ for both populations (Figure~\ref{fig:BH-BH-without-CE-key-parameters-impact}).

For BBH mergers, Figure~\ref{fig:BH-BH-without-CE-parameter-axis} shows that changing the remnant-mass prescription can alter the without-CE contribution significantly by $\sim10$--$50\%$ in several studies \citep{RomanGarza:2020, Broekgaarden2022, Boesky:2024gw, Shao2021}, but with no clear consistent correlation between remnant-mass prescription and CE formation-channel contribution. However, we also find that the \citet{Briel:2022}, \citet{Broekgaarden2022}, and the $\sigma=265\kms$ models of \citet{Boesky:2024gw} show that changing the remnant-mass prescription alters the without-CE fraction by only $\lesssim5\%$. This wide range of responses suggests that remnant-mass prescriptions do not influence formation channels in isolation. Instead, their effect is mediated through their coupling to other aspects of binary evolution, most notably natal kicks, but also mass-transfer physics and stellar-evolution prescriptions, such that the same remnant-mass variation can produce markedly different outcomes even within otherwise similar BBH populations.

For BHNS mergers, Figure~\ref{fig:BH-NS-without-CE-parameter-axis} shows that most studies find larger without-CE contributions when adopting the delayed remnant-mass prescription of \citet{Fryer:2012} rather than the rapid prescription. The primary exception is \citet{RomanGarza:2020}, who find a slightly larger without-CE contribution for the rapid model. The magnitude of the effect also varies substantially across studies: while some report only weak sensitivity to the remnant-mass prescription ($\lesssim 5\%$ changes in the without-CE fraction; e.g., \citealt{Boesky:2024gw, Broekgaarden2022}), others find variations as large as $\sim 50\%$.

Taken together, the BBH and BHNS results indicate that the impact of remnant-mass prescriptions on formation-channel demographics is highly framework dependent. Although remnant prescriptions directly modify compact-object masses and fallback fractions, their effect on the relative importance of CE and without-CE pathways emerges through their coupling to other ingredients of the binary-evolution model, including natal kicks, mass-transfer physics, and stellar evolution. Consequently, trends inferred from remnant-mass variations within a single population-synthesis framework may not generalize across codes or even across different compact-object populations.

\item \textbf{\textit{Variations in the assumed star formation history primarily affect the normalization of the merger rate.}}
Finally, changes in the adopted star formation history mainly affect the normalization of the merger rate, often by up to an order of magnitude, while leaving the relative formation-channel contributions largely unchanged. 
This is seen for the models of  \citep{Broekgaarden2022} and  \citet{vanSon2023} in Figures~\ref{fig:BH-BH-without-CE-parameter-axis} and~\ref{fig:BH-NS-without-CE-parameter-axis}. 
This is because the choice for the star formation history model primarily `reweights' the underlying population across cosmic time and metallicity, rather than fundamentally changing the balance between CE and without-CE formation pathways \citep[cf.][]{deSa:2024initial, Willcox:2025, Levina:2026}.

\end{itemize}

{We caution that the between-code differences identified here need not reflect physical assumptions alone: nominally identical prescriptions may be implemented differently, and numerical or algorithmic choices can themselves generate substantial differences (cf. inter-code comparisons such as \citealp{toonen2014popcorn, Belczynski:2021, Guerrero:2026}). Because the constituent studies were not designed as a controlled inter-code experiment, our compilation cannot separate implementation-driven differences from genuinely physical ones, which is itself an argument for dedicated benchmark comparisons (Section~\ref{sec:discussion-recommendations-pop-synth}). These codes were also developed by different groups, at different times, and often for different original goals; many have since been adapted to gravitational-wave populations, but these historical differences help explain why similar assumptions can yield different outcomes.}

{The frameworks also differ in a more basic way. Most are rapid population-synthesis codes (e.g.\ COMPAS, StarTrack, SeBa, BSE, binary\_c, MOBSE) that evolve stars through analytic or interpolated fitting formulae, whereas POSYDON and BPASS rely on detailed stellar-structure calculations, either through pre-computed MESA binary grids (POSYDON) or by running STARS models per system (BPASS).\footnote{The two earliest POSYDON-based studies in our compilation \citep{Bavera:2021, RomanGarza:2021} used a pre-release version combining MESA grids for binary-interaction phases with the rapid COSMIC (SSE) code elsewhere, whereas \citet{Xing:2024-BHNS-allZ} uses the grid-based POSYDON v1 framework \citep{Fragos+2023ApJS}.} This has two practical consequences. First, the number of models per study is uneven: BPASS in particular contributes few variations, not because its parameter space is narrow but because detailed calculations make large grids costly, which should be kept in mind when comparing point density across codes in Figures~\ref{fig-comparison-bbh-simulation-silo}--\ref{fig-comparison-bhns-simulation-silo}. Second, and more importantly, the binary-interaction physics setting the CE versus without-CE division is often handled differently. Rapid codes decide Roche-lobe overflow stability through parameterized criteria (a critical mass ratio $q_{\rm{c}}$ or adiabatic response exponent $\zeta$; see Appendix~\ref{app:fc_data-specs-acronyms}), typically keyed to donor type, whereas POSYDON and BPASS know the donor's internal structure at the onset of interaction and follow the stability and response directly. This is a qualitative rather than quantitative difference: agreement or disagreement in the CE fraction between a detailed and a rapid code can reflect the treatment of stellar structure itself, not just nominal parameter choices. 
We return to mass-transfer stability in Section~\ref{sec:discussion-mass-transfer-dominant-uncertainty} and CE prescriptions in Section~\ref{sec:discussion-CE-prescriptions}.}

Taken together, our results demonstrate that the contribution of CE and without-CE formation pathways remains highly sensitive to the underlying assumptions of stellar and binary evolution models, with substantial correlations and degeneracies between different model parameters. Moreover, trends inferred from varying a single parameter within a single population-synthesis framework do not necessarily generalize across codes, to models in which multiple parameters vary simultaneously, or even across different compact-object populations. \textit{Consequently, caution is warranted when interpreting formation-channel trends from a limited set of simulations or from any single population-synthesis framework.}

\section{Discussion}
\label{sec:fc-discussion}
Our compilation reveals a central challenge in interpreting gravitational-wave sources: merger-rate predictions alone do not uniquely constrain the evolutionary pathways of compact-object binaries. Across population-synthesis studies, models that reproduce similar intrinsic merger rates can correspond to fundamentally different formation histories, reflecting a strong degeneracy between rates and formation channels. 
At the same time, the degree of uncertainty varies across compact-object classes, with BBH and BHNS systems showing large diversity in formation pathways, while BNS mergers display strong agreement and are almost universally associated with CE evolution. In the following, we interpret these results in terms of the underlying binary-interaction physics and discuss how future simulation work in combination with future observations can break these degeneracies.

\begin{figure*}
    \centering
    \includegraphics[width=1\linewidth]{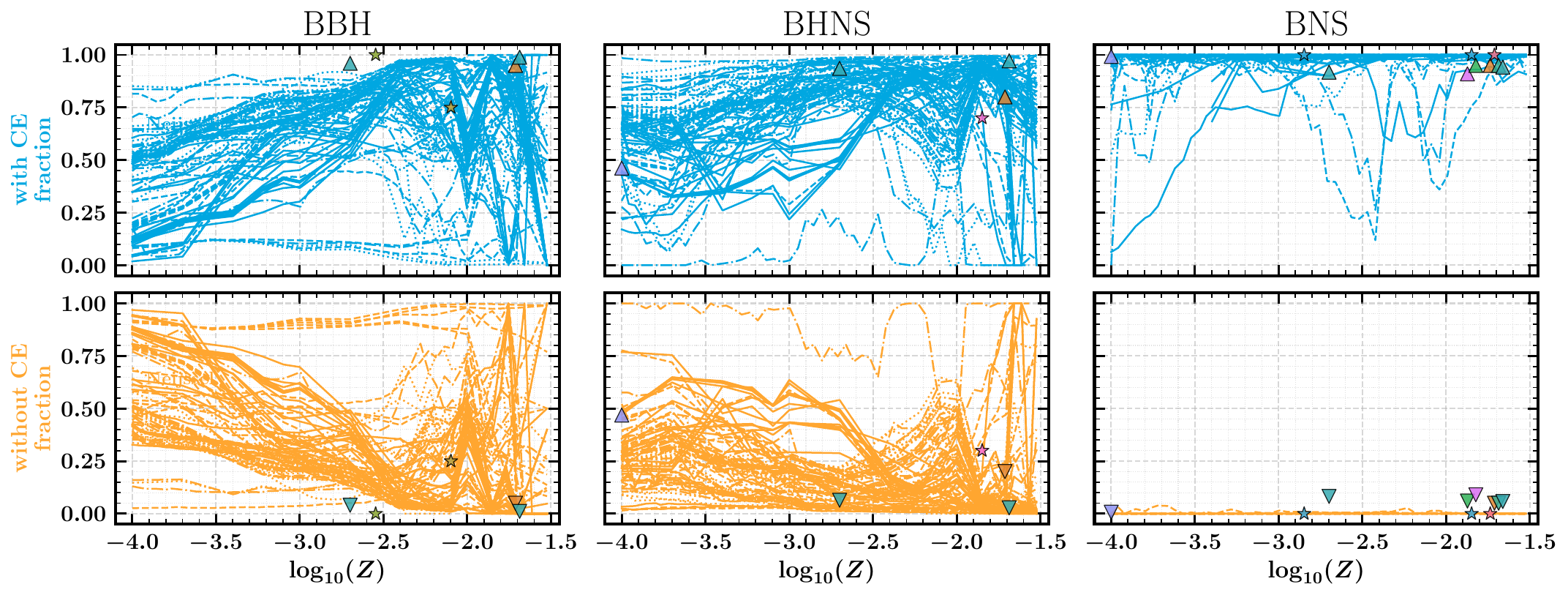}
\caption{
Formation-channel fractions as a function of metallicity for compact-object binaries across multiple population-synthesis studies and literature constraints. 
Models are shown from \citet{Ioro:2023sevn}, \citet{Broekgaarden2022}, \citet{vanSon:2024}, and \citet{Neijssel:2019} (a single BBH model), while colored markers indicate literature constraints at specific metallicities compiled in Table~\ref{tab:fc_Z_statements_bns}; arrows denote upper or lower limits and stars indicate approximate values. 
Columns correspond to the compact-object merger type (\ac{BBH}, \ac{BHNS}, and \ac{BNS}), while the top and bottom rows show the fractions of systems forming with and without a \ac{CE} phase, respectively.
The \citet{vanSon:2024} BBH models include contributions from chemically homogeneous evolution (\ac{CHE}), although these do not significantly alter the overall trends shown here. Some \citet{Broekgaarden2022} models additionally contain a non-negligible ``other'' channel for which the presence or absence of a \ac{CE} phase is not specified (leading to some with CE and without CE contributions adding up to less than 1). 
Low-number statistics typically introduces increased stochastic scatter at $\log_{10}(Z) \gtrsim -2$ for BBH and BHNS models.
All metallicities are converted to $\log_{10}(Z)$, and complementary fractions are inferred where needed assuming 
$f_{\rm with\,CE} + f_{\rm without\,CE} = 1$.
The figure highlights the large diversity in predicted \ac{BBH} and \ac{BHNS} formation-channel fractions across metallicity and population-synthesis frameworks. In contrast, all studies consistently predict that \ac{BNS} formation is strongly dominated by channels involving a \ac{CE} phase, with typically $\lesssim 10\%$ of systems forming without \ac{CE} evolution across the explored metallicity range.
For \ac{BBH} and \ac{BHNS} mergers, the relative contribution of without-\ac{CE} formation channels generally increases toward the lowest metallicities ($\log_{10}(Z) \lesssim -2.5$), particularly for \ac{BBH} systems, with the exception of a few models where the without contribution is always small.
See Appendix~\ref{sec:appendix-formation-channel-contributions-efficiency-metallicity} for additional discussion.  See  \href{https://floorbroekgaarden.github.io/Rates_of_Formation_Channels/interactive_figures_and_tables/formation_channel_rates_table.html}{interactive figure} and  \href{https://github.com/FloorBroekgaarden/Rates_of_Formation_Channels}{GitHub} for details and code.}
    \label{fig:formation-efficiency-results}
    \end{figure*}

%%%%%%%%%%%%%%%%%%%%%%%%%%
\subsection{A Robust Result: BNS Formation Requires CE}
\label{sec:discussion-bns-robust-result}

In contrast to the large diversity observed for BBH and BHNS systems, the formation of BNS mergers is strikingly consistent across models. All studies find that BNS systems form mainly through evolutionary pathways that include at least one CE phase, with only a negligible fraction (at most $\lesssim10\%$) forming without CE \citep{Boesky:2024gw, Broekgaarden2022, Romagnolo:2023, Pellouin:2025}. 
Although our compilation in Figure~\ref{fig:level-1-bns} includes a limited number of studies, primarily because many works do not report both intrinsic merger rates and formation channels for BNS, this conclusion is robust. The dominance of CE-mediated formation channels for BNS mergers is consistently reported across binary population-synthesis studies, including those beyond our compilation  such as studies that only report formation channel contributions for simulations at a fixed metallicity (see Table~\ref{tab:fc_Z_statements_bns} for an overview).

To highlight this, Figure~\ref{fig:formation-efficiency-results} shows the formation-channel fractions as a function of metallicity, drawing on formation efficiency data from \citet{Ioro:2023sevn}, \citet{Broekgaarden2022}, and \citet{vanSon:2024}, supplemented by the BBH formation efficiency of \citet{Neijssel:2019} and literature constraints at individual metallicities (Table~\ref{tab:fc_Z_statements_bns}; including \citealt{Belczynski:2002, Andrews:2015, Chruslinska:2018, Dominik:2012, VignaGomez:2018, Tanaka:2023, Mestichelli:2025, vanSon:2024, Ioro:2023sevn, Chattaraj:2026}). Across all studies and metallicities, \ac{BNS} mergers are consistently dominated by channels involving at least one \ac{CE} phase, while \ac{BBH} and \ac{BHNS} systems show large diversity in their formation-channel fractions.

This result of CE for BNS mergers can be understood from simple binary-evolution considerations. At some stage, BNS progenitors must evolve through a NS–star binary configuration \citep[e.g.][]{Tauris:2015, tauris2017formation}. Forming a merging system requires a sufficiently tight orbit prior to the second supernova. However, the typical mass ratios in such systems (e.g., $\sim 1.4\,\Msun$ neutron star with a $\gtrsim 7\,\Msun$ companion) strongly favor dynamically unstable mass transfer, leading to a CE phase \citep[e.g.][]{Gallegos-Garcia:2023, Chattaraj:2026}. Avoiding CE would require a narrow and fine-tuned region of parameter space involving low-mass companions or unusually massive neutron stars, making such pathways impossible or extremely rare \citep{vanSon:2024, Gallegos-Garcia:2023, Chattaraj:2026}. 

On top of this, the stellar structure of typical BNS progenitors further favors CE-mediated evolution. Compared to the more massive stars that form BHs, NS progenitors more readily develop extended convective envelopes during post-main-sequence evolution. Such donors are generally more susceptible to dynamically unstable mass transfer than radiative-envelope stars (they tend to expand more (often) in response to mass loss), increasing the likelihood of CE evolution \citep[e.g.][]{Ge:2015, Temmink:2023}. In addition, these stars generally have smaller ratios of core mass to total mass, corresponding to more massive envelopes and a stronger tendency toward unstable mass transfer.
At the same time, once a CE phase is initiated, envelope ejection may be more readily achieved in these lower- and intermediate-mass giants than in the much more tightly bound envelopes of massive black-hole progenitors \citep[e.g.][]{Marchant:2021, Temmink:2023}. Together, these effects further bias merging BNS formation toward evolutionary pathways that involve at least one CE episode.

Only a few CE-free pathways have been identified for the formation of merging BNSs. 
In \citet{Gallegos-Garcia:2023}, all merging BNSs with a $1.4\,\Msun$ NS companion undergo a CE phase, while a small subset of systems with a $2\,\Msun$ NS companion can merge after only stable mass transfer, provided the NS receives a sufficiently large natal kick. 
Because these calculations begin from pre-existing NS-star binaries, rather than from zero-age binaries, the formation efficiency of the full evolutionary pathway remains difficult to assess. 
Similarly, \citet{vanSon:2024} find that BNS mergers are formed almost exclusively through the CE channel; only their no-natal-kick model shows a non-zero, but negligible, contribution from the only-stable-mass-transfer channel, and no representative evolutionary sequence is given for these systems. 
This is consistent with the arguments of \citet{Chattaraj:2026}, who find that the with-\ac{CE} channel overwhelmingly dominates BNS formation at solar metallicity ($\gtrsim95\%$), but also show that $\sim5\%$ of BNS systems can form without any mass transfer at all (which is considered in our taxonomy under without-CE) while the channel with only stable mass transfer is intrinsically inefficient, contributing only a very small fraction ($\lesssim1\%$) of merging BNS systems in their simulations.
Together, these results further support the conclusion that merging BNS formation is overwhelmingly \ac{CE} dominated.
The studies in Figure~\ref{fig:level-1-bns} that report non-negligible BNS contributions from without-\ac{CE} channels \citep{Boesky:2024gw, Broekgaarden2022} do not provide detailed descriptions of the corresponding formation pathways. Both studies also assume that Case~BB mass transfer, a phase thought to play a key role in BNS formation \citep[e.g.][]{Andrews:2015, VignaGomez:2018, Chattaraj:2026}, always proceeds stably, which may overestimate the fraction of BNS systems forming without a \ac{CE} phase.
Thus, while without-CE BNS formation may occur in finely tuned regions of parameter space, current population-synthesis studies generally support the conclusion that merging BNS formation is overwhelmingly CE-dominated.

This robust requirement of CE evolution for BNS formation provides an important anchor point: while the relative importance of CE and stable mass-transfer channels remains highly uncertain for BBH and BHNS systems, CE evolution appears to be an essential ingredient for producing merging BNS systems. 
Consequently, BNS mergers may provide a particularly valuable laboratory for constraining CE physics. This is further strengthened by the fact that alternative formation channels beyond isolated binary evolution generally face greater challenges in reproducing the observed BNS merger rates than they do for BBH mergers \citep{MandelBroekgaarden:2021}.
Because their formation pathways almost universally require a CE phase, the observed BNS merger rate directly traces the efficiency and success of CE evolution. In particular, a deficit of BNS mergers relative to BBH or BHNS populations \citep[e.g.][]{GWTC-5:populations, Fishbach:2026} could indicate inefficient envelope ejection or a high failure rate of CE interactions.
This indicates that relative merger rates between BNS, BHNS, and BBH populations may provide a complementary constraint on \ac{CE} physics, helping to break degeneracies that cannot be resolved using merger rates of individual populations alone.

A similar perspective emerges from studies of double white dwarfs\footnote{And other compact object binaries involving a WD \citep[e.g.][]{Toonen:2018}.}, another compact-binary population whose close and merging systems are widely believed to require at least one episode of CE evolution to shrink the binary to sufficiently short orbital periods \citep[e.g.,][]{Nelemans:2001popsynth, Nelemans:2003, Toonen:2012}. Classical formation scenarios for close and merging double white dwarfs often invoke two successive CE phases \citep[e.g.][]{Toonen:2012}, although more recent work suggests that the first mass-transfer episode can frequently proceed through stable, non-conservative Roche-lobe overflow, followed by a later CE phase \citep{Woods:2012, Li:2023-mt-stability}, {although the stability and nature of this first mass-transfer phase remains debated \citep[e.g.][]{Nelemans:2025}}.
Taken together, these studies suggest that the principal uncertainty for close and merging double white dwarfs is not whether CE evolution occurs, but rather how many CE episodes take place and at which evolutionary stages they occur.

%%%%%%%%%%%%%%%%%%%%%%%%%%%%
\subsection{Breaking the Degeneracy: Observational Signatures of Formation Channels}
\label{sec:discussion-observational-signatures}

The degeneracy between merger rates and formation pathways we find implies that rates alone cannot uniquely constrain binary evolution. Robust inference therefore requires the joint interpretation of multiple observables, as different formation channels imprint correlated signatures in masses, mass ratios, delay times, and spins. We discuss below how different observables may distinguish between formation channels with and without a CE phase, focusing on specifically channels with SMT (so not the `no mass transfer' level 2 channels) and CE channel. Qualitative trends from the simulation literature are highlighted in Tables~\ref{tab:smt_statements}--\ref{tab:ce_statements}.

\paragraph{Masses}
A common trend across studies is that SMT formation channels preferentially produce the most massive \ac{BBH} systems \citep{Neijssel:2019, Belczynski:2022formationChannel, vanSon:2022, vanSon:2022-nopeaks, Hendriks:2023, Li:2025, Willcox:2025} as the more massive stars respond typically more stable during mass transfer episodes \citep[e.g.][]{Temmink:2023, vanSon:2022}. 
In contrast, the contribution of \ac{CE}-driven channels to the mass distribution is less consistent across studies. Some find that CE pathways dominate the lower-mass population \citep{Li:2025, Hendriks:2023}, while others attribute low-mass \ac{BBH} systems primarily to without CE channels \citep[e.g.][]{vanSon2023}, or find that both with CE and without-CE channels contribute comparably \citep[e.g.][]{Willcox:2025}.

\paragraph{Mass Ratios}
The mass-ratio distribution provides a complementary diagnostic, though with more complex and sometimes non-monotonic behavior. SMT channels can produce relatively equal-mass systems ($q \sim 0.8$--$1$) in the absence of strong kicks \citep{Briel:2026}, but can also yield intermediate mass ratios  ($q \sim 0.4$--$0.8$)  \citep{vanSon:2022, Briel:2026, Olejak:2024, Banerjee:2024, Zevin:2020} or contribute to the most unequal BBH mass ratios \citep{Banerjee:2024}, depending on the model assumptions including efficiency of mass accretion, BH kicks, and angular-momentum loss. 
CE channels are found to produce broader mass ratio distributions, but are also found to produce preferably the equal mass ($q\sim 1$) BBH mergers \citep[e.g.][]{Oh:2023, Zevin:2020, vanSon:2022}. These trends emphasize that the mass-ratio distribution is not a simple discriminator, but instead encodes the interplay between the with and without CE channels and the underlying model assumptions.

\paragraph{Delay times}
Delay-time distributions might provide one of the clearest separations between formation channels. SMT channels generically produce long delay times ($\gtrsim 0.1$--$1\,\mathrm{Gyr}$), reflecting the limited orbital tightening achievable through stable mass transfer, particularly when angular-momentum loss is inefficient \citep{Gallegos-Garcia2021, Gallegos-Garcia2024, vanSon:2022-nopeaks, Olejak:2024, Oh:2023, Briel:2026, Klencki:2026-smt}. In contrast, CE evolution can shrink binaries efficiently, producing typically shorter delay times ($\lesssim  1\,\mathrm{Gyr}$) and dominating rapidly merging populations \citep{Neijssel:2019, vanSon:2022}, though this might strongly depend on the assumed CE efficiency \citep[e.g.][]{Bavera:2021, Boesky:2024gw}.   
For BHNS, the SMT channel might lead to shorter periods and delay times at solar metallicity \citet{Xing:2024-BHNS-solarZ}.
However, this distinction is itself sensitive to angular-momentum loss prescriptions: more efficient angular-momentum loss during SMT can partially compensate for the lack of a CE phase and produce shorter delay times \citep{Olejak:2024} (though see \citealt{Klencki:2026-smt}) whilst different assumptions in the CE efficiency can drastically change the BBH delay time from the with-CE channel \citep{Boesky:2024gw}. This highlights that delay times are not purely a channel diagnostic, but also probe the underlying physics of mass transfer.

\paragraph{Spins}
Spin distributions provide an additional, partially independent probe of binary evolution. SMT channels typically produce low or modest spins, as post-interaction binaries remain relatively wide and tidal spin-up is inefficient \citep{Klencki:2026-smt}. In contrast, CE channels can lead to higher spins for the second-born compact object due to tighter post-CE orbits and enhanced tidal interactions \citep[][though see \citealt{Detmers:2008}.]{Bavera:2021}. 

At the same time, SMT channels can produce a range of effective spins depending on accretion efficiency and angular-momentum transport, and may contribute to both low-spin and moderate-spin populations \citep{Olejak:2024, Briel:2026, Xu:2025}. As with masses and delay times, the spin distribution reflects not only the formation channel, but also the underlying assumptions about mass transfer, rotation, and angular-momentum transport within stars \citep[][]{Belczynski:2020bigBHpaper,FullerLu:2022, Gottlieb:2023, Jacquemin-Ide:2024}. Even without tidal spin-up of the Wolf-Rayet stars during a post-interaction stage, tides during the SMT phase itself could be enough produce moderate spins \citep{Xu:2025}"

%%%%%%%%%%%%%%%%%%%%%%%%%%%%%%%%%%%
\paragraph{Multi-messenger and intermediate populations.}
A key avenue to break the degeneracies in the key parameters affecting BBH and BHNS formation channels (as identified in Figure~\ref{fig:BH-BH-without-CE-key-parameters-impact}) is to connect compact-object mergers to their progenitor populations. 
Systems such as high-mass X-ray binaries, stripped stars, Wolf--Rayet binaries, and supernova progenitors directly probe the phases of mass transfer and envelope ejection that set the relative importance of CE for forming tight compact object binaries  \citep[e.g.,][]{Vinciguerra:2020, GallegosGarcia:2022, Rocha:2024, Zapartas:2025, VivancoCadiz:2026}. 
{Among these, stripped helium stars are a particularly promising probe: as the direct stripped progenitors of many compact objects in interacting binaries, their observed masses, orbital properties, and companion types will directly constrain the mass-transfer and stripping physics that governs the formation of double compact object systems. Once nearly absent from observational samples, they are now being discovered in growing numbers \citep[e.g.][]{Drout:2023, Gotberg:2023,  Ludwig:2026, Picco:2026-helium, Wang:2026-helium}.}
In particular, these intermediate populations can constrain the physical parameters identified in Section~\ref{sec:results-Simulation-Variations-and-Parameter-Dependencies}; including mass-transfer stability criteria ($q_{\rm c}$), CE efficiency ($\alpha_{\rm CE}$), mass transfer efficiency, and angular-momentum loss during non-conservative transfer, which ultimately control both the formation-channel fractions and their observable signatures \citep[e.g.,][]{Marchant:2024,  Lechien:2025, Sen:2026, vanSon:2026}. 
Future work should therefore move beyond merger-only constraints and adopt a multi-phase, multi-messenger approach, quantifying the contribution of the formation channels shown in Figure~\ref{fig:formation-channel-cartoon} not only to gravitational-wave mergers but also to their electromagnetic progenitor and intermediate populations. Examples include studies exploring formation pathways to astrometric \textit{Gaia} binaries \citep{Chawla2022} and Wolf--Rayet star populations \citep{Korb:2025}. Jointly modeling these populations offers an avenue for constraining the uncertain binary-evolution assumptions from Figure~\ref{fig:BH-BH-without-CE-key-parameters-impact} across multiple stages of compact-object formation. 

{A further avenue is to test a single binary-evolution framework against a broad range of observables simultaneously, which can expose model deficiencies that any one observable would miss. The \textsc{BPASS} models, for example, have been compared against resolved eclipsing binaries, supernova progenitors and rates, massive-star type ratios, X-ray binary populations, and integrated galaxy spectra \citep[e.g.][]{Eldridge:2017, Briel:2022,  EldridgeStanway:2022, Stevance:2023, Byrne:2025}, and similar efforts have been made for most population synthesis codes. Such joint analyses illustrate how intermediate and multi-messenger observations can inform and constrain binary-evolution models beyond merger rates alone.}

A complementary  avenue to break these degeneracies is provided by electromagnetic transients that directly probe phases of unstable mass transfer, envelope ejection, and stellar coalescence. In particular, luminous red novae (LRNe) are now robustly associated with stellar mergers and CE events (e.g., \citealt{Soker&Tylenda06,Tylenda+11}), tracing the onset of dynamical instability and the subsequent inspiral in real time \citep{Ivanova:2013LRN,2016MNRAS.455.4351P,MacLeod+17,Howitt:2020,Blagorodnova+21,Twum2026}. Observations and modeling show that LRNe arise from envelope ejection following unstable mass transfer, often preceded by sustained mass loss through the outer Lagrange point, and encode ejecta masses, velocities, and energetics that directly reflect the efficiency of envelope removal \citep{MatsumotoMetzger:2022, Twum2026}. As such, LRNe provide one of the few direct empirical probes of CE physics, constraining the conditions under which binaries merge versus survive envelope ejection, the same processes that dominate the uncertainty in formation-channel predictions in Figures~\ref{fig:level-1-BBH}--\ref{fig:level2-bns}.

Fast blue optical transients (FBOTs; \citealt{Drout+14}), and in particular the luminous FBOT (LFBOT) subclass (e.g., AT2018cow; \citealt{Prentice+18,Margutti+19,Ho+19,Perley+19}), may probe a complementary regime involving compact objects embedded within stellar envelopes (e.g., \citealt{Soker+19}). A promising scenario is that LFBOTs arise from delayed mergers between a black hole or neutron star and the helium core of a companion, either following a classical CE phase \citep{Metzger22}, or after an extended phase of stable mass transfer that terminates in a delayed dynamical instability \citep{KlenckiMetzger25}. In this picture, the multiwavelength transient is powered by highly super-Eddington accretion onto the compact object and interaction with the circumstellar medium shaped by the early mass-transfer phase \citep{KlenckiMetzger25,Aspegren&Kasen26}. Importantly, these events trace systems that lie near the boundary between stable and unstable mass transfer, i.e., those governed by the critical mass-ratio and angular-momentum-loss prescriptions that drive the diversity between CE and without-CE channels as shown in Figures~\ref{fig-comparison-bbh-simulation-silo} and ~\ref{fig:BH-BH-without-CE-key-parameters-impact}. LFBOTs may therefore represent ``failed'' gravitational-wave progenitors, directly sampling binaries that would otherwise contribute to the compact-object merger population if they avoided late-stage instability \citep{KlenckiMetzger25}.

Finally, longer-lived electromagnetic signatures such as ultra-luminous X-ray sources (ULXs; \citealt{Kaaret+17,Pavlovskii:2017}) and associated energetic synchrotron radio nebulae (``hyper-nebulae''), provide a time-integrated probe of hyper-Eddington mass transfer onto compact objects. During phases of extreme accretion, powerful jets and disk winds inflate extended nebulae of relativistic particles, which can remain observable for $\gtrsim 10^3$--$10^6$ yr and trace the cumulative energy injection from mass transfer episodes \citep{Sridhar&Metzger22}. Population models show that these systems arise across both stable and unstable mass-transfer channels, spanning a wide range of accretion rates and evolutionary states \citep[][]{Podsiadlowski:2003, Wiktorowicz:2017, Misra:2024}. As a result, ULX and hypernebulae provide a statistical probe of the prevalence, duration, and energetics of mass-transfer phases and  possible super-Eddington accretion, which are some of the key uncertainties identified in Figures~\ref{fig-comparison-bbh-simulation-silo} and ~\ref{fig:BH-BH-without-CE-key-parameters-impact} for the formation channel contributions.   Taken together, these electromagnetic counterparts offer a multi-phase, multi-timescale view of binary interaction physics, from pre-instability mass loss to final merger or survival, and thus provide a future pathway to break the degeneracies between CE and without-CE formation channels.

Given the robust result that merging BNS systems predominantly form through evolutionary pathways involving at least one episode of CE evolution (Section~\ref{sec:discussion-bns-robust-result} and Figures~\ref{fig:level-1-bns} and ~\ref{fig:formation-efficiency-results}), the observed abundance of BNS mergers provides a particularly direct empirical constraint on CE physics. Unlike BBH and BHNS populations, where the relative importance of with CE and without CE pathways remains highly uncertain, the formation efficiency of merging BNSs appears to be tightly coupled to the survival and outcome of CE evolution. Consequently, independent measurements of the neutron star merger rate and its electromagnetic counterparts offer a powerful avenue for constraining CE physics and calibrating population-synthesis predictions.

Kilonovae, thermal transients powered by the radioactive decay of freshly synthesized r-process material ejected during compact-object mergers, provide one such probe of the BNS merger population \citep[e.g.,][]{MetzgerBerger:2011, Salafia:2022, Rastinejad:2025}. However, despite their importance, the intrinsic kilonova rate remains only weakly constrained. To date, only a small number of confirmed kilonovae have been identified, and no compelling kilonova has yet been discovered in a blind wide-field optical survey independent of a gravitational-wave or gamma-ray trigger \citep[][]{Andreoni:2020}. Short gamma-ray bursts, widely interpreted as relativistic jets launched during neutron star mergers, provide a complementary tracer of the same underlying population \citep[e.g.,][]{Berger:2014, Fong:2015}. Comparisons between short gamma-ray bursts-inferred rates and gravitational-wave measurements suggest broad consistency, but remain subject to substantial uncertainties associated with jet opening angles, beaming corrections, and observational selection effects \citep[e.g.,][]{Fong:2015, Kunnumkai:2026}.

Recent gravitational-wave population analyses have further suggested that the intrinsic BNS merger rate may be lower than originally inferred from the first detections, potentially introducing tension with independent constraints from short gamma-ray bursts, Galactic double neutron stars, and r-process enrichment \citep[e.g.,][]{Fishbach:2026, GWTC-5:populations}, although see \citet{Kunnumkai:2026} for alternative interpretations. Together, these multi-messenger observations provide a critical normalization of the BNS merger rate and, by extension, of the efficiency of the evolutionary pathways that produce merging neutron stars  \citep[e.g.][]{Schiebelbein-Zwack:2026}. Because our results indicate that the BNS formation efficiency and rate  is substantially more sensitive to CE assumptions than to metallicity-dependent variations  (Figures~\ref{fig:level-1-bns} and \ref{fig:formation-efficiency-results}, cf. \citealt{vanSon:2024}), observational constraints on the absolute BNS merger rate might translate comparatively directly into constraints on CE survival and envelope ejection physics.
Future improvements in blind kilonova searches, short gamma-ray bursts population studies, and joint gravitational-wave--electromagnetic observations will therefore provide an important route toward breaking degeneracies in binary-evolution models by anchoring population-synthesis predictions to the absolute BNS merger rate \citep[e.g.][]{VanBemmel:2025, Nicholl:2025}.

%%%%%%%%%%%%%%%%%%%%%
\paragraph{White dwarf populations} 
Given the striking contrast between the formation channels of BNS mergers and those of BBH and BHNS systems (Figure~\ref{fig:summary-without-CE-channel}), an important next step is to investigate whether similar trends emerge in other populations containing low-mass compact remnants, particularly binaries containing one or more white dwarfs \citep[e.g.,][]{Nelemans:2001popsynth, Nelemans:2003, Toonen:2012, Toonen:2018, Roy:2026-dwd, Grondin:2026}. Although this review focuses on BBH, BHNS, and BNS systems, white dwarf binaries provide a complementary and often more directly observable laboratory for calibrating the physical processes that govern binary evolution.
In particular, because white dwarf binaries are orders of magnitude more numerous than compact-object mergers and are observed throughout the Milky Way by large spectroscopic, photometric, and astrometric surveys \citep[e.g.,][]{2010MNRAS.402..620R, 2014A&A...570A.107R, Zorotovic:2020, 2021MNRAS.506.5201R, 2024ApJ...976..102G, 2024MNRAS.527.6100N, 2025ApJS..279...47L}, they provide powerful constraints on the outcomes of mass transfer and CE evolution.

In particular, the observed orbital-period and separation distributions of post-CE white dwarf binaries provide important empirical constraints on CE physics. The existence of very short-period systems \citep[e.g.,][]{2011A&A...536A..43N, 2021MNRAS.501.1677H, 2022MNRAS.512.1843H, 2024MNRAS.533..324L} demonstrates that CE evolution can drive substantial orbital contraction, while the observed diversity of post-CE separations \citep[e.g.,][]{Zorotovic:2010, Zorotovic:2022, Scherbak:2023, 2024PASP..136h4202Y, 2024MNRAS.52711719Y, Grondin:2026} provides valuable constraints on the efficiency of envelope ejection and on the possible contribution of additional energy sources, such as recombination energy, convection, or accretion-powered outflows. More broadly, comparisons between observed white dwarf populations and binary-population synthesis models have emerged as one of the most important empirical tests of CE evolution, offering an opportunity to constrain the same uncertain physics that shapes the formation channels of gravitational-wave sources. {Such comparisons are already being used to constrain CE physics \citep{Tang:2024, vanZeist:2024}, since the double-white-dwarf population, the Galactic SN~Ia rate, and the LISA-resolved population all depend sensitively on the same CE evolution that governs merging compact-object formation.}

%%%%%%%%%%%%%%%
\paragraph{Observational constraints from gravitational waves on formation-channel fractions.}
Given the interest in understanding the formation channels of BBH, BHNS, and BNS sources and the role of CE versus SMT channels as shown in Figure~\ref{fig:level-1-BBH}--\ref{fig:level2-bns}, it is promising that there is also growing number of studies that have attempted to infer the relative contributions of CE- and SMT-mediated pathways directly from gravitational-wave observations. These studies typically combine population-synthesis expectations with hierarchical Bayesian inference, using the joint distributions of masses, spins, mass ratios, and redshifts to distinguish between (or infer) formation channels \citep[e.g.,][]{Zevin:2017, Zevin:2020, Bavera:2021, FishbachVanSon:2023, Colloms:2025, Alvarez-Lopez2025, Godfrey:2026}. Current catalogs provide evidence for multiple BBH subpopulations, including features such as a $\sim10,M_\odot$ mass peak, structure in the mass-ratio distribution, correlations involving $\chi_{\rm eff}$, and an asymmetric $\chi_{\rm eff}$ distribution, all of which may encode contributions from different formation pathways \citep[e.g.,][]{GWTC-5:populations, Godfrey:2026}. In particular, several studies have connected low-mass BBH features and mass-ratio structure to isolated binary evolution and, in some cases, to stable mass-transfer pathways \citep[e.g.,][]{Bavera:2021, vanSon:2022, vanSon:2022-nopeaks, Broekgaarden:2022,  Godfrey:2026, Willcox:2025}. 

However, CE and SMT channels often occupy overlapping regions of observable parameter space (as outlined in our Section~\ref{sec:discussion-observational-signatures}), and inferred branching fractions remain sensitive to assumptions about mass-transfer stability, angular-momentum loss, CE survival, natal spins, and the set of channels included in the inference \citep[e.g.,][]{Zevin:2020, Bavera:2021, Marchant:2021, Colloms:2025}. Thus, while gravitational-wave observations are beginning to empirically constrain the relative importance of CE and SMT evolution, current constraints remain broad and model dependent. Future catalogs, together with more flexible and higher-dimensional population models, will be required to turn these emerging signatures into robust constraints on compact-binary formation physics.
This overlap reflects a deeper physical degeneracy in binary evolution modeling: qualitatively different mechanisms for orbital shrinkage and angular-momentum redistribution can produce compact-object populations with broadly similar observable properties. Consequently, inferring formation channels from gravitational-wave observations alone may remain fundamentally challenging unless additional observables or independent constraints on binary interaction physics become available.

\subsection{What constitutes a Formation Channel?}
\label{sec:discussion-what-constitutes-a-formation-channel}

A fundamental ambiguity in the literature, evident from the diverse taxonomies shown in Figure~\ref{fig:formation-channel-cartoon} and Figures~\ref{fig:level2-bbh}--\ref{fig:level2-bns}, is the lack of a universally accepted definition of what constitutes a distinct formation channel.  Different studies adopt markedly different classification schemes, often tailored to the specific physical processes or questions under investigation. 

For example, some works define (sub)channels based on the sequence of binary interactions—tracking whether CE, SMT, or no mass transfer occurs before or after the first supernova \citep{Briel:2021, Oh:2023}. Others distinguish channels based on the nature of the CE phase itself, such as single-core versus double-core CE \citep[e.g.][]{Belczynski:2002, VignaGomez:2018}, or the evolutionary state (i.e., structure or stellar type) of the donor at the onset of the CE \citep[e.g.][]{Ruiter:2019, VignaGomez:2020, Chattaraj:2026, DorozsmaiToonen:2022}. 

Additional approaches emphasize the number of mass-transfer episodes, including repeated CE or SMT phases—an aspect that is often neglected but can be critical for binary tightening \citep[e.g.][]{Dominik:2013, Stevance:2023}. For BNS formation in particular, several studies highlight the role of specific physical ingredients, such as case BB mass transfer and ultra-stripped supernovae \citep{Tauris:2013, Tauris:2015, tauris2017formation, VignaGomez:2018, Andrews:2015, Xing:2024-BHNS-allZ}, electron-capture supernovae \citep{Kruckow:2018, Tanaka:2023, VignaGomez:2018}, or accretion-induced collapse \citep{Dominik:2012, Ruiter:2019, Chruslinska:2018, Pellouin:2025}. Other formation channel distinctions that have been pointed out include systems that initiate mass transfer only after the first supernova  \citep{Willcox:2025, Mestichelli:2025}, quasi-chemically homogeneous evolution \citep{Eldridge:2019}, pathways leading to very short delay times \citep{Andrews2020}, or systems that survive via favorable (“lucky”) natal kicks \citep{Broekgaarden2022}. 

Taken together, these examples illustrate that what is labeled a ``formation channel'' is not uniquely defined, but instead reflects a choice of which physical processes are considered most informative for organizing evolutionary pathways and addressing the central question of compact-binary astrophysics: how do binaries evolve into sufficiently tight compact-object systems that can merge within a Hubble time? In practice, the appropriate level of granularity depends on the scientific objective, whether the goal is to isolate key physical mechanisms (e.g., CE evolution or supernova physics), trace evolutionary sequences, or connect theoretical predictions to observable populations.

Beyond differences in nomenclature, these classifications often reflect deeper assumptions about the underlying physics of binary interactions. For example, the distinction between stable and unstable mass transfer depends sensitively on how different studies model angular-momentum loss, accretion efficiency, donor-envelope response, and the transition to dynamical instability. In many cases, binaries may evolve through intermediate regimes that are difficult to classify cleanly as either fully stable mass transfer or canonical CE evolution. This includes systems undergoing thermal-timescale transfer, delayed dynamical instability, or grazing-envelope evolution. As a result, formation-channel boundaries are not always physically unique and may depend on the adopted population-synthesis framework and classification criteria.

It is therefore perhaps unsurprising that, despite compact-object mergers representing intrinsically rare evolutionary outcomes, the literature identifies a wide diversity of pathways leading to BBH, BHNS, and BNS mergers. At the same time, this diversity is itself an important result, as it demonstrates that different physical assumptions and modeling choices can open, suppress, or reshape distinct routes to merger. Consequently, even the seemingly simple question, \textit{``how do merging BBH, BHNS, or BNS systems form?''}, does not admit a unique answer within current binary-evolution frameworks

%%%%%%%%%%%
\subsection{Mass Transfer and Binary Interactions as the Dominant Uncertainty}
\label{sec:discussion-mass-transfer-dominant-uncertainty}

Our results in Figures~\ref{fig-comparison-bbh-simulation-silo}--\ref{fig:BH-NS-without-CE-parameter-axis} reinforce the emerging view that mass transfer, particularly its stability, is a key uncertainty in modeling the formation of merging compact-object binaries

\paragraph{Mass-transfer stability}
Mass-transfer stability is a key uncertainty, particularly for the SMT channel. By shifting systems between SMT and CE pathways, Figure~\ref{fig-comparison-bbh-simulation-silo} shows that different stability prescriptions can substantially alter the relative importance of formation channels \citep{DorozsmaiToonen:2022}. In extreme cases, permissive stability criteria can produce populations dominated by without-CE evolution \citep{Li:2023-mt-stability, Olejak:2021CE}. \citet{vdH:2017} first pointed out that the greater stability of mass transfer from radiative-envelope donors compared to convective giants opens the possibility of forming gravitational-wave sources through sequences of stable mass-transfer episodes involving BH accretors. Subsequent studies have shown that many systems evolving through the SMT channel lie close to the boundary between stable and unstable mass transfer, highlighting the importance of accurately determining this stability criterion \citep[e.g.][]{Pavlovskii:2016, Marchant:2021,Gallegos-Garcia2021,Picco:2024,Klencki:2026-smt}.

Determining the location of this stability boundary remains an area of active research. A growing body of work demonstrates that mass-transfer stability depends sensitively on the donor's internal structure, evolutionary state, mass, and metallicity, rather than being well described by simple prescriptions based on stellar type or fixed critical mass ratios \citep{Hjellming:1987, Passy2012, Pavlovskii:2015,Pavlovskii:2017, Ge:2015,Ge:2020,Ge:2024,Marchant:2021,Klencki:2020,Klencki:2021,Renzo:2021,Renzo:2023,Temmink:2023,Temmink:2025}. Detailed stellar-evolution calculations show that the critical mass ratio for stability can vary substantially with the donor's internal structure, particularly for radiative-envelope stars \citep{Ge:2015,Ge:2020a,Pavlovskii:2017,Klencki:2026-smt}. At the same time, stability depends not only on the donor but also on the response of the accretor. For systems with stellar accretors, the ability of the companion to accommodate rapid mass accretion may itself strongly influence whether mass transfer remains stable \citep{Henneco:2024,Lau:2024,Schurmann:2024_stability}. This suggests that the physics governing mass transfer before and after formation of the first compact object may differ substantially, an effect that is only beginning to be incorporated into rapid population-synthesis models \citep{Schurmann:2025_combine,KlenckiMetzger25}.

These results challenge the simplified stability prescriptions commonly adopted in rapid population-synthesis codes \citep{Hurley:2002}, which may either overestimate or underestimate the prevalence of SMT depending on the adopted assumptions \citep{Gallegos-Garcia2021,Zhao:2024,Schurmann:2025_combine}. 
{This distinction maps directly onto the rapid-versus-detailed divide among the codes compiled here. Rapid codes often impose the stability boundary through tabulated critical mass ratios or adiabatic exponents \citep{Soberman:1997}, whereas detailed codes such as \textsc{POSYDON} and \textsc{BPASS} let it emerge from the donor's modeled structural response. 
Because this boundary sets the division between the CE and without-CE channels, differences in how it is determined can produce systematic offsets between rapid and detailed codes even when their nominal parameters agree. \citep[e.g.][]{Olejak:2024, Agrawal:2023}}.
Progress will likely require a combination of improved 3D simulations of binary mass transfer \citep[e.g.][]{Ryu:2025}, better modeling of stellar envelope physics \citep[e.g.][]{Jiang:2023}, and a more complete understanding of the structural response of both donors and accretors \citep{Renzo:2021,Renzo:2023,Schneider:2024,Xu:2025,Klencki:2026-smt}. Observational constraints may soon come from growing samples of massive post-interaction binaries, including recently discovered populations of stripped-star systems \citep{Drout:2023,Gotberg:2023,Villasenor:2023,Ramachandran:2023,Ramachandran:2024,Marchant:2024}.

%%%%%%%%%%%%%%%%%%%%%%%%
\paragraph{Orbital evolution and angular-momentum loss}
In addition to the stellar response, mass-transfer stability depends sensitively on the orbital evolution driven by the transfer process itself. One of the primary uncertainties is how much mass escapes the binary and how much angular momentum it carries away \citep[e.g.][]{Sarna:1993, TaurisvdH:2006, deMink:2007}, causing some of the uncertainty in BBH, BHNS formation channels as shown in Figure~\ref{fig:BH-BH-without-CE-key-parameters-impact}. Many prescriptions for non-conservative mass transfer assume isotropic re-emission from the vicinity of the donor and/or accretor, while additional mass-loss channels such as L2 outflows can remove substantially more angular momentum and drive more efficient orbital tightening \citep{vanSon:2022-nopeaks, Willcox:2025, LuWenbin2023, Picco:2024, Olejak:2024, Gallegos-Garcia2024}. In some cases, these outflows may remain partially bound to the system and form circumbinary structures that further modify the long-term orbital evolution of the inner binary \citep[e.g.][]{Siwek+2023MNRAS,Valli+2024AA,Wei:2024}. These uncertainties directly affect whether SMT channels can produce binaries sufficiently compact to merge within a Hubble time, as well as the resulting properties of the merging population, including mass ratios and orbital periods \citep[e.g.][]{Olejak:2024}.

Importantly, orbital evolution and mass-transfer stability are not independent uncertainties, but are physically coupled. Enhanced angular-momentum loss generally accelerates orbital shrinkage and causes the Roche lobe to contract more rapidly relative to the donor star, thereby reducing the stability of mass transfer \citep[e.g.][]{Picco:2024,Willcox:2025,Ge:2024,Klencki:2026-smt}.\footnote{Possible ways of accounting for this effect in rapid population synthesis include the $\zeta$ formalism \citep{Willcox:2025}, interpolation within grids of detailed binary-evolution calculations \citep{Ge:2020}, or stability criteria based on minimum orbital separations rather than critical mass ratios \citep{Klencki:2026-smt}.} This creates a feedback loop: stronger angular-momentum loss can help binaries reach the compact separations required for merger, while simultaneously shrinking the region of parameter space in which stable mass transfer is possible. As a result, varying angular-momentum loss and mass-transfer stability independently may overestimate the true uncertainty associated with the SMT channel.

A useful calibration point is provided by compact-object + massive-star binaries, which represent an intermediate evolutionary stage in both SMT and CE pathways \citep[e.g.][]{Picco:2024,Korb:2025}. For example, \citet{Klencki:2026-smt} find that varying the angular-momentum loss prescription substantially changes which BH + massive-star systems are expected to become SMT progenitors, while having a much smaller effect on the overall extent of the SMT parameter space. This behavior arises naturally from the coupling between orbital shrinkage and stability, which imposes a fundamental limit on the amount of tightening achievable through SMT evolution. While only a small number of such systems are currently known, upcoming astrometric, spectroscopic, and time-domain surveys are expected to increase the sample to tens or even hundreds of well-characterized BH + massive-star binaries \citep{Langer:2020,Janssens:2022,ElBadry:2024,Shenar:2024,Abrams:2025}. If models can robustly predict what fraction of these systems evolve into merging compact-object binaries, the observed population could provide a powerful empirical normalization for the contribution of the SMT channel.

%

%%%%%%%%%%%
\paragraph{Common-envelope efficiency and binding energy.}
In addition to uncertainties in stable mass transfer, the efficiency of CE ejection remains one of the dominant uncertainties in compact-binary formation. Variations in the CE efficiency parameter $\alphaCE$ and the envelope-structure parameter $\lambda$ can significantly affect both merger rates and the relative contributions of different formation channels as shown in Figure~\ref{fig-comparison-bbh-simulation-silo}--\ref{fig:BH-NS-without-CE-parameter-axis}. Because CE evolution is often responsible for the most dramatic orbital shrinkage in isolated binary evolution, these parameters directly influence which binaries survive the CE phase and emerge sufficiently compact to merge within a Hubble time. Importantly, the effects of CE efficiency are often degenerate with other uncertain aspects of binary evolution, complicating attempts to constrain CE physics using merger rates alone \citep[e.g.][]{Boesky:2024gw,Sgalletta:2026}.

However, uncertainty in CE evolution is not limited to the values of $\alphaCE$ and $\lambda$. The widely used $\alpha$--$\lambda$ formalism is a simplified description of a complex hydrodynamic process involving envelope ejection, angular-momentum transport, energy dissipation, recombination, jets, and potentially multiple episodes of interaction \citep[][]{ivanova2013common, Ivanova:2020book}. Consequently, calibrating a single CE-efficiency parameter for an entire binary population is unlikely to capture the full diversity of CE outcomes. The effective efficiency of envelope ejection may depend on a wide range of factors, including the evolutionary state of the donor, the structure and binding energy of the envelope, the binary mass ratio, additional energy sources, and the possibility of jet-assisted envelope removal \citep[e.g.][]{LivioSoker:1988,ivanova2013common,Ivanova:2020book,Grichener:2018,Soker+19,Lau:2021,Lau2022,HiraiMandel:2022,Twum2026,Marchant2021}. As a result, different physical assumptions can lead to similar effective values of $\alphaCE$ while corresponding to fundamentally different evolutionary pathways.
{A related and still-uncertain ingredient is the treatment of stripped helium (Wolf--Rayet) stars, the typical products of successful envelope ejection. Because their strong, metallicity-dependent winds can widen the post-CE orbit, uncertainties in Wolf--Rayet mass-loss rates feed directly into whether a system that survives a CE phase remains tight enough to merge, and hence into the effective without-CE versus CE contribution to the merging population \citep[e.g.][]{Laplace:2020, Schneider:2024, Merritt:2025, Gilkis:2026, Romagnolo:2026,Romagnolo_2026_AtlasII}. This sensitivity is amplified by the fact that helium-star wind prescriptions, which are one of the major drivers in the determination of black-hole masses \citep{Romagnolo_2026_AtlasII}, differ substantially across the codes compiled here \citep[e.g.][]{Laplace:2020, Merritt:2025, Romagnolo:2026}.}

These challenges suggest future progress will require moving beyond the interpretation of $\alphaCE$ as a single fundamental parameter and instead developing prescriptions that more directly capture the underlying physics of CE evolution. Hydrodynamic simulations are beginning to provide important insights into envelope ejection, energy transport, and post-CE orbital evolution, although significant challenges remain in bridging the gap between these calculations and population-synthesis models. We discuss these developments in more detail in Section~\ref{sec-discussion:hydrodynamic-constraints}.

\subsection{Common-Envelope Prescriptions Are Not a Single Physical Model}
\label{sec:discussion-CE-prescriptions}

{Beyond the degeneracies in the value of \alphaCE{} discussed above, a more basic point is that our division of formation pathways into those that do and do not involve a \ac{CE} phase, adopted throughout this work for clarity, obscures a wide diversity in how \ac{CE} evolution itself is modeled  \citep[e.g.][]{ivanova2013common, Ivanova:2020book, HiraiMandel:2022, Fragos:2019, Nelemans:2025}. 
As a result, the label ``CE'' does not correspond to a single physical prescription with a unique orbital outcome; rather, it encompasses a range of formalisms that can shrink the orbit by very different amounts for the same input binary \citep[see discussion in][]{Iaconi:2019, HiraiMandel:2022, Nelemans:2025}.}

{The most widely used description is the energy-based $\alpha$-$\lambda$ formalism \citep[e.g.][]{Webbink:1984, deKool:1990}, in which a fraction $\alphaCE$ of the released orbital energy is available to unbind the envelope, whose binding energy is parameterized through the structural factor $\lambda$. 
Here the post-\ac{CE} separation follows from an energy balance, and both $\alphaCE$ and $\lambda$ carry substantial uncertainty. 
An alternative is the angular-momentum-based $\gamma$ formalism
\citep{Nelemans:2000, Nelemans:2005}, in which the envelope is assumed to carry away a specific angular momentum set by a parameter $\gamma$, rather than being ejected through an energy budget; some detailed frameworks effectively adopt a $\gamma$-like treatment, for example \textsc{BPASS}, whose \ac{CE} prescription has been shown to behave more like the $\gamma$ formalism than the energy-based $\alpha$--$\lambda$ prescription \citep[e.g.][]{Eldridge:2017, Stevance:2023}. 
The two formalisms can predict markedly different post-interaction orbits for the same system, and neither is a first-principles description: both stand in for a multidimensional hydrodynamic process (Section~\ref{sec-discussion:hydrodynamic-constraints}).
Note that the $\gamma$ parameter used in this \ac{CE} context is conceptually distinct from the $\gamma$-like prescriptions for angular-momentum loss during \emph{stable} mass transfer discussed in
Section~\ref{sec:results-Simulation-Variations-and-Parameter-Dependencies}, although both encode how much angular momentum leaves the binary.}

{Because these prescriptions differ in their orbital consequences, two studies that both report a system as undergoing ``CE'' may nonetheless disagree substantially on whether that system survives as a tight merging binary. 
This is a further reason why the with-\ac{CE} fraction alone does not uniquely determine the merger rate, and why cross-study comparisons of \ac{CE} contributions must be interpreted with the underlying \ac{CE} prescription in mind. 
It also underlies the framework-dependent behavior we find for $\alphaCE$ variations in Section~\ref{sec:results-Simulation-Variations-and-Parameter-Dependencies}: the ``sweet spot'' of $\alphaCE \sim 0.5$--$2$ is specific to the energy formalism, and codes adopting different \ac{CE} treatments need not show the same dependence.}

\subsection{Hydrodynamic Constraints on Common-Envelope Prescriptions}
\label{sec-discussion:hydrodynamic-constraints}
%%%%%%

One important point that emerges from comparing BBH, BHNS, and BNS formation channels in our study is that many commonly adopted CE prescriptions should not be viewed simply as interchangeable parameter choices, but as approximations to a fundamentally unresolved hydrodynamic problem. Quantities such as the CE efficiency parameter $\alphaCE$, the envelope binding-energy parameter $\lambda$, critical mass-ratio stability criteria, and prescriptions tied to convective or radiative envelope structure are widely used throughout population-synthesis calculations. However, these quantities ultimately act as effective descriptions of multidimensional processes involving inspiral-driven energy deposition, angular-momentum transport, shocks, envelope restructuring, recombination physics, and the dynamical response of the stellar interior.

Over the past decade, hydrodynamic simulations have significantly improved our understanding of CE evolution, particularly for highly extended RGB, AGB, and red-supergiant donors \citep{RickerTaam2012,Passy2012,Ohlmann2016,Iaconi2017,Iaconi2018,MacLeod:2018, De:2020, Reichardt:2020, Sand2020,Chamandy2020}. Despite large differences in numerical methods, resolution, and donor structure, these simulations reveal several remarkably consistent features. In most calculations, the interaction begins with a rapid plunge-in phase in which orbital energy and angular momentum are transferred to the envelope through shocks, tidal torques, and hydrodynamic drag. Large-scale spiral shocks and asymmetric outflows develop quickly, leading to substantial envelope expansion. However, many simulations find that the inspiral slows considerably after the initial plunge, often entering a prolonged phase of slower orbital evolution characterized by circulation flows, repeated passages through shocked gas, and continued redistribution of energy and angular momentum over many dynamical timescales \citep{Ivanova2016,Iaconi2017,Chamandy2020}. 

A major lesson emerging from these studies is that the outcome of the interaction is not determined solely by the orbital energy released during the initial inspiral, but by the subsequent hydrodynamic evolution of the disturbed envelope. In many simulations, the initial plunge-in alone is insufficient to fully unbind the envelope, and the longer-term redistribution of energy and momentum through shocks, turbulent flows, fallback, envelope inflation, and recombination-assisted expansion becomes essential in determining the final ejecta mass and orbital separation \citep{Ivanova2013Review,Ivanova2015,MacLeod:2018, Fragos:2019, De:2020, Lau2022,Everson2025}. Hydrodynamic calculations also increasingly emphasize the importance of donor structure in regulating CE outcomes. Successful envelope ejection and compact binary survival appear to occur preferentially for highly evolved donors with extended weakly bound envelopes and steep density gradients between the core and envelope \citep{MacLeod2015,LawSmith2020,Marchant2021, Klencki:2021}. In these systems, the outer envelope binding energy per unit mass can become comparable to the available recombination-energy reservoir, allowing recombination-assisted outflows to contribute significantly during the late inspiral and expansion phases \citep{Ivanova2015,Lau2022}.

Importantly, these studies also show that the envelope binding energy is not static during the interaction. As orbital energy and angular momentum are deposited into the envelope, the donor dynamically restructures, altering the density profile, entropy stratification, and effective binding energy of the remaining gas as the inspiral proceeds \citep{2026arXiv260527579H}. This restructuring is difficult to capture in rapid prescriptions that rely on precomputed stellar profiles or fixed $\lambda$ parameterizations. Consequently, differences between analytic stellar fits, interpolated stellar tracks, hybrid MESA-based approaches, and fully resolved stellar models do not merely correspond to calibration differences, but often reflect qualitatively different assumptions regarding stellar structure and the hydrodynamic evolution of the envelope itself. In practice, different prescriptions for the envelope structure and core-envelope boundary can produce binding energies that differ by factors of several and in some cases by nearly an order of magnitude, leading to substantially different predictions for CE survival and merger outcomes. Prescriptions based on fixed $\lambda$ values, tabulated $\lambda$ calculations \citep[e.g.,][]{XuLi2010}, enthalpy-based formalisms \citep{Ivanova2011}, recombination-inclusive energy budgets \citep{Webbink2008,Ivanova2015,Lau2022}, and hydrodynamic inspiral calculations therefore encode distinct physical assumptions regarding what constitutes available energy for envelope ejection and what conditions define successful binary survival.

Another important uncertainty concerns the final inspiral or ``parking'' conditions near the donor core. In many population-synthesis implementations, the stellar core is effectively treated as a point mass and the inspiral termination is determined through simplified energetic arguments. Hydrodynamic studies, however, suggest that the final separation depends sensitively on the detailed structure of the core-envelope transition region and on how the inspiraling companion couples to the surrounding gas near the core \citep{LawSmith2020,CohenSoker2023}. Entropy gradients, pressure support, and unresolved gas dynamics near the inner envelope can therefore materially influence whether inspiral stalls, survives, or proceeds to merger. 

Taken together, these results suggest that the diversity of formation-channel predictions across population-synthesis studies reflects not only differences in binary-evolution parameters, but also differing implicit assumptions about the hydrodynamic evolution of the CE interaction itself.

\subsection{Limitations of the Current Compilation}

Our compilation focuses on intrinsic formation-channel predictions from isolated binary evolution and therefore excludes several classes of studies that could provide complementary constraints.
First, we do not include studies that report results with detection-weighted populations, or without channel-resolved rates. While such studies provide valuable insights, they are difficult to incorporate into a homogeneous comparison due to differences in assumptions about metallicity, selection effects, and detector sensitivity.
Second, certain formation channels, such as chemically homogeneous evolution or residual (other) channels, are treated inconsistently across studies. A more systematic inclusion of these channels will be necessary for a complete mapping between theoretical predictions and observations.
Third, we rely on published intrinsic merger rates rather than re-analyzing simulation outputs directly. A fully self-consistent re-analysis of publicly available datasets, using uniform assumptions for cosmic star formation history and channel classification, would provide a more robust comparison but is beyond the scope of this work.
Finally, the classification of subchannels varies significantly across studies, with different works adopting distinct definitions based on interaction sequences, CE types, or evolutionary phases.

% \subsection{Future Directions}

\subsection{Recommendations for Population-Synthesis Studies}
\label{sec:discussion-recommendations-pop-synth}
Based on this meta-analysis, we highlight several practical recommendations for future population-synthesis studies. 
{We emphasize that the aim of these recommendations is greater comparability and transparency, not uniformity: the diversity of approaches documented in this work reflects genuine, unresolved uncertainties in massive-binary physics and differing modeling philosophies across the community, and this diversity is itself scientifically valuable. The goal should be to make differences between codes \emph{interpretable} and traceable to specific physical or implementation choices, rather than to suppress them}

\begin{itemize}

    \item \textit{{Adopt transparent (and, if possible, consistent) formation-channel definitions.}} 
    Clearly define formation channels and avoid ambiguous terminology such as “only stable mass transfer,” particularly in cases where systems may not undergo mass transfer at all (see Section~\ref{sec:method-Without-CE-versus-SMT-terminology}. Explicitly state the selection criteria used to classify systems into different channels.

    \item \textit{{Ensure reproducibility through open data and post-processing pipelines.}} 
    Make simulation outputs and post-processing pipelines publicly available to enable re-analysis under consistent assumptions. This is especially critical for formation-channel studies, where differences in classification schemes can significantly affect inferred contributions.

    \item \textit{{Systematically explore key parameter uncertainties.}} 
    Prioritize variations in parameters that strongly impact formation-channel contributions, particularly mass-transfer stability criteria and processes governing angular-momentum loss (e.g. $\gamma$). This includes the mass-transfer efficiency ($\beta$), CE efficiency ($\alpha_{\rm CE}$),  supernova prescriptions, supernova kick magnitudes (e.g. $\sigma$),  envelope binding-energy assumptions, and mass transfer stability criteria. Additionally the remnant mass prescription and stellar winds also can significantly affect the formation channel contributions (Figure~\ref{fig:BH-BH-without-CE-key-parameters-impact}). In particular, future work should continue connecting rapid population-synthesis prescriptions to multidimensional hydrodynamic simulations of binary interactions, since many phenomenological parameters likely encode complex processes involving donor structure, orbital evolution, and time-dependent envelope response that are not accurately captured by the simplified parameter approximations.

    \item \textit{{Expand formation-channel analyses across populations.}} 
    Extend detailed formation-channel studies to a broader set of simulations, especially for \ac{BHNS} and \ac{BNS} populations, where systematic comparisons remain limited. Future work should additionally extend formation-channel analyses beyond merging compact-object binaries to other astrophysical populations, including intermediate evolutionary stages of gravitational-wave sources. This will help map how different physical assumptions impact different compact-object populations, including whether certain parameters affect BBH, BHNS, and BNS systems differently or alter their relative formation efficiencies and channel contributions in distinct ways.

    \item \textit{{Move beyond merger rates.}} 
    Investigate the observable imprints of different formation channels beyond rates such as their impact on mass, spin, and redshift distributions, to enable comparisons with current and future gravitational-wave observations and help quantify and characterize distinguishable observable features for these channels.

    \item {\textit{Develop community benchmark suites and coordinated inter-code comparisons.} A key limitation of any cross-study compilation, including ours, is that nominally identical physical assumptions (e.g. a given CE efficiency, mass-transfer stability criterion, or wind prescription) may be implemented differently across codes, so that apparent agreement or disagreement can reflect implementation and numerical choices rather than the underlying physics. Disentangling these requires controlled inter-code comparisons in which the same standardized binary models (spanning a grid of initial masses, mass ratios, separations, and metallicities) are evolved independently by each framework and compared directly, at the level of both individual systems and integrated populations. Such benchmark efforts (analogous to, e.g., \citealt[][]{toonen2014popcorn, Belczynski:2021, Agrawal:2020MNRAS, Agrawal:2023, Smith:2026, Guerrero:2026}, and the LISA Synthetic UCB Catalogue Group's efforts and standardized formats for white dwarf simulations) would help isolate implementation differences that are otherwise difficult to diagnose. Further extending these efforts would be a valuable step. }

\end{itemize} 

\medskip
{These recommendations differ substantially in the effort they require, and we do not expect all of them to be equally feasible for every group. We therefore suggest a rough prioritization. The most immediately actionable steps (``low-hanging fruit'') are those that require little additional computation: adopting more transparent formation-channel definitions, publicly releasing simulation outputs and post-processing pipelines, and adopting a common output format such as \textsc{BinCodex}. These can be undertaken by essentially any group and would already substantially improve comparability across studies. 
A second tier involves running a code comparison by participating in coordinated inter-code comparisons based on shared sets of test binaries; this is more straightforward for rapid population-synthesis codes but likely more demanding for others with larger computing times per binary.}

\bigskip

Overall, our results support a growing consensus that the key question is no longer whether CE or only stable mass transfer channels dominate, but rather how the interplay between these channels, and the uncertainties in mass transfer physics, shapes the observable population of gravitational-wave sources.

%%%%%%%%%%%%%%%%%%%%%%%%%
\section{Conclusions}
\label{sec:fc-conlcusion}

In this work, we compiled and systematically compared formation-channel predictions from more than one hundred population-synthesis simulations to assess the role of common-envelope (CE) evolution in forming gravitational-wave sources within the isolated binary evolution paradigm. Our main conclusions are:

\begin{itemize}

\item \textit{\textbf{Formation channels are not uniquely defined across the literature.}} The diversity of Level 2 classifications (Figure~\ref{fig:formation-channel-cartoon}) demonstrates that formation channels represent bookkeeping frameworks rather than uniquely defined physical entities. Establishing common taxonomies,  explicitly reporting channel fractions, and making post-processing formation channel scripts publicly available will be important for future cross-code comparisons and interpretation of gravitational-wave observations.

\item \textit{\textbf{BBHs and BHNS can form both mostly with or without CE and merger rates alone do not constrain formation pathways.}}  
Figures~\ref{fig:level-1-BBH} and \ref{fig:level-1-bhns} and the global overview in Figure~\ref{fig:summary-without-CE-channel} demonstrate that BBHs and BHNS can form predominantly through either CE or without-CE pathways, while still producing comparable merger rates.  
This demonstrates that qualitatively different mechanisms for orbital shrinkage and angular-momentum redistribution can reproduce similar compact-object merger populations. Consequently, merger rates alone do not uniquely constrain the underlying physics of binary interactions.

\item \textit{\textbf{BNS formation robustly requires CE evolution, unlike BBH and BHNS.}}  
In contrast to BBH and BHNS systems, Figure~\ref{fig:level-1-bns} shows that BNS mergers are consistently dominated ($\gtrsim 90$--$100\%$) by channels involving at least one CE phase across all compiled simulations. This is further confirmed in Figure~\ref{fig:formation-efficiency-results} that shows BNS formation across metallicity has negligble contribution from without CE channels.  This qualitative difference implies that CE physics may play a critical role in setting the relative rates of BNS versus BBH/BHNS mergers, and that constraints on CE efficiency from observations could directly impact population ratios and their relative rates. 

\item \textit{\textbf{A small number of key physical assumptions dominate  formation-channel uncertainty.}}
Figures~\ref{fig-comparison-bbh-simulation-silo}--\ref{fig:BH-NS-without-CE-parameter-axis} show that parameter variations within a given population-synthesis framework typically explore only a limited region of formation-channel parameter space, while the largest differences often arise between population-synthesis codes. At the same time, Figure~\ref{fig:BH-BH-without-CE-key-parameters-impact} demonstrates that a small set of assumptions governing binary interactions can nevertheless substantially alter formation-channel contributions. In particular, prescriptions for mass-transfer stability, angular-momentum loss, mass-transfer efficiency ($\beta$), supernova remnant-mass prescriptions,
the CE efficiency ($\alpha_{\rm CE}$),  and natal kicks have the largest impact  (often by $\gtrsim 30\%$) on the balance between CE and without-CE pathways. Importantly, these parameters often act non-linearly and in a correlated fashion, producing competing effects on binary survival and orbital tightening. Consequently, trends inferred from one-at-a-time parameter variations do not necessarily generalize across population-synthesis frameworks.

\item \textit{\textbf{The diversity of CE and without-CE contributions is not driven by a single alternative pathway.}}  
The Level~2 decomposition (Figures~\ref{fig:level2-bbh}--\ref{fig:level2-bns}) show that large variations in the Level~1 CE fraction generally arise from a redistribution across multiple evolutionary sequences rather than the emergence of a single dominant alternative channel.
This highlights the high-dimensional nature of binary evolution and implies that disagreements between population-synthesis studies cannot generally be reduced to a single physical mechanism or formation pathway{, and may in addition reflect deep structural differences between codes, including their stellar-evolution treatments and implementation choices, rather than any single alternative evolutionary channel}.

\item \textit{\textbf{Merger rates alone are fundamentally insufficient to determine formation pathways.}}  
Given that qualitatively different CE and without-CE populations can reproduce similar merger rates (Figures~\ref{fig:level-1-BBH} and \ref{fig:level-1-bns}), breaking these degeneracies will require additional observables, including masses, mass ratios, spins, delay times, and redshift evolution. Equally important will be connecting gravitational-wave sources to intermediate populations such as stripped stars, BH+massive-star binaries, X-ray binaries, and supernova progenitors, which directly probe the binary-interaction physics that governs compact-object formation (see Section~\ref{sec:discussion-observational-signatures}).

\item \textbf{Future progress requires moving towards a community-wide framework for formation-channel studies.}
Our results highlight the need for coordinated efforts to map the uncertainty space of massive binary evolution and improve the comparability of formation-channel predictions across studies (Section~\ref{sec:discussion-recommendations-pop-synth}). In particular, future work should (i) adopt transparent and, where possible, consistent formation-channel definitions, (ii) make simulation outputs and post-processing pipelines publicly available to enable reproducible formation-channel analyses, (iii) systematically explore key uncertainties governing binary interactions, especially mass-transfer stability, angular-momentum loss, CE evolution, and supernova physics (iv) extend detailed formation-channel studies across BBH, BHNS, BNS, and intermediate binary populations, and (v) move beyond merger-rate comparisons by quantifying the observable signatures of different formation pathways in masses, spins, and redshift distributions. Such efforts will be essential for robustly connecting gravitational-wave observations to the physical processes shaping massive binary evolution across cosmic time.
\end{itemize}

To facilitate future comparisons and reproducibility, we make all compiled formation-channel data publicly available at \citet{Broekgaarden:Zenodo-Common-Common-Envelope}  and provide additional interactive figures and online tables to explore the data. These resources provide a community benchmark for exploring formation-channel predictions across population-synthesis studies, visualizing the impact of different physical assumptions, and comparing results across codes within a unified taxonomy.

Taken together, our results suggest that the central challenge in interpreting gravitational-wave populations is not simply determining the value of a particular binary-evolution parameter, but disentangling fundamentally different physical pictures of how massive binaries shrink and survive binary interactions. Within current isolated binary-evolution frameworks, BBH and BHNS mergers can arise through pathways spanning the full range from CE-dominated to without-CE–dominated evolution, whereas BNS formation remains strongly tied to CE evolution. This contrast provides a valuable opportunity: by combining gravitational-wave observations with electromagnetic constraints on intermediate binary populations and continued advances in theoretical modeling, future studies may be able to directly determine which binary-interaction processes dominate the formation of compact-object mergers. Realizing this goal will require moving beyond merger-rate comparisons alone toward a unified framework that systematically quantifies formation-channel contributions and their uncertainties across population-synthesis models.

\section*{Acknowledgments}
The authors thank Rhea Kumar, Laya Binu and Manasvini Komandur for useful discussions. 
FSB and SAL acknowledge support from NASA HPOSS grant 80NSSC25K7555 under award number 316592-00001. 
FSB acknowledges support from NSF Astronomy and Astrophysics Research Grant under award number 2606407. 
AL acknowledges support of the City University of New York Graduate Center M.S. Program in Astrophysics and the Simons Foundation Presidents Discretionary Fund.
SMG acknowledges the support of the Natural Sciences and Engineering Research Council of Canada (NSERC) and is partially funded through an NSERC Postdoctoral Fellowship (PDF). BDM acknowledges support from NASA (grant 80NSSC24K0408) and the National Science Foundation (grant AST-2406637). The Flatiron Institute is supported by the Simons Foundation.  d
TBS acknowledges support from the National Science Foundation Graduate Research Fellowship Program under Grant No. 1839285. 
AR acknowledges financial support from the European Research Council for the ERC Consolidator grant DEMOBLACK, under contract no. 770017.
LMS acknowledges support from the Alexander von Humboldt Foundation.
AR and LMS acknowledge support from the German Excellence Strategy via the Heidelberg Cluster of Excellence (EXC 2181 - 390900948) STRUCTURES.
This work benefited from discussions during the CCA Stable Mass Transfer workshop in NYC 2024.

\section*{Software Acknowledgement}

This work made use of the following software packages: \textsc{astropy} \citep{astropy:2013,astropy:2018,astropy:2022,astropy_17756022}, \textsc{Jupyter} \citep{2007CSE.....9c..21P,kluyver2016jupyter}, \textsc{matplotlib} \citep{Hunter:2007}, \textsc{numpy} \citep{numpy}, \textsc{python} \citep{python}, \textsc{scipy} \citep{2020SciPy-NMeth, scipy_14880408},  \textsc{PyTables} \citep{pytables}, and WebPlotDigitizer \citep{rohatgi2020webplotdigitizer}.  Claude code and ChatGPT were used to help with converting our static figures using \texttt{plotly} \citep{plotly, plotly_14366349} into interactive figures to help with data visualization.  
This research has made use of the Astrophysics Data System, funded by NASA under Cooperative Agreement 80NSSC21M00561. 
Software citation information aggregated using \textsc{\href{https://www.tomwagg.com/software-citation-station/}{The Software Citation Station}} \citep{software-citation-station-paper,software-citation-station-zenodo}.

This work made use of publicly available data and code for creating the \ac{BBH}, \ac{BHNS}, and \ac{BNS} formation efficiencies shown in Figures~\ref{fig:formation-efficiency-results} and~\ref{fig:formation-efficiency-results-detailed}: data from \citet{van_son_2024_14508864} for the formation efficiency results of \citet{vanSon:2024}; data from \citet{iorio_2023_7794546} to reproduce the formation efficiencies of \citet{Ioro:2023sevn} (using the plotting code at \url{https://gitlab.com/iogiul/iorio22_plot/-/tree/v3/formation_channels}); and data from \citet{Broekgaarden:Zenodo-DCO-formation-channels} for the formation efficiencies based on \citet{Broekgaarden:2021efa, Broekgaarden2022}. The formation efficiency of \citet{Neijssel:2019} for \ac{BBH} systems was digitized from their Figure~1 using WebPlotDigitizer \citep{rohatgi2020webplotdigitizer}. Literature constraints at individual metallicities were compiled from the references listed in Table~\ref{tab:fc_Z_statements_bns}.

\section*{Data Availability}
All data and code to reproduce all results and figures in this work are publicly available at \citet{Broekgaarden:Zenodo-Common-Common-Envelope} (Archived code with DOI) and on  \href{https://github.com/FloorBroekgaarden/Rates_of_Formation_Channels}{this GitHub repo}. Interactive tables and figures can be accessed \href{https://floorbroekgaarden.github.io/Rates_of_Formation_Channels/interactive_figures_and_tables/formation_channel_rates_table.html}{online}.

\section*{Author contributions}
FSB led the research (plots, catalogs, metadata, data analysis) and the bulk of the
manuscript work (abstract, introduction, method, results, discussion, conclusion, and appendixes). AL and SL provided extensive help with understanding the results and shaping the results throughout the manuscript. JK, KAR, LvS, SMG, MG, BDM,  ER, and AT provided critical feedback and/or contributed significant text throughout the research process that helped fundamentally shape the paper.  MS, JH, TBS, AR, EB, and LMdS provided useful feedback, comments, and suggestions for the research direction, figures, and tables. 

% add data statements for Lieke, Iorio's and Neijssel data.(Zenodo links from appendix text)

% \bibliographystyle{apsrev4-1}
\bibliographystyle{mnras}
% You should give the same name for your .bbl as your main .tex
% since it is a requirement for posting on ArXiv.
\bibliography{biblio, biblio2, BDM}

\appendix
\twocolumngrid

\section{Appendix A: Data Retrieval}
\label{app:fc_data-retrieval}
This appendix describes the process used to collect the formation-channel contributions and merger rates presented in this study. 
We compile data from recent population synthesis studies of isolated binary evolution that report intrinsic merger rates and explicitly quantify the relative contributions of different formation channels, with particular attention to the occurrence of \ac{CE} phases in the formation of \ac{BBH}, \ac{BHNS}, and \ac{BNS} systems. 
We restrict our sample to studies published within the past decade that (i) provide cosmologically integrated intrinsic merger rates and (ii) report a breakdown of these rates by distinct formation channels.
We do not include studies that only present formation-channel fractions at fixed metallicity, or that focus exclusively on detectable (observer-frame) rates. Incorporating such results would require re-computing cosmological rate integrations under a consistent set of assumptions (most notably regarding detector sensitivity and selection effects), which is beyond the scope of this work but constitutes an important direction for future study. A list of excluded studies is provided in the following Appendix section. For studies that do not tabulate rates for the full set of channels in our taxonomy, we adopt a uniform bookkeeping convention: any unreported channel is assigned a rate of zero for that study. This choice enables a consistent cross-study compilation while explicitly reflecting the information content of the original publication; it may therefore be interpreted as a lower limit on the contribution from unreported channels.
All included datasets are listed alphabetically below.
Throughout this appendix, we adopt the formation-channel classification
scheme introduced in the main text (see Figure~\ref{fig:formation-channel-cartoon}).
At the first level, we distinguish between systems that do and do not undergo a \ac{CE} phase during their evolution.
Where available, we further adopt the more detailed sub-classifications reported by the original authors (level~2 classification).
We list the studies alphabetically, and in brackets list the acronyms used in the figures.

\paragraph{ \citet{Bavera:2021} [Ba21]}
\citet{Bavera:2021} compute \ac{BBH} merger rates by coupling detailed binary stellar evolution simulations performed with \textsc{MESA} to the rapid population synthesis code \textsc{COSMIC} via an early version of the \textsc{POSYDON} framework. 
Their study explores a range of population synthesis assumptions varying \ac{CE} parameters, mass transfer and accretion parameters, and initial conditions  and reports BBH rates subdivided into the classic \ac{CE} and \ac{SMT} formation channels, which correspond to the `with CE' and `without CE' categories in Fig~\ref{fig:formation-channel-cartoon}, respectively.
We extract total merger rates from Tables~1, 3, and~4 of \citet{Bavera:2021}. From Table~1, which explores variations in the CE efficiency parameter $\alpha_{\mathrm{CE}}$, we adopt the seven CE rate estimates listed in the first seven rows and pair each with the constant SMT rate given in the eighth row in order to compute the relative CE and SMT contributions.
The lower six rows of Table~1 provide three additional total BBH rate model estimates (CE+SMT combined).  
Tables~3 and~4 contribute three and two additional models for which the authors report total BBH rate estimates, respectively.  In each case, we exclude the fiducial models listed in the first row of these tables, as they duplicate the fiducial model already included from Table~1.
In total, this yields 15 distinct BBH merger-rate estimates and their corresponding CE and SMT contributions.

\paragraph{\citet{Boesky:2024popsynth, Boesky:2024gw} [Bo24]}
\citet{Boesky:2024popsynth,Boesky:2024gw} use the \textsc{COMPAS} population synthesis code to compute 21 model variations: 12 exploring different values of the \ac{CE} efficiency parameter ($\alpha$) and binding energy parameter ($\beta$), and 9 varying the remnant mass prescription and the \ac{SN} natal-kick magnitude distribution.  
We extract formation channel contributions from  Figure~12 of \citet{Boesky:2024gw} (at redshift $z\sim0$)  and from the authors through private communication.  
The published figure includes a non-negligible  contribution for some models from an ``other'' formation channel, without explicitly stating whether these systems experience a \ac{CE} phase.  
After inspecting the underlying model data, we verified that all systems classified as ``other'' do undergo a CE episode (but, for example, do not undergo SMT prior to the first SN hence the `other' classification).  
Accordingly, we assign these systems to the ``with CE'' category in the Level 1 classification, and label them as ``other with CE'' in the Level~2 classification.

\paragraph{\citet{Briel:2022} [BRI22]}
\citet{Briel:2022} employ the \textsc{BPASS} population synthesis framework to investigate the formation of BBH mergers, with a particular focus on distinguishing systems formed through the \ac{CE} and \ac{SMT} channels.
 Formation-channel contributions for their fiducial model are presented in Figure~7 of their study.
For the Level~1 classification, we assign all systems that experience at least one \ac{CE} episode (i.e., those labeled as undergoing CEE in Figure~7) to the ``with CE'' category, and classify the remaining systems as ``without CE''. For the Level~2 classification, we adopt the authors’ more detailed taxonomy, which distinguishes systems according to whether \ac{CE}, \ac{SMT}, or no mass transfer occurs in the primary and/or secondary component.
We obtain the total BBH merger rate for the fiducial model from Table~1 of \citet{Briel:2022}, which reports $\Rbhbh = 6.5\,\Gpcyr$. The authors also explore variations in the supernova-engine prescription, adopting the ``DELAYED'' and ``RAPID'' remnant mass prescription models of \citet{Fryer:2012}. As these results are not explicitly tabulated in the published article, we obtained the corresponding merger rates and formation-channel contributions via private communication. The resulting total BBH merger rates are $\Rbhbh = 57.1\,\Gpcyr$ for the ``RAPID'' model and $\Rbhbh = 48.0\,\Gpcyr$ for the ``DELAYED'' model. All data derived from these models are included in our accompanying data release.

\paragraph{\citet{Broekgaarden:2021iew} [BRO22]}
\citet{Broekgaarden:2021iew} employ an earlier version of the \textsc{COMPAS} population synthesis framework and explore a range of star-formation history and stellar evolution assumptions, yielding 560 distinct binary population-synthesis realizations.
For our analysis, we use the BBH, BHNS, and BNS merger rates from their 20 primary population-synthesis model variations. We extract the corresponding formation-channel fractions and total merger rates assuming the star-formation history model $\mathrm{xyz}=312$, which best matches the gravitational-wave observed merger rates \citep{Broekgaarden:2021efa}. These data are obtained from the authors’ publicly available repository \citep{Broekgaarden:Zenodo-DCO-formation-channels}.\footnote{\url{https://github.com/FloorBroekgaarden/DCO_FormationChannels}} Unless stated otherwise, we foremost include the $\mathrm{xyz}=312$ model, as the other cosmic star formation model variations probe uncertainties in the star-formation history rather than in binary evolution. For the detailed comparisons shown in {Figure~\ref{fig-comparison-bbh-simulation-silo}}, however, we include the full set of model variations including variations in the cosmic star formation history model. It can be seen that the choice of star formation history model impacts the formation channel contribution slightly (changing it by 20\%) and especially impacts the total (BBH) rate. 
The authors classify binaries into several formation channels. Systems that undergo a \ac{CE} phase are subdivided into the \emph{classic CE} channel, characterized by stable mass transfer prior to the formation of the first compact object followed by a CE phase after the SN (i.e. the `SMT $<$ SN $<$ CE' channel) and two additional CE-related channels: the \emph{single-core CE} and \emph{double-core CE} channels. These distinguish whether one or both stars, respectively, have evolved off the main sequence (and thus developed a core–envelope structure) before the first compact object forms.
The stable mass-transfer channel comprises systems that undergo stable mass transfer both before and after the formation of the first compact object, analogous to the `SMT $<$ SN $<$ SMT' channel. Systems that experience stable mass transfer only before or only after compact-object formation are grouped into the \emph{other} category, corresponding to `other no CE' in Fig.~\ref{fig:formation-channel-cartoon}.
We adopt these definitions for our Level~2 classification. For the Level~1 categorization, we combine the three CE-related channels (classic, single-core, and double-core CE) into a single ``with CE'' group and classify all remaining systems as ``without CE''.

\paragraph{\citet{DorozsmaiToonen:2022} [DT22]}
\citet{DorozsmaiToonen:2022} use the \textsc{SeBa} population-synthesis code to study the formation channels, properties, and merger rates of \ac{BBH} systems, exploring the impact of variations in several uncertain physical parameters, including the mass-transfer stability criterion, mass-transfer efficiency, specific angular-momentum loss, and stellar-wind strength.
The authors consider three principal formation channels: a stable mass-transfer channel and two subtypes of the CE channel, distinguished by the evolutionary state of the donor star during the second mass-transfer episode. In subtype~5b, the donor has a radiative envelope (rCEE), whereas in subtype~5c the donor has a deep convective envelope (cCEE).
We extract the corresponding formation-channel rates from the legends of their figures. For our Level~2 classification, we treat channels~5b and~5c as distinct CE subchannels, but for the Level~1 classification we combine them into a single ``with CE'' category.

\paragraph{\citet{Hendriks:2023} [Hen23]}
\citet{Hendriks:2023} use the \textsc{binary\_c} population-synthesis code to model \ac{BBH} mergers and to investigate how different prescriptions for pair-instability supernovae affect the resulting merger population.
From their Figure~4 and the accompanying text, we extract the total BBH merger rate and the relative contributions of the different formation channels for the fiducial model.
The reported total BBH merger rate, obtained by combining the SMT and CE channels, is approximately $25\,\GpcminThree\yearmin$, with a contribution of about $10\,\GpcminThree\yearmin$ from the SMT channel.
We therefore classify their SMT channel as ``without CE'' and assign the remaining contribution to the ``with CE'' category at Level~1.

\paragraph{\citet{Li:2025} [Li25]}
\citet{Li:2025} study the formation channels, mass distributions, and merger rates of \ac{BBH} systems using the binary population-synthesis code \textsc{MOBSE}, explicitly incorporating chemically homogeneous evolution (CHE) alongside the more traditional CE and SMT isolated-binary evolution pathways.
The authors explore a grid of population-synthesis models in which key uncertain physical parameters are varied, including the angular-momentum loss prescription during non-conservative mass transfer, the \ac{CE} ejection efficiency $\alpha_{\rm CE}$, the Wolf--Rayet wind mass-loss multiplier, the natal kick velocity dispersion, and the mass-transfer stability criterion.
Across all models, \ac{BBH} mergers are classified into three mutually exclusive formation channels: (i) a CHE channel, in which both stars evolve chemically homogeneously without undergoing any Roche-lobe overflow; (ii) a CE channel, in which at least one dynamically unstable mass-transfer episode leads to envelope ejection; and (iii) a SMT channel, in which both interaction phases proceed stably without a CE phase.
We extract the intrinsic merger rates at redshift $z=0.2$ for each formation channel directly from the legends of Figure~4 in \citet{Li:2025}, which report channel-specific merger rates (in units of $\mathrm{Gpc^{-3}\,yr^{-1}}$) for each model in the parameter study.
For our analysis, we treat CHE, CE, and SMT as three distinct Level~1 formation channels; no further subdivision of the CE or SMT channels is applied.
The total merger rate for each model is taken as the sum of the three reported channel contributions, consistent with the authors’ definitions\footnote{Note that due to small rounding errors, the total BBH merger rates reported in each figure in \citet{Li:2025} is not the sum of the subchannel rates. We take the sum of the subchannel rates for that reason as the total BBH rate to be consistent throughout. }.

\paragraph{\citet{Olejak:2021CE} [OL21]}
\citet{Olejak:2021CE} use the \textsc{StarTrack} population-synthesis
code to investigate the formation channels of \ac{BBH} mergers within
the isolated binary evolution scenario.
They consider three \textsc{StarTrack} models: M380.B (the fiducial
model), M480.B (which adopts a revised criterion for \ac{CE}
development), and M481.B (which further modifies this criterion via an
additional stability flag).
Model M480.B introduces a revised \ac{CE} onset criterion that requires
four simultaneous conditions to be satisfied for a \ac{CE} phase to
occur. In particular, this prescription allows \ac{SMT} to proceed for
donors within a specified radius range even for systems with strongly
unequal mass ratios at the onset of mass transfer (up to $q<8$), based
on the grid of donor-radius ranges presented by \citet{Pavlovskii:2017}.
The authors extend this grid with a fitted prescription applicable to a
broader range of masses and compact-object accretors.
Model M481.B is identical to M480.B, but includes an additional ``MT
switch'' flag that forces systems undergoing stable mass transfer to enter a
\ac{CE} phase if the donor radius exceeds twice its Roche-lobe radius.
An overview of these assumptions is provided in Figure~9 of
\citet{Olejak:2021CE}.
We extract formation-channel contributions for the three models from
Table~4 (\ac{BBH}) and Table~5 (\ac{BHNS}) of \citet{Olejak:2021CE}.
For \acp{BBH}, we interpret channel~1 in Table~4 as the
`SMT$+$SMT' channel (stable mass transfer both before and after
the formation of the first compact object).
Channels~2, 3, and~4 are grouped into the classic \ac{CE} channel
(`SMT$+$CE'), as they all involve \ac{SMT} prior to the first
supernova (SN1) followed by a \ac{CE} phase.
Channel~5 corresponds to the double-core \ac{CE} channel (DCCE).
Channel~6 is assigned to the ``other'' category, as no detailed
evolutionary pathway is specified.
For \ac{BHNS} systems, we group channels~1, 3, and~4 into the classic
\ac{CE} channel (`SMT$+$CE'), since they involve \ac{SMT} prior to
SN1 followed by a \ac{CE} phase.
Channel~2 is classified as the only \ac{SMT} channel
(`SMT$+$SMT').
Channel~5 is assigned to the ``other'' category due to the lack of
further specification.
For the Level~1 classification, we assign all channels involving at
least one \ac{CE} episode to the ``with CE'' category, and all channels
involving only stable mass transfer to the ``without CE'' category.

\paragraph{\citet{Pellouin:2025} [PE25]}
\citet{Pellouin:2025} use COSMIC to study the contribution of different
isolated binary evolution formation channels to the formation of \ac{BNS}.
They consider three main channels:
(i) their `standard channel', corresponding to the classic \ac{CE} channel
(SMT$+$CE in Figure~\ref{fig:formation-channel-cartoon});
(ii) their \emph{equal mass-ratio at ZAMS channel}, which is extremely
similar to the double-core \ac{CE} channel; and
(iii) an accretion-induced collapse (AIC) channel.
We extract the \ac{BNS} formation-channel rates from the $z \simeq 0$
results shown in their Figure~10 and convert them to units of
$\GpcminThree\yearmin$.
This yields a total local BNS merger rate of approximately
$298\,\GpcminThree\yearmin$, with contributions of
$287\,\GpcminThree\yearmin$ from the double-core CE channel and
$11\,\GpcminThree\yearmin$ from the AIC channel.
Following the authors' statement that they do not find any BNS forming
without at least one CE phase, we classify the AIC channel as an ``other''
channel \emph{with CE} in our taxonomy.
100\% of their \ac{BNS} form with at least one CE, meaning that the
contribution from channels without a CE is zero.

\paragraph{\citet{Romagnolo:2023} [RO23]}
\citet{Romagnolo:2023} use StarTrack to investigate how imposing an upper limit on stellar radial expansion affects the formation of \ac{BNS}, \ac{BHNS}, and \ac{BBH} mergers. 
We obtain their \ac{BBH}, \ac{BHNS}, and \ac{BNS} merger rates and associated formation-channel fractions via private communication, as the channel breakdown is not reported in the published paper. 
The total merger rates differ slightly (few percent) from those listed in their Table~3 due to statistical sampling noise in the authors’ merger-rate estimation.
For the formation channels, \citet{Romagnolo:2023} largely adopt the same classification scheme as \citet{Broekgaarden:2021efa}, with a few notable differences. 
In particular, their ``only stable mass transfer'' channel includes systems with no mass transfer either before or after the first supernova (though this is a negligible fraction). 
We therefore classify this entire category as a ``no CE'' channel, without subdividing into only SMT subchannels.
They further distinguish a classic channel, a double-core \ac{CE} channel, and a single-core \ac{CE} channel. 
Although their single-core \ac{CE} definition does not formally require the \ac{CE} to occur prior to the first supernova, in practice this is true for the vast majority of systems, as post-SN1 systems are otherwise likely to disrupt. 
We therefore map their \ac{CE} channels onto our ``single-core'' and ``double-core'' subchannels. 
Finally, due to their channel definitions, no systems are assigned to an ``other'' category in their classification.

\paragraph{\citet{Romagnolo:2025} [RO25]}
\citet{Romagnolo:2025} use StarTrack to investigate how different CE survival criteria, based on the presence of a convective envelope, affect the formation of compact-object mergers.
We adopt their reported  BBH, BHNS, and BNS merger rates and formation channels classification from private communication, as the latter are not explicitly presented in the paper. The total merger rates used here differ slightly from those in their Table~2 due to statistical sampling noise in the authors’ merger-rate calculations.
The models in this study explore alternative prescriptions for unsuccessful (``pessimistic'') CE evolution.
The fiducial model assumes the standard SSE-based prescription in which Hertzsprung-gap (HG) donors are not allowed to survive a CE phase. Two alternative prescriptions replace this stellar-type criterion with envelope-structure criteria derived from detailed stellar-evolution calculations. The first, $M_{\alpha,\mathrm{ML}\,1.5}$, uses fits from \citet{Klencki:2020convective} to determine whether the donor envelope is convective. The second, $M_{\alpha,\mathrm{ML}\,1.82,\mathrm{MLpp}}$, employs new fits based on \textsc{MESA} models provided by the authors. 
For each of these prescriptions, the authors additionally consider a variant in which the maximum stellar radius in SSE is artificially capped (RMAX). We retrieve and classify formation channels for all these models following the same procedure described in the previous subsection for \citet{Romagnolo:2023}.

\paragraph{\citet{RomanGarza:2021} [RG21]}
\citet{RomanGarza:2021} compute \ac{BBH} and \ac{BHNS} merger rates by coupling detailed binary stellar evolution simulations performed with \textsc{MESA} to the rapid population synthesis code \textsc{COSMIC} via an early version of the \textsc{POSYDON} framework. They compute rates for nine population-synthesis model variants that differ in their supernova remnant-mass prescriptions and natal-kick velocity assumptions. 
We extract their \ac{BBH} and \ac{BHNS} merger rates from their Table~3, adopting the combined ``CE+SMT'' rate as the total merger rate and the individual ``CE'' and ``SMT'' contributions as proxies for the classic CE channel (level~I) and the stable mass-transfer channel (level~II), respectively. The authors do not further subdivide these channels.

\paragraph{\citet{Sgalletta:2025} [Sg25]}
\citet{Sgalletta:2025} use SEVN to compute \ac{BBH} merger rates, with particular emphasis on the impact from different cosmic star formation history assumptions. 
They use the same formation channels as considered in \citet{Broekgaarden:2021efa, Broekgaarden:2021iew} based on only stable mass transfer, the classic CE channel, and the single-core CE and double-core CE channels. We retrieve their formation channel rates from digitizing Figure~6, which shows the contributions of the formation channels to the BBH merger rate. We take the rate at redshift $z\sim 0$ using a plot digitizer. Note that in the Figure it is clear that the classic CE channel dominates the formation of BBH mergers in all model variations. Interestingly, they find that the only SMT channel can especially contribute (and become dominant) at high redshifts $z \gtrsim 4$ for a few of the submodels ($\alphaCE=3$ and $\alphaCE=5$).

\paragraph{\citet{Shao:2021} [SL21]}
\citet{Shao:2021} use an updated version of BSE to investigate the impact of different supernova prescriptions on the formation and merger rates of compact-object binaries, including \ac{BBH} and \ac{BHNS} systems. We extract the intrinsic merger rates for \ac{BBH} and \ac{BHNS} systems from their Table~1 for the three supernova models considered (rapid, delayed, and stochastic prescriptions). The corresponding formation channel breakdown into stable mass transfer and CE pathways is obtained via private communication, as these are not explicitly tabulated in the paper.

\paragraph{\citet{vanSon:2022} [vSon22]}
\citet{vanSon:2022} use \textsc{COMPAS} to compute \ac{BBH} merger rates, with particular emphasis on the relative contributions of the only SMT and CE formation channels. We extract the formation-channel fractions from the right-hand panel of their Figure~4, adopting the values in the lowest-redshift bin. Although this figure presents the \emph{detectable} merger rate, it assumes a perfect detector and therefore corresponds to the underlying astrophysical rate at low redshift. This yields an only SMT fraction of $\simeq 0.37$ and a CE fraction of $\simeq 0.63$ in the lowest-redshift bin.
We adopt their total local merger rate of $R_{\mathrm{BBH}} \simeq 73\,\Gpcyr$ at $z \approx 0$ from the main text (Section~7.3).

\paragraph{\citet{vanSon:2023sfrd} [vSon23]}
\citet{vanSon:2023sfrd} use \textsc{COMPAS} to study how different metallicity-dependent star-formation histories affect the \ac{BBH} merger rate in both the only SMT and CE channels. We extract their \ac{BBH} merger rates and channel contributions from their Figures~3 and~4, summing the stable channel and CE channel components to obtain the total rate. Unless stated otherwise, we adopt the fiducial model, as the model variations primarily probe uncertainties in the star-formation history rather than in binary evolution. For the detailed comparisons shown in {Figure~\ref{fig-comparison-bbh-simulation-silo}}, however, we include the full set of model variations including variations in the cosmic star formation history model.

\paragraph{\citet{Xing:2024-BHNS-allZ} [Xing24]}
\citet{Xing:2024-BHNS-allZ} use the \textsc{POSYDON} binary population synthesis framework to study the formation of \ac{BHNS} mergers across a broad range of metallicities.
We extract the total BHNS merger rates from Table~1, using the reported $\mathcal{R}_{\rm NSBH}$ values from the published version of the paper.\footnote{Note that their earlier arXiv version reports slightly different NSBH merger rates.}
This yields four models: a fiducial model, a variant with $\alpha_{\rm CE}=2$, and two variants with modified core-collapse kick dispersions $\sigma_{\rm cc}$.
The authors additionally explore versions of each model with the maximum neutron-star mass reduced from the default $2.5\,M_\odot$ to $2.0\,M_\odot$.
However, formation-channel contributions for these submodels are not tabulated, and inspection of Figure~3 indicates that this variation has a negligible effect on the relative channel fractions.
We therefore exclude these submodels from our analysis.

\section{Appendix B: Additional Studies Not Included}
We briefly summarize additional simulations from the literature that report formation-channel contributions but are excluded from the main compilation because they consider only a fixed metallicity, do not provide cosmologically integrated intrinsic merger rates, or lack the quantitative channel breakdown required by our unified taxonomy. See also Section~\ref{sec:fc-discussion} for more details.

\paragraph{\citet{Andrews:2015}}
\citet{Andrews:2015} perform a large population-synthesis study of BNS using \textsc{StarTrack}, comparing their model predictions to the observed Galactic BNS population in the orbital period-eccentricity plane. They find that all (short-period) BNS systems form through binary evolution channels that include a \ac{CE} phase following the first \ac{SN}.
They identify three dominant formation channels, distinguished by (i) whether the system undergoes stable Case~BB mass transfer after the \ac{CE} phase and (ii) whether the second neutron star forms via an iron core-collapse or an electron-capture \ac{SN}. Channels that include Case~BB mass transfer produce systematically tighter binaries, while channels involving electron-capture supernovae, and hence low natal kicks, give rise to systems with lower eccentricities.
At (approximately) solar metallicity, they find that channels involving Case~BB mass transfer dominate the formation of BNS systems. Their results further highlight that electron-capture supernovae are required to reproduce the observed population of low-eccentricity systems, and they highlight the coupled role of post-\ac{CE} mass transfer and natal kick physics in shaping the observable DNS population.
We do not include this study in our formation-channel comparison because it does not report formation-channel fractions for intrinsic merger rates and focuses instead on a fixed metallicity.

\paragraph{\citet{Banerjee:2024}} \citet{Banerjee:2024} use the rapid population-synthesis code \textsc{BSE} to model the properties of \ac{BBH} mergers formed through isolated binary evolution, with a focus on the spin–mass ratio correlation. They explore both  \ac{CE} and \ac{SMT} pathways by constructing models that are dominated by one channel or the other by changing the critical mass ratio \citep[cf.][]{Olejak:2021CE}. In particular, their default models yield $\sim90\%$ of \ac{BBH} mergers through \ac{CE} evolution, while models adopting extreme mass-transfer stability assumptions (e.g., $q_{\rm c}=8$) instead produce $\sim90\%$ of systems through \ac{SMT}.
We do not include this study in our formation-channel comparison because it does not report formation-channel fractions for intrinsic merger rates based on self-consistent model populations. Instead the analysis focuses on normalized distributions at fixed metallicities rather than intrinsic rates.

\paragraph{\citet{Belczynski:2002}}
\citet{Belczynski:2002} use the \textsc{StarTrack} population-synthesis code to model the formation of \ac{BBH}, \ac{BHNS}, and \ac{BNS} systems at solar metallicity $Z=0.02$. They present representative formation pathways in their Table~3, classifying channels based on the number and type of CE episodes, including single-core CE, double-core CE, and CE with significant accretion during the \ac{CE} (`hyper-critical accretion'). In these models, merging systems predominantly form through channels involving at least one \ac{CE} phase, highlighting the importance of CE evolution in producing sufficiently tight binaries. 
We do not include this study in our formation-channel overview because it does not report systematic formation-channel fractions for intrinsic merger rates across its model grid.

\paragraph{\citet{Belczynski:2022formationChannel}}
\citet{Belczynski:2022formationChannel} investigate \ac{BBH} formation using the \textsc{StarTrack} population-synthesis code, exploring variations in the initial mass function and supernova engine prescriptions (delayed and rapid). They focus on the relative contributions of SMT and CE channels, showing that \ac{SMT}-formed systems tend to produce slightly more massive \ac{BBH} mergers than those formed through \ac{CE} evolution.
They further demonstrate that assumptions about the onset and development of \ac{CE} evolution significantly affect both the dominant formation pathway (i.e., \ac{CE} versus Roche-lobe overflow) and the resulting \ac{BBH} mass distribution. In particular, they find that \ac{BBH} formation at low metallicities ($Z \lesssim 0.4\,Z_\odot$) is dominated by the \ac{CE} channel (with a factor $\sim 3\times$ higher formation efficiencies), while at higher metallicities $Z \gtrsim 0.4\,Z_\odot$ the contributions from \ac{SMT} and \ac{CE} become comparable. We do not include this study in our formation-channel overview because the absence of a quantitative decomposition into channel contributions as a function of merger rate prevents a direct comparison with the framework adopted in this work. We note that the models considered are the same as those presented in \citet{Olejak:2021CE} which are included.

\paragraph{\citet{Briel:2026}} \citet{Briel:2026}  present a full ZAMS-to-merger population study of BBH formation through the SMT channel using POSYDON with detailed binary grids across eight metallicities. 
They discuss details of the SMT channel, but do not present intrinsic merger rates, which is why we do not include this study in our key results. 
They find that SMT BBH mergers arise predominantly from short-period progenitors ($P_{\rm ZAMS}\lesssim 10$ days) in which \emph{both} the initial STAR+STAR interaction and the later STAR+BH interaction occur while the donor is on the main sequence (i.e., Case A in both phases), with only a limited contribution from wider Case B/C systems. In their fiducial (zero natal-kick) models, SMT does not produce BBH mergers above $Z\gtrsim 0.2\,Z_\odot$ because wind-driven mass loss widens the orbit, preventing merger within a Hubble time. The resulting SMT population favors near-unity mass ratios (via mass-ratio reversal) and shows a bimodal $\chi_{\rm eff}$ distribution with peaks at $\chi_{\rm eff}=0$ and $\chi_{\rm eff}\simeq 0.1$--0.15; importantly, the $\chi_{\rm eff}=0$ peak appears only at low metallicity ($Z<0.01,Z_\odot$), while at higher $Z$ only the $\chi_{\rm eff}\sim 0.15$ peak remains. They also find characteristically long delay times: even at $Z=10^{-4}Z_\odot$ the minimum delay time is $\sim0.2\,\mathrm{Gyr}$, and the delay-time distribution shifts to even longer times with increasing metallicity, implying an intrinsic SMT BBH merger-rate density that rises toward the present day ($z=0$). Finally, introducing a mass-scaled natal-kick prescription leaves the main SMT population largely intact but can add eccentricity at second collapse, enabling an additional low-mass, unequal-mass-ratio sub-population (and allowing SMT mergers at $Z>0.2\,Z_\odot$) that would otherwise be too wide to merge within a Hubble time.

\paragraph{\citet{Chattaraj:2026} }
\citet{Chattaraj:2026}  use detailed binary-evolution models with POSYDON to study the formation of BNS systems at solar metallicity, focusing on the role of CE evolution and its dependence on donor structure and evolutionary stage. In particular, they find that BNS almost always form through two distinct CE subchannels split by Case B and Case C mass transfer that lead to different outcomes in terms of whether binaries merge within a Hubble time, and explore how assumptions about envelope ejection efficiency and supernova physics affect BNS merger rates and properties. They find that $\gtrsim$95\% of all BNS systems form through these two CE subchannels and that such systems always undergo a stable case BB mass transfer phase and an ultra-stripped SN after the CE phase.  Up to 5\% of BNS form mostly through a non-interacting channel, and more rare ($\lesssim1\%$) a channel with only stable mass transfer, and the double-core CE channel ($\lesssim2\%$). 
We do not include their results in our formation-channel overview because the study does not report formation-channel fractions and only has intrinsic merger rates based on a single (solar) metallicity, making it difficult to consistently compare with the studies included in our compilation.

\paragraph{\citet{Chruslinska:2018}}
\citet{Chruslinska:2018} use StarTrack to investigate the formation channels of \acp{BNS}, focusing primarily on individual metallicity environments. For solar metallicity, they find that the dominant formation pathway is the ``classic'' \ac{CE} channel, which typically includes a subsequent phase of stable mass transfer onto the compact object (i.e., Case~BB mass transfer). This model is an order of magnitude more common than any other subchannels they identify.% .
They also identify subdominant channels, including formation through accretion-induced collapse, where stable mass transfer prior to the first supernova leads to the formation of a white dwarf that subsequently collapses into a neutron star. This pathway can contribute at most  10\%  of all BNS but 
most of them (90\%) require CE with HG donor before
the formation of the second NS.
So combined, the CE channel likely forms at least $\gtrsim95\%$, if not more, of their BNS at solar metallicity, which we add to Table~\ref{tab:fc_Z_statements_bns}. 
Additionally, they discuss alternative channels involving two \ac{CE} phases after the first supernova, as well as channels with \ac{CE} evolution prior to the first supernova, although the latter are found to be rare and at least an order of magnitude less efficient than the classic channel.
We do not include this study in our main comparative analysis because the results are primarily presented for individual metallicities (e.g., solar versus low metallicity), and the work does not provide a systematic breakdown of formation channel contributions integrated over cosmic star-formation and metallicity evolution. In particular, the relative contributions of different channels to the cosmological \ac{BNS} merger rate are not quantified in a way that can be consistently mapped onto our unified taxonomy.

\paragraph{\citet{DallAmico:2025accertionCHE}}
\citet{DallAmico:2025accertionCHE} investigate the formation of \ac{BBH}, \ac{BHNS}, and \ac{BNS} mergers through accretion-induced chemically homogeneous evolution (CHE). They find that, in standard binary evolution channels, a \ac{CE} phase is typically essential to shrink the orbit and enable mergers, although this is not quantified within their framework.
In contrast, models including accretion-induced CHE produce binaries that often avoid \ac{CE} evolution altogether, as the secondary stars remain compact and do not fill their Roche lobes, thereby suppressing mass transfer phases. However, the majority of these CHE systems have inspiral times exceeding a Hubble time, resulting in a strongly reduced number of merging double compact objects.
We do not include this study in our formation-channel comparison because it does not provide a systematic breakdown of channel contributions within a simulation to the merging population that can be consistently mapped onto our unified taxonomy.

\paragraph{\citet{Dominik:2012,Dominik:2013}}
\citet{Dominik:2012,Dominik:2013} present an extensive suite of population-synthesis simulations using \textsc{StarTrack}, exploring a wide range of binary-evolution assumptions. They identify representative formation pathways (Tables~4 and 5 of \citealt{Dominik:2012}) at fixed metallicities, highlighting characteristic evolutionary sequences for \ac{BBH}, \ac{BHNS}, and \ac{BNS} systems. Across their models, merging systems overwhelmingly originate from channels that involve at least one \ac{CE} phase. 
Quantitatively, at solar metallicity, at least $94.2\%$ (BNS), $97.2\%$ (BHNS), and $98.9\%$ (BBH) of merging systems form through channels that include a CE phase, with the remaining fraction categorized as ``other'' (i.e., not assigned to a specific formation pathway). At subsolar metallicity ($0.1,\Zsun$), the CE fraction remains similarly high: $\geq 91.8\%$ (BNS), $93.6\%$ (BHNS), and $96\%$ (BBH). We add these fractions to Table~\ref{tab:fc_Z_statements_bns}. The identified channels span a range of CE configurations, including CE episodes prior to the first supernova, double-core CE, pathways involving multiple CE phases, and involving accretion-induced collapse (AIC). Notably, none of the reported channels correspond to purely stable mass transfer sequences or to systems that avoid mass transfer altogether (although these could contribute to `other').
We do not include these studies in our formation-channel overview because they do not report quantitative formation-channel fractions for intrinsic merger rates across their model grid.

\paragraph{\citet{Ioro:2023sevn}}
\citet{Ioro:2023sevn} explore the formation and properties of \ac{BBH}, \ac{BHNS}, and \ac{BNS} mergers using an extensive suite of \textsc{SEVN} simulations spanning a wide range of physical assumptions. They present a detailed breakdown of formation-channel contributions to the formation efficiency as a function of metallicity (their Figures~14 and 16), adopting a Level~2 classification that distinguishes between classic CE (SMT+CE), only SMT (SMT+SMT), single-core and double-core CE, and an ``other'' category.

We do not include these results in our main overview, as the authors do not report formation-channel contributions in terms of intrinsic merger rates. Instead, we incorporate their metallicity-dependent formation-channel fractions using the publicly available dataset from \citet{iorio_2023_7794546}, accessed via the plotting code at \url{https://gitlab.com/iogiul/iorio22_plot/-/tree/v3/formation_channels}, and include them in Figures~\ref{fig:formation-efficiency-results} and~\ref{fig:formation-efficiency-results-detailed}.

Across their broad model variations—including changes in $\alpha_{\rm CE}$, chemically homogeneous evolution, supernova natal kicks, remnant-mass prescriptions, and envelope binding energies—they find a robust and consistent result: \ac{BNS} mergers form exclusively through channels involving at least one CE phase. In contrast, \ac{BHNS} and \ac{BBH} systems exhibit a strong dependence on model assumptions and metallicity, with contributions from both CE and without-CE channels.

\paragraph{\citet{Kruckow:2018}}
\citet{Kruckow:2018} used the \textsc{Combine} framework to study the formation of gravitational-wave sources from isolated binary evolution. We exclude their results from our cosmological rate comparison because their analysis is limited to a single metallicity ($Z=0.0088$) and includes all compact-object binaries, including systems with delay times exceeding a Hubble time.
Importantly, they state that their formation-channel distribution of merging systems differs significantly from that of the full formation population for which they present quantified formation channel breakdowns in their study. While their Appendix~C indicates that BBHs predominantly form without mass transfer (channel D) in the overall population, the subset of merging systems is instead dominated by channels involving stable mass transfer (channel A) and CE evolution (channel C). 
For merging BHNS systems, channels involving stable mass transfer (channel A) and classical CE evolution (channels B/C) appear to dominate. In the case of binary neutron star mergers, the classical CE channel (B/C) is the primary formation pathway. The authors do not provide precise quantitative breakdowns for the merging population, and we therefore treat these conclusions as qualitative.

\paragraph{\citet{Mestichelli:2025}}
\citet[][]{Mestichelli:2025} model the formation of \ac{BHNS} and \ac{BNS} mergers from metal-free Population~III and early Population~II progenitors using the \textsc{SEVN} binary population-synthesis code. They compute merger-rate densities by convolving their merger catalogs with metallicity-dependent cosmic star-formation histories using CosmoRate. Their analysis focuses exclusively on low-metallicity environments, adopting $Z \sim 10^{-11}$ for Population~III and $Z \sim 10^{-4}$ for Population~II star models.

The authors define formation channels (their Table~2) based on whether systems undergo stable mass transfer  and/or CE evolution relative to the formation of the first compact remnant: Ch.~0 (no interaction), Ch.~1 (SMT+CE), Ch.~2 (SMT+SMT), Ch.~3 (single-core CE), Ch.~4 (double-core CE), and Ch.~5 (interaction only after the first remnant; CE/SMT not specified).
Because the study is restricted to fixed low-metallicity simulations and does not provide a quantitative breakdown of formation channels for the intrinsic merger population, we do not include it in our main results. However, the authors report channel contributions to the merger formation efficiency $\eta$ (Tables~4 and 6), which we include in Appendix~\ref{sec:appendix-formation-channel-contributions-efficiency-metallicity}. For consistency with our classification, we group Channels~1, 3, and 4 as `with CE', Channels~0 and 2 as `without CE', and Channel~5 as `other' (CE not specified).
Under this mapping, \ac{BHNS} systems show a substantial contribution from both pathways: $46.2\%$ ($36\%$) form through channels with CE and $46.9\%$ ($55.9\%$) without CE for Population~II (Population~III), respectively, with the remaining $6.7\%$ ($8\%$) in the `other' category.
In contrast, \ac{BNS} formation is overwhelmingly dominated by CE evolution: $99.3\%$ ($91.9\%$) of Population~II (Population~III) systems form through channels involving at least one CE phase, with the remainder attributed to the `other' category, which likely also includes CE evolution. No contribution from `without CE' channels is reported for BNS systems.

\paragraph{\citet{Neijssel:2019}}
\citet{Neijssel:2019} use \textsc{COMPAS} to model the rates and properties of detectable \ac{BBH}, \ac{BHNS}, and \ac{BNS} populations under different cosmic star-formation histories. They find that the detectable \ac{BBH} population is strongly dominated by channels involving only stable mass transfer (i.e., without a \ac{CE} phase), contributing nearly $80\%$ of mergers (their Figure~12), compared to $15.5\%$ from the classic \ac{CE} channel, $0.3\%$ from double-core \ac{CE}, and the remainder from other pathways.

We do not include this work in our main overview, as formation channel fractions are reported only for the detectable population rather than intrinsic merger rates. In addition, more recent \textsc{COMPAS} studies with comparable fiducial assumptions provide directly comparable intrinsic channel breakdowns and are therefore prioritized in our compilation.

However, \citet{Neijssel:2019} also present \ac{BBH} formation-channel contributions as a function of metallicity in their Figure~1, expressed as formation efficiencies (mergers per unit star-forming mass). We include these in Figures~\ref{fig:formation-efficiency-results} and~\ref{fig:formation-efficiency-results-detailed} by digitizing their Figure~1 using WebPlotDigitizer \citep{rohatgi2020webplotdigitizer}. From these results, they find that $\sim 30\%$--$50\%$ of \ac{BBH} systems form through channels without \ac{CE}, with the remainder forming through \ac{CE}-involving pathways, and only a weak dependence on metallicity.

\paragraph{\citet{Olejak:2024} }
\citet{Olejak:2024} use StarTrack to investigate the spin and mass ratio properties of \ac{BBH} population from the \ac{CE} and \ac{SMT} channels. We do not include their study as they do not report BBH merger rates or formation channel contributions. Their models are similar to the ones used in \citet{Olejak:2021CE} which are included in our main overview. The default model in \citet{Olejak:2024} finds a dominant contribution from the \ac{SMT} channel. For the same reason we also do not include \citep{Olejak:2021iux}. 

\paragraph{\citet{Oh:2023}} \citet{Oh:2023} use the rapid population-synthesis code \textsc{COSMIC} to investigate the evolutionary pathways leading to asymmetric \ac{BBH} and \ac{BHNS} mergers, with a particular focus on the role of natal kicks and the timing of mass-transfer phases (e.g., whether systems undergo \ac{CE} before or after the first supernova). They identify several formation channels and show that highly asymmetric systems are preferentially produced through channels involving \ac{CE} before the first supernova. 
We do not include this study in our formation-channel comparison because it does not report formation-channel fractions for intrinsic merger rates, and therefore cannot be consistently incorporated into our quantitative channel accounting. In addition, the simulations are performed at only three discrete metallicities, which limits direct comparison with studies that explore a broader metallicity range or integrate over cosmic star formation.

\paragraph{\citet{Riley:2020} }
\citet{Riley:2020} 
implement \ac{CHE} in the rapid population-synthesis code COMPAS and, under a consistent set of assumptions, compute \ac{BBH} merger rates including the redshift-dependent merger-rate density and detectability for CHE and ``conventional'' isolated binaries, finding CHE could contribute up to $\sim$70\% of isolated-binary BBH detections. Their CHE systems do not undergo mass transfer that could harden the binary except for RLOF at ZAMS, which is assumed to be an over-contact phase leading to mass equilibration and CHE. 
We do not include their rates in our formation-channel overview because, while they model both dynamically stable mass transfer and CE evolution in the non-CHE channel, they do not report a quantitative breakdown of what fraction of their BBH progenitors experienced CE versus only stable MT, which is the split we need for our channel accounting.

\paragraph{\citet{Stevance:2023}}
\citet[][]{Stevance:2023} use \textsc{BPASS} to explore isolated binary evolution pathways leading specifically to \ac{BNS} systems analogous to GW170817. Because the study focuses on reproducing a single event rather than providing a comprehensive formation-channel breakdown for the full BNS population, we do not include it in our main analysis.
They find that all GW170817-like systems in their models undergo at least one \ac{CE} phase, with many experiencing two or even three CE episodes (see their Figure~2). The identified pathways include variants of the classic CE channel (stable mass transfer prior to the first supernova followed by a CE phase), but the dominant contribution arises from systems that undergo a CE phase both before and after the formation of the first compact object, with or without additional stable mass transfer.
We note that \textsc{BPASS} adopts a different treatment of CE evolution compared to most population-synthesis codes and is based on the \textsc{STARS} stellar evolution tracks, which may contribute to differences in inferred channel contributions. Interestingly, the authors still find that all formation channels for GW170817 include at least one \ac{CE} phase consistent with other work.

\paragraph{\citet{Tanaka:2023}}
\citet[][]{Tanaka:2023} use the \textsc{binary\_c} population-synthesis code to study \ac{BNS} merger formation across a wide range of model assumptions, varying the \ac{CE} efficiency parameter $\alpha$, the core-collapse kick dispersion $\sigma_{\rm cc}$, and metallicity. Their simulations are limited to three discrete metallicities and do not provide a detailed breakdown of formation-channel contributions to the intrinsic merger rates, and we therefore do not include this study in our main analysis.
Nonetheless, the authors report that a large majority of BNS systems—between $81\%$ and $100\%$—undergo at least one \ac{CE} phase involving a neutron star during their evolution (but they don't specify whether this is for specific metallicity grids or for the entire population). They also identify a small contribution from accretion-induced collapse, with up to $0.7\%$ of all BNS systems and up to $1.7\%$ of merging BNS systems forming one of the neutron stars via a white dwarf accretion-induced collapse channel. These fractions are reported for individual simulations and are not weighted to represent the intrinsic merging population.

\paragraph{\citet{Ugolini:2025}}
\citet{Ugolini:2025} use \textsc{SEVN} to evolve a large grid of isolated binary simulations of \ac{BBH} systems, exploring how variations in stellar evolution and binary interaction physics (including \ac{CE} efficiency, supernova prescriptions, and the initial mass function) shape the resulting black hole mass distribution. In particular, they demonstrate that both stable mass transfer and \ac{CE} evolution can contribute to features in the \ac{BBH} mass spectrum, such as the $\sim 35\,M_\odot$ peak, with their relative importance depending sensitively on the assumed \ac{CE} efficiency. The \ac{CE} channel seems to produce the low mass BH peak below $\lesssim 10\Msun$, but this is based on samples from the simulation (not intrinsic rate). 
We do not include this study in our comparative analysis because it is a preliminary work and does not calculate intrinsic BBH merger rates required for our comparison.

\paragraph{\citet{vanSon:2022-nopeaks}}
\citet{vanSon:2022-nopeaks} use the rapid population-synthesis code \textsc{COMPAS} to investigate the formation of \ac{BBH} and \ac{BHNS} mergers through the stable mass transfer (\ac{SMT}) channel, with a particular focus on understanding the origin of features in the low-mass end of the black hole mass distribution $(\lesssim 10\Msun)$. They derive an analytical expression for the minimum primary black hole mass set by stability criteria during binary evolution and explore how variations in key physical assumptions (e.g., mass-transfer efficiency, core mass fraction, and supernova prescriptions) impact the resulting merger rates and mass distributions. Their models are explicitly constructed to include only systems that undergo exclusively stable mass transfer, thereby isolating the SMT channel and excluding binaries that experience a \ac{CE} phase or other evolutionary pathways.
We do not include this study in our formation-channel comparison because it does not provide formation-channel fractions or total intrinsic merger rates across multiple channels. Instead, the reported rates correspond solely to the SMT channel by design, and therefore cannot be directly compared to studies that model the relative contributions of different formation pathways within a self-consistent population.

\paragraph{\citet{vanSon:2024}}
\citet{vanSon:2024} present an extensive suite of \textsc{COMPAS} simulations to investigate the formation efficiency of \ac{BBH}, \ac{BHNS}, and \ac{BNS} mergers as a function of metallicity. They report formation-channel contributions in terms of formation efficiencies (their Figure~4), which we reproduce using the publicly available dataset \citep{van_son_2024_14508864} and include in Figures~\ref{fig:formation-efficiency-results} and~\ref{fig:formation-efficiency-results-detailed}. Their results show that \ac{BNS} systems form almost exclusively through channels involving a \ac{CE} phase across all metallicities, \ac{BHNS} systems are predominantly formed via \ac{CE} channels, and \ac{BBH} systems can receive a substantial contribution from channels without \ac{CE}, including \ac{CHE}. We do not include this study in our main tables and figures, as it does not provide a consistent breakdown of formation channels in terms of intrinsic merger rates.

\paragraph{\citet{VignaGomez:2018}}
\citet[][]{VignaGomez:2018} use \textsc{COMPAS} simulations at a single (solar-like) metallicity to investigate the rates and properties of \ac{BNS} mergers. In their fiducial model ($Z = 0.0142$), they identify two dominant formation pathways: the classic channel, accounting for $\sim70\%$ of systems, and the double-core \ac{CE} channel, contributing $\sim21\%$. The remaining $\sim9\%$ of systems form through a range of alternative pathways (likely still involving \ac{CE} phases). Their total contribution from channels with \ac{CE} is thus at least $\gtrsim91\%$ at $Z = 0.0142$. 
Because the analysis is restricted to a single metallicity and does not incorporate cosmological weighting, we do not include these results in our comparison of intrinsic merger-rate predictions.

\paragraph{\citet{VignaGomez:2020}}
\citet[][]{VignaGomez:2020} use \textsc{COMPAS} to investigate the formation channels and \ac{CE} properties of binaries that produce \ac{BNS} mergers at solar metallicity ($Z = 0.0142$). They focus in particular on how different double neutron star subpopulations—classified by the donor type during the \ac{CE} phase—map onto distinct evolutionary pathways, such as the classic \ac{CE} channel and the double-core \ac{CE} channel.
In their fiducial model, $\sim69\%$ of BNS systems form through the classic channel and $\sim14\%$ through the double-core \ac{CE} channel (see their Section~3.1). The remaining $\sim17\%$ arise from variations of these dominant pathways, typically involving modifications to the sequence of mass-transfer episodes or the omission of specific phases. While not always explicitly classified, these alternative channels likely also involve at least one \ac{CE} phase. Taken together, this implies that at least $\gtrsim 83\%$ of BNS systems in their simulations form through channels involving \ac{CE} evolution.
Because the study is restricted to a single metallicity and does not provide cosmologically weighted merger rates, we do not include it in our main analysis.

\paragraph{\citet{Xing:2024-BHNS-solarZ}} \citet{Xing:2024-BHNS-solarZ} use detailed binary-evolution models to investigate the formation of compact-object binaries, with a particular focus on the role of mass transfer and \ac{CE} evolution in shaping their properties. They identify characteristic evolutionary pathways and examine how different mass-transfer assumptions influence the resulting populations.
For their solar-metallicity simulation, they find that $\sim70\%$ of \ac{BHNS} mergers form through channels involving a \ac{CE} phase (primarily the classic CE channel), while the remaining $\sim30\%$ form through exclusively stable mass transfer, with most of these systems undergoing two stable mass-transfer episodes; one before SN1 and one after SN1. They also find distinct (effective) spin properties and period distributions between the CE and SMT populations, which we summarize in Section~\ref{sec:fc-discussion}.
We do not include this study in our formation-channel comparison, as it is restricted to a single (solar) metallicity and does not provide channel fractions for intrinsic merger rates across a broader model grid. This limits a consistent comparison with studies that explore metallicity-dependent trends and report quantitative channel contributions. We instead include the companion study \citet{Xing:2024-BHNS-allZ}, which extends this analysis to multiple metallicities and provides a more suitable basis for comparison within our framework.

\paragraph{\citet{Zevin:2020}}
\citet[][]{Zevin:2020} use the rapid population-synthesis code \textsc{COSMIC} to investigate the formation of GW190814-like compact binaries. They identify two main evolutionary pathways: Channel~A, which proceeds predominantly through stable mass transfer, and Channel~B, which resembles the classic \ac{CE} channel. These pathways highlight the role of supernova physics and natal kicks in producing highly asymmetric merging systems.
They find that at low metallicities both channels contribute comparably to the formation of GW190814-like \ac{BBH} systems, whereas at higher metallicities ($Z \gtrsim \Zsun/30$) the only stable mass transfer channel (Channel~A) becomes dominant, accounting for $\sim73\%$ of such systems.
We do not include this study in our formation-channel comparison because the authors focus on GW190814-like \ac{BBH} rates but do not present a formation channel breakdown for the intrinsic BBH merger rate.

%%%%
\section{Appendix C: Population-Synthesis Parameters, Acronyms, and Assumptions}
\label{app:fc_data-specs-acronyms}

This appendix summarizes the key population-synthesis parameters, assumptions, and acronyms used throughout this work and in the compiled literature sample. Our meta-analysis combines results from a wide range of studies that adopt different physical prescriptions, parameterizations, and nomenclature. To enable a consistent comparison, we standardize terminology and group related assumptions (e.g., mass-transfer stability, angular-momentum loss, and supernova physics) under common labels used in our figures. 
For each parameter, we provide a brief description of its physical meaning, indicate the symbol used in this work, and list representative studies that vary it.  We note that many of these parameters likely encode broader assumptions about stellar structure and hydrodynamic envelope response that impact the importance of CE in forming gravitational-wave sources (see Section~\ref{sec-discussion:hydrodynamic-constraints} for a discussion).

\paragraph{Stellar-evolution track labels (SSE, MESA, PARSEC, STARS, hybrid)}
The ``track'' column in Figures~\ref{fig:level-1-BBH}--\ref{fig:level2-bns}
indicates the stellar-evolution framework used to model single-star
structure and evolution within each population-synthesis code.
\textsc{SSE} refers to the analytic fitting formulae of \citet{Hurley:2002},
which parameterize stellar radii, luminosities, and core masses as a
function of initial mass, metallicity, and age; this framework is used in
\textsc{COMPAS}, \textsc{StarTrack}, \textsc{SeBa}, \textsc{BSE},
\textsc{binary\_c}, and \textsc{MOBSE}.
\textsc{MESA} denotes grids of detailed stellar-evolution models computed
with the \textsc{MESA} code \citep[Modules for Experiments in Stellar
Astrophysics;][]{Paxton:2011}, and is used in \textsc{POSYDON}
\citep{Fragos+2023ApJS,Andrews+2025ApJS, Xing:2024-BHNS-allZ}.
\textsc{PARSEC} denotes grids based on the PARSEC stellar tracks \citep{Bressan:2012}, used in \textsc{SEVN} \citep[e.g.][]{Sgalletta:2025}.
\textsc{STARS} refers to the Cambridge \textsc{STARS} stellar-evolution code \citep{Eldridge:2004-STARS}, whose single-star models form the basis of the binary population-synthesis code \textsc{BPASS} \citep{Eldridge:2017, StanwayEldridge:2018}.
\textsc{hybrid} (appearing for the early \textsc{POSYDON} models of \citealt{Bavera:2021} and \citealt{RomanGarza:2021}) indicates a combination of \textsc{MESA} grids for the binary interaction phases with the rapid \textsc{COSMIC} code for phases not covered by the grids.
These distinct stellar-evolution frameworks encode different assumptions
about stellar radii, core-mass growth, wind mass loss, and the onset of
envelope inflation, all of which directly affect whether a binary undergoes
Roche-lobe overflow and whether that overflow is stable.

\paragraph{$\alphaCE$ (common-envelope efficiency)}
The CE efficiency parameter $\alphaCE$ (or $\alpha$) sets how effectively orbital energy is used to unbind the donor envelope during a CE phase. Higher $\alphaCE$ values correspond to more efficient envelope ejection and typically wider post-CE separations, while low $\alphaCE$ promotes tighter binaries or mergers during the CE phase. 
This parameter is denoted by $\alpha$ in our figures. It is widely varied across studies, including \citet{Broekgaarden:2022, Boesky:2024gw, Briel:2022, Bavera:2021, Xing:2024-BHNS-allZ, Li:2025}, and \citet{Sgalletta:2025}.

\paragraph{$\beta$ (mass-transfer efficiency)}
The mass-transfer efficiency $\beta$ describes the fraction of transferred mass that is accreted by the companion during Roche-lobe overflow between two stars. It controls both the mass-ratio evolution and the amount of angular momentum lost from the system, thereby affecting the stability of mass transfer and orbital evolution. 
This parameter is denoted by $\beta$ in our figures. It is varied in, for example, \citet{Broekgaarden:2022, Boesky:2024gw}, and \citet{DorozsmaiToonen:2022}. In these studies it is only used for mass transfer onto a star (for compact objects Eddington limited accretion is assumed).

\paragraph{$\gamma$ (angular-momentum loss parameter)}
The angular-momentum loss parameter $\gamma$ sets the specific angular momentum carried away by mass lost from the binary, but its physical meaning depends on the adopted prescription. In the $\gamma$-formalism (e.g., \citet{DorozsmaiToonen:2022}), $\gamma$ directly parameterizes the specific angular momentum of the ejected material relative to the binary orbit, with larger $\gamma$ corresponding to more efficient orbital angular-momentum loss and thus stronger orbital shrinkage during non-conservative mass transfer. In contrast, other implementations (e.g., \citet{Li:2025}) adopt alternative prescriptions in which $\gamma$ can take negative or varying values depending on the assumed mass-loss channel, effectively modifying whether mass loss extracts or redistributes angular momentum within the system. 
As a result, variations in $\gamma$ can qualitatively change the orbital evolution during stable mass transfer, determining whether binaries tighten sufficiently to merge within a Hubble time or instead widen and avoid merger. This parameter is denoted by $\gamma$ in our figures and is explicitly varied (though in different underlying prescriptions) in \citet{DorozsmaiToonen:2022} and \citet{Li:2025}.

%%%%%%%%%%%%%%%%%%%
\paragraph{$q_{\rm c}$ and $\xi$ (mass-transfer stability criteria)}
The stability of mass transfer is commonly parameterized either in terms of
a critical mass ratio $q_{\rm c}$, or through the \textsc{zeta
prescription}, which uses the $\zeta$-based stability criterion
\citep[following][]{Soberman:1997} with response coefficients such as the
adiabatic mass--radius exponent $\xi \equiv \mathrm{d}\ln R /
\mathrm{d}\ln M$ (e.g.\ $\xi_{\rm ad,rad}$ in \citet{DorozsmaiToonen:2022}
for radiative envelopes).
In the $q_{\rm c}$ formalism, mass transfer becomes dynamically unstable
when the donor-to-accretor mass ratio exceeds a threshold value, leading to
a CE phase.
In contrast, $\xi$-based prescriptions compare the donor's radial response
to mass loss with the response of the Roche lobe, providing perhaps a more
physically motivated, structure-dependent stability criterion.
Different studies adopt different implementations: some assume fixed values
of $q_{\rm c}$, while others use prescriptions based on detailed stellar
models \citep[e.g.,][]{Claeys:2014, Ge:2020} or directly compute $\xi$ as
in \citet{DorozsmaiToonen:2022}.
These choices can shift systems between stable and unstable mass-transfer
regimes and therefore strongly affect the relative importance of formation
channels with and without CE evolution.
This parameter is denoted by $q_{\rm c}$ or $\xi$ in our figures and is
varied in studies such as \citet{Bavera:2021}, \citet{Olejak:2021CE},
\citet{Li:2025}, and \citet{DorozsmaiToonen:2022}.
 
The following distinct $q_{\rm c}$ labels appear as varied parameters in
Figures~\ref{fig:level-1-BBH}--\ref{fig:level2-bns}:
\begin{itemize}
  \item \textsc{$q_{\rm c}$ Belcz}: the $q_{\rm c}$ prescription of
    \citet{Belczynski:2008}, as implemented and described in
    \citet{Bavera:2021} (their Table~1).
    In this prescription, the critical mass ratio is set to a fixed
    value that depends on the stellar type of the donor, following the
    original \textsc{StarTrack} convention \citep{Belczynski:2008};
    for giant donors the adopted value is effectively
    $q_{\rm c} \approx 3$--$4$ depending on evolutionary phase.
    Used in \citet{Bavera:2021} (\textsc{BA21}).
 
  \item \textsc{$q_{\rm c}$ Claeys}: the $q_{\rm c}$ prescription of
    \citet{Claeys:2014}, which provides fitting formulae for the
    critical mass ratio as a function of the donor's stellar type and
    mass, derived from detailed stellar-structure arguments.
    Compared to the \citet{Belczynski:2008} prescription, the Claeys
    values tend to be somewhat lower for giant donors, leading to a
    higher incidence of CE evolution.
    Used in \citet{Bavera:2021} (\textsc{BA21}).
 
  \item \textsc{$q_{\rm c}$ Ge}: the $q_{\rm c}$ prescription of
    \citet{Ge:2020}, which is based on adiabatic mass-loss calculations
    of detailed stellar models and provides the adiabatic mass--radius
    exponent $\xi_{\rm ad}$ as a function of donor mass, evolutionary
    state, and metallicity.
    These calculations generally yield higher stability thresholds
    than either the \citet{Belczynski:2008} or \citet{Claeys:2014}
    prescriptions for radiative-envelope donors, and therefore produce
    a larger fraction of systems evolving through stable mass transfer.
    Used in \citet{Li:2025} (\textsc{Li25}).
 
  \item \textsc{$q_{\rm c} < 8$}: a fixed threshold of $q_{\rm c} = 8$,
    meaning that Roche-lobe overflow is treated as stable for any
    donor-to-accretor mass ratio below~8.
    This highly permissive criterion substantially expands the
    parameter space in which binaries avoid CE evolution.
    Used in \citet{Olejak:2021CE} (\textsc{OL21}; their model M480.B),
    which additionally uses the \citet{Pavlovskii:2017} donor-radius
    grid to determine stability across a range of masses and
    compact-object accretors.
    The related label \textsc{$q_{\rm c} < 8$ CE switch} refers to
    model M481.B of \citet{Olejak:2021CE}, which applies the same
    $q_{\rm c} = 8$ threshold but additionally forces systems into a
    CE phase if the donor radius exceeds twice its Roche-lobe radius
    during the stable mass-transfer episode.
\end{itemize}
 
Beyond the explicitly varied $q_{\rm c}$ labels above, several codes
use fixed stability prescriptions that are not varied but differ
between frameworks and therefore contribute to the cross-study
diversity.
The \textsc{zeta prescription} \citep[following][]{Soberman:1997},
used as the default in \textsc{COMPAS} (\textsc{BRO22}, \textsc{vSon22},
\textsc{vSon23}) and the basis for the $\xi_{\rm ad,rad}$ parameter in
\textsc{SeBa} (\textsc{DT22}), determines stability by comparing the
adiabatic mass--radius exponent of the donor,
$\zeta_{\rm ad} = (\mathrm{d}\ln R / \mathrm{d}\ln M)_{\rm ad}$,
to the Roche-lobe response \citep[see][for a detailed account in
COMPAS]{Willcox:2023}.
Crucially, $\zeta_{\rm ad}$ is not a numerical derivative of the
\textsc{SSE} stellar tracks; instead it is taken from the analytical
polytropic solutions of \citet{Soberman:1997} and
\citet{Hjellming:1987}, with the \textsc{SSE}-determined stellar type
at the onset of Roche-lobe overflow used to select the appropriate
value ($\zeta_{\rm ad}>0$ for radiative donors, favouring stability;
$\zeta_{\rm ad}<0$ for convective giants, favouring instability).
In \textsc{SeBa}, the radiative-envelope exponent $\xi_{\rm ad,rad}$
is treated as a \emph{free parameter} by \citet{DorozsmaiToonen:2022}
(\textsc{DT22}), varied between 4 and 7.5; higher values substantially
increase the without-CE fraction (Figure~\ref{fig:BH-BH-without-CE-parameter-axis}).
The remaining codes in our compilation ---
\textsc{SEVN}/\textsc{PARSEC} (\textsc{Sg25}), \textsc{MOBSE}
(\textsc{Li25}), \textsc{binary\_c} (\textsc{Hen23}), and \textsc{BSE}
(\textsc{SL21}) --- similarly rely on analytic fitting formulae and
an explicit $q_{\rm c}$ or $\zeta$ threshold, without self-consistently
computing the donor's structural response.
The two exceptions in our compilation are \textsc{POSYDON}
\citep[\textsc{BA21}, \textsc{RG21}, \textsc{Xing24};][]{Fragos+2023ApJS,
Andrews+2025ApJS} and \textsc{BPASS} \citep[\textsc{BRI22};][]{Eldridge:2017},
which both avoid a fixed threshold: \textsc{POSYDON} interpolates
pre-computed \textsc{MESA} binary grids that self-consistently capture
the donor response as a function of mass, mass ratio, separation, and
metallicity, while \textsc{BPASS} runs detailed \textsc{STARS} models
for each individual system \citep[see footnote~2 of][for both
exceptions]{Bavera:2021}.

Taken together, the MT stability prescriptions used across the compiled
studies span a wide range --- from simple fixed $q_{\rm c}$ thresholds
and fitting-formula approaches through physically motivated
adiabatic-response criteria to fully self-consistent detailed
stellar-evolution models --- and this diversity is one of the primary
drivers of the spread in predicted with-CE and without-CE
formation-channel fractions
(Figures~\ref{fig-comparison-bbh-simulation-silo}--\ref{fig:BH-NS-without-CE-parameter-axis}).

\paragraph{$T_{\rm eff}$ (convective-envelope proxy)}
The effective temperature $T_{\rm eff}$ is used in several studies as a proxy for the evolutionary stage at which a star develops a convective envelope, which in turn affects the stability of mass transfer.
Variations in this threshold can shift systems between stable and unstable mass-transfer regimes.
This parameter appears in four distinct flavors in Figures~\ref{fig:level-1-BBH}--\ref{fig:level2-bns}:
\begin{itemize}
  \item \textsc{Teff-K}: boundary from \citet{Klencki:2021}, based on
    detailed stellar-evolution models for the onset of a deep convective
    envelope; used in \citet{DorozsmaiToonen:2022} and \citet{Romagnolo:2025}.
  \item \textsc{Teff-IT}: boundary from \citet{Ivanova:2004}, which adopts $T_{\rm eff} = 10^{3.73}$~K as the  dividing line between radiative and convective donors; used in
    \citet{DorozsmaiToonen:2022}.
  \item \textsc{Teff-MESA}: boundary based on new fits from \textsc{MESA} stellar models developed in \citet{Romagnolo:2025}.
\end{itemize}
Note that the online table and the study \citet{Romagnolo:2025} sometimes refers to the label \textsc{Teff-B} (referring to the same
$T_{\rm eff} = 10^{3.73}$~K criterion as adopted in
\citealt{Belczynski:2008}); the equivalent prescription
within \textsc{SeBa} is labeled \textsc{Teff-IT} following the original reference, we use this notation instead.

\paragraph{Supernova model: delayed, rapid, MM SN, N20 SN}
Supernova (SN) prescriptions determine remnant masses and natal kicks,
affecting binary survival and post-SN orbital properties.
Common variations included in our compilation are:
\begin{itemize}
  \item \textsc{rapid} and \textsc{delayed}: the rapid and delayed
    core-collapse remnant-mass prescriptions of \citet{Fryer:2012},
    which differ in whether the explosion occurs on a short ($\lesssim
    250$~ms) or longer timescale, leading to different mass-gap outcomes
    between neutron stars and black holes.
  \item \textsc{MM SN}: the stochastic explosion model of
    \citet{MandelMueller:2020} (Mandel \& M{\"u}ller 2020), in which
    the remnant mass and kick are drawn from a parameterized
    neutrino-driven wind prescription.
  \item \textsc{N20 SN}: the neutrino-driven engine model based on the
    ``N20'' progenitor of \citet{Sukhbold:2016} as implemented in
    \citet{RomanGarza:2021}.
    In this prescription, low-mass black holes form through partial
    fallback and receive non-zero natal kicks, while higher-mass
    progenitors collapse directly.
  \item \textsc{BPASS}: the remnant-mass prescription internal to
    \textsc{BPASS} \citep{Eldridge:2017}, used in \citet{Briel:2022}.
\end{itemize}
These assumptions are labeled as SN model variations in our figures and
are explored in \citet{Broekgaarden2022}, \citet{Briel:2022},
\citet{Boesky:2024gw}, and \citet{Shao:2021}.

\paragraph{$\sigma$ (natal kick dispersion), no BH kick, ECSN kicks}
The parameter $\sigma$ sets the dispersion of the natal kick velocity
distribution imparted to compact objects at formation, commonly drawn from
a Maxwellian distribution \citep[e.g.,][]{Hobbs:2005}.
Larger $\sigma$ values increase binary disruption probabilities and alter
merger rates and formation channels.
This parameter is denoted by $\sigma$ in our figures and is varied in
\citet{Broekgaarden2022}, \citet{Boesky:2024gw}, \citet{Xing:2024-BHNS-allZ},
and \citet{Li:2025}.
 
Several related sub-variants also appear in the figures:
\begin{itemize}
  \item \textsc{no BH kick}: a model in which black holes receive zero
    natal kick velocity at formation, regardless of their mass.
    This increases binary survival rates and is explored in
    \citet{Broekgaarden2022} and \citet{RomanGarza:2021}.
  \item \textsc{ECSN low kick}: a model in which neutron stars formed
    via electron-capture supernovae (ECSN) receive a low natal kick,
    drawn from a separate low-$\sigma$ distribution
    \citep[e.g.,][]{Podsiadlowski:2004}.
    ECSN are thought to occur for progenitors near the boundary of
    core collapse ($\sim 8$--$10\,\Msun$) and typically eject only
    a small amount of mass, resulting in low kicks and nearly circular
    post-SN orbits.
    Explored in \citet{RomanGarza:2021}.
  \item \textsc{ECSN 265 kick}: a model in which electron-capture SN
    kicks are drawn from the same high-$\sigma = 265\,\kms$ distribution
    used for iron core-collapse supernovae, rather than from a separate
    low-kick prescription.
    Explored in \citet{RomanGarza:2021}.
\end{itemize}

% ------------------------------------------------------------------
\paragraph{$f_{\rm WR}$ (Wolf--Rayet wind scaling)}
The parameter $f_{\rm WR}$ scales the mass-loss rates of Wolf--Rayet
stars, influencing stellar masses prior to collapse and thus
compact-object masses and orbital evolution.
This parameter is denoted by $f_{\rm WR}$ in our figures and is varied in
\citet{Broekgaarden2022} and \citet{Li:2025}.

% ------------------------------------------------------------------
\paragraph{$M_{\rm NS,max}$ (maximum neutron-star mass)}
The assumed maximum neutron-star mass determines whether compact remnants
are classified as neutron stars or black holes, thereby affecting the
relative rates of BNS, BHNS, and BBH systems.
This parameter is denoted by $M_{\rm NS,max}$ in our figures and is varied
in \citet{Broekgaarden2022}.

% ------------------------------------------------------------------
\paragraph{Pair-instability supernova (PISN) treatment}
Assumptions about pair-instability and pulsational pair-instability
supernovae affect the upper mass spectrum of black holes and the formation
efficiency of massive BBHs.
These assumptions are varied in \citet{Broekgaarden2022}.
 
% ------------------------------------------------------------------
\paragraph{$\dot{m}_{\rm Edd}$ (Eddington-limited accretion)}
The Eddington accretion rate $\dot{m}_{\rm Edd}$ sets the maximum rate at
which compact objects can accrete matter.
In binary evolution models, this limit effectively caps the accretion onto
the compact object during Roche-lobe overflow, such that any excess mass is
lost from the system.
Different studies adopt different treatments of this limit, including
strict Eddington-limited accretion or allowing for super-Eddington
accretion with multiplicative factors
(e.g., $10^3$--$10^6 \times \dot{M}_{\rm Edd}$).
This parameter is denoted by $\dot{M}_{\rm Edd}$ in our figures and is
varied in studies such as \citet{Bavera:2021}.

% ------------------------------------------------------------------
\paragraph{$R_{\max}$ (maximum stellar radius / radial expansion limit)}
The parameter $R_{\max}$ sets the maximum radius that a star can reach
during its evolution, effectively controlling the extent of radial
expansion phases (e.g., giant branches).
This directly impacts the likelihood and timing of Roche-lobe overflow, and
therefore whether binaries interact, undergo stable mass transfer, or enter
a CE phase.
\citet{Romagnolo:2023} and \citet{Romagnolo:2025} adopt different
implementations for $R_{\max}$, including limits derived from different
stellar-evolution models (e.g., METISSE, MESA with and without MLT++)
or treatments of envelope inflation; see \citet{Romagnolo:2023} and
\citet{Romagnolo:2025} for details.
In the figures, \textsc{Rmax} appended to a model label indicates that
this radial cap is applied.
 
% ------------------------------------------------------------------
\paragraph{Case BB CE (Case BB mass-transfer stability)}
The label \textsc{case BB CE} (appearing in \textsc{BRO22} rows) refers to
model variants in which Case~BB mass transfer --- mass transfer from a
helium-rich (stripped) donor star onto a compact object after the first
supernova --- is assumed to be dynamically \emph{unstable}, leading to a
second CE phase.
In the default \textsc{COMPAS} model, Case~BB mass transfer is assumed to
always proceed stably; the \textsc{case BB CE} variant relaxes this
assumption, substantially reducing the fraction of BNS and BHNS systems
that avoid a CE phase.
This variant is explored in \citet{Broekgaarden2022} and directly affects
the BNS formation channel fractions in Figure~\ref{fig:level-1-bns}.
 
% ------------------------------------------------------------------
\paragraph{no MT ZAMS (suppressed mass transfer at zero-age main sequence)}
The label \textsc{no MT ZAMS} (appearing in \textsc{BA21} rows) refers to
a model variant in which mass transfer at or very near the zero-age main
sequence --- typically arising from contact or over-contact systems ---
is suppressed.
In the default \citet{Bavera:2021} framework, such systems are allowed to
undergo mass transfer and potentially equilibrate their masses, which can
seed the chemically homogeneous evolution (CHE) pathway.
The \textsc{no MT ZAMS} variant prevents this, removing the initial
mass-transfer episode and thereby suppressing the CHE and related
short-period formation channels.
 
% ------------------------------------------------------------------
\paragraph{Opt CE (Optimistic common-envelope assumption)}
The optimistic CE assumption refers to models in which binaries are allowed
to survive CE phases that would otherwise be assumed to lead to mergers
under more conservative criteria.
In particular, this includes allowing systems undergoing CE initiated by
Hertzsprung-gap (HG) donors to successfully eject the envelope and form a
bound post-CE binary, despite the lack of a well-defined core-envelope
structure in such stars.
This assumption increases the number of systems that survive CE evolution,
thereby boosting the overall merger rate and enhancing the contribution of
formation channels involving CE phases.
This assumption is denoted as \textsc{opt CE} in our figures and is explored in studies such as \citet{Broekgaarden2022}. \\

% ------------------------------------------------------------------
\paragraph{ $\log U$ (initial orbital-separation prior)}
In \citet{Bavera:2021}, the authors present formation channels for models
varying the initial distribution for the orbital separation.
By default, the separation is drawn from a power-law distribution
motivated by observations of massive binaries
\citep[e.g.,][]{Sana2012}.
The variant labeled \textsc{log U prior} instead uses a distribution that
is uniform in the logarithm of the separation (\"Opik's law), which
increases the relative contribution of wide binaries and thereby modifies
the likelihood of Roche-lobe overflow and CE evolution.
This parameter is denoted as ``$\log U$ prior'' in our figures and is
varied in \citet{Bavera:2021}.

\begin{table*}
\centering
\caption{Literature compilation of formation-channel contributions for BNS,
BHNS, and BBH mergers from isolated binary evolution studies that report
results at a fixed metallicity rather than as cosmologically integrated
intrinsic merger rates, and are therefore not included in the main
compilation of Figures. Entries are sorted by metallicity within
each compact-object class. The metallicity $Z$ is given as an absolute
value; many studies model either $\Zsun = 0.02$ (e.g. StarTrack) or 
 $\Zsun = 0.0142$ (COMPAS, POSYDON; \citealt{Asplund:2009}). Where
a study reports only a lower or upper limit on the CE fraction, this is
indicated with $\gtrsim$ or $\lesssim$ accordingly; entries marked ``rest
not specified'' indicate that the remaining fraction is not explicitly
assigned to a with-CE or without-CE channel by the original authors.}
\label{tab:fc_Z_statements_bns}
\small
\setlength{\tabcolsep}{5pt}
\renewcommand{\arraystretch}{1.15}
\begin{tabular}{l l l l}
\multicolumn{4}{l}{\textbf{BNS} formation channels }\\
\hline
Study  & $Z$ & CE statement & notes   \\
\hline
\citet{Mestichelli:2025}         & $10^{-11}$          & $\geq 91.9\%$ \withCE    & rest not specified \\
\citet{Mestichelli:2025}         & $10^{-4}$           & $\geq 99.3\%$ \withCE    & rest not specified \\
\citet{Dominik:2012}             & $2\times10^{-3}$    & $\geq 91.8\%$ \withCE    & rest not specified \\
\citet{Gallegos-Garcia:2023}     & $1.42\times10^{-3}$ & $\sim 100\%$ \withCE     & rare SMT channel for $M_{\rm NS}=2\,\Msun$ and large natal kicks \\
\citet{Belczynski:2002}          & $2\times10^{-2}$    & $\gtrsim 95\%$ \withCE   & including a double-core CE channel \\
\citet{Andrews:2015}             & $2\times10^{-2}$    & $\sim 100\%$ \withCE     & case BB / electron-capture SN \\
\citet{Chruslinska:2018}         & $2\times10^{-2}$    & $\gtrsim 95\%$ \withCE   & \\
\citet{Dominik:2012}             & $2\times10^{-2}$    & $\gtrsim 94.2\%$ \withCE & \\
\citet{VignaGomez:2018}          & $1.42\times10^{-2}$ & $\gtrsim 91\%$ \withCE   & remaining $9\%$ CE not specified \\
\citet{Gallegos-Garcia:2023}     & $1.42\times10^{-2}$ & $\sim 100\%$ \withCE     & stable-MT-only BNS mergers absent for fiducial $1.4\,\Msun$ NSs \\
\citet{Chattaraj:2026}           & $1.42\times10^{-2}$ & $\gtrsim 95\%$ \withCE   & CE always with case BB MT and USSN, case B/C MT subchannels \\
\citet{Chattaraj:2026}           & $1.42\times10^{-2}$ & $\lesssim 6\%$ \withoutCE & $5\%$ can form without MT at all, $1\%$ forms through the SMT channel \\
\hline
        & & & \\
\multicolumn{4}{l}{\textbf{BHNS} formation channels }\\
\hline
Study  & $Z$ & CE statement & other important fc   \\
\hline
\citet{Mestichelli:2025}         & $10^{-11}$          & $\geq 36\%$ \withCE      & small fraction not specified \\
\citet{Mestichelli:2025}         & $10^{-11}$          & $\geq 55.9\%$ \withoutCE & small fraction not specified \\
\citet{Mestichelli:2025}         & $10^{-4}$           & $\geq 46.2\%$ \withCE    & small fraction not specified \\
\citet{Mestichelli:2025}         & $10^{-4}$           & $\geq 46.9\%$ \withoutCE & small fraction not specified \\
\citet{Dominik:2012}             & $2\times10^{-3}$    & $\geq 93.6\%$ \withCE    & \\
\citet{Belczynski:2002}          & $2\times10^{-2}$    & $\gtrsim 80\%$ \withCE   & including single-core CE channel \\
\citet{Dominik:2012}             & $2\times10^{-2}$    & $\gtrsim 97.2\%$ \withCE & \\
\citet{Xing:2024-BHNS-solarZ}    & $1.42\times10^{-2}$ & $\sim 70\%$ \withCE      & \\
\citet{Xing:2024-BHNS-solarZ}    & $1.42\times10^{-2}$ & $\sim 30\%$ \withoutCE   & \\
\hline
  & & & \\
\multicolumn{4}{l}{\textbf{BBH} formation channels }\\
\hline
Study & $Z$ & CE statement & other important fc   \\
\hline
\citet{Dominik:2012}                      & $2\times10^{-3}$    & $\geq 96\%$ \withCE          & \\
\citet{Belczynski:2022formationChannel}   & $\lesssim8\times10^{-3}$  & $\sim 75\%$ \withCE    & CE $3\times$ higher than SMT channel \\
\citet{Belczynski:2022formationChannel}   & $\gtrsim8\times10^{-3}$   & \withoutCE{} higher  & without CE channel dominates over CE channel at higher $Z$ \\
\citet{Briel:2026}                        & $\gtrsim2.84\times10^{-3}$ & $0\%$ \withoutCE      & without SN natal-kicks SMT does not produce BBH mergers \\
\citet{Belczynski:2002}                   & $2\times10^{-2}$    & $\gtrsim 95\%$ \withCE       & includes single-core CE before SN1 \\
\citet{Dominik:2012}                      & $2\times10^{-2}$    & $\gtrsim 98.9\%$ \withCE     & \\
\hline
\end{tabular}
\end{table*}

% \newpage 
\section{Appendix D: Formation Channel Contribution to Formation Efficiency}
\label{sec:appendix-formation-channel-contributions-efficiency-metallicity}

\begin{figure*}
    \centering
    \includegraphics[width=1\linewidth]{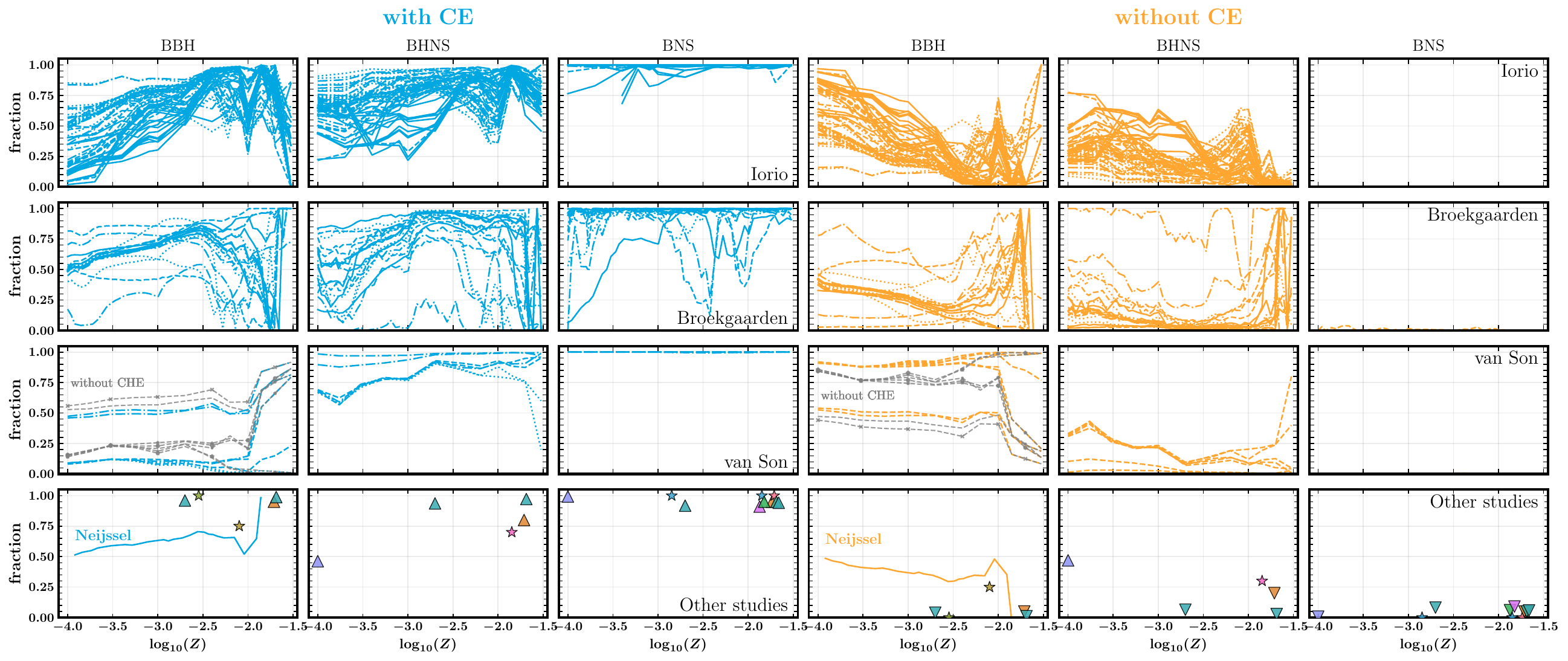}
    \caption{
    Formation-channel fractions as a function of metallicity for compact-object binaries across multiple population-synthesis studies and literature constraints.
    Rows corresponds (top to bottom) to  \citet{Ioro:2023sevn} \citep[using publicly available data from][]{iorio_2023_7794546}, \citet{Broekgaarden2022} \citep[using publicly available data from][]{Broekgaarden:Zenodo-DCO-formation-channels}, \citet{vanSon:2024} \citep[using publicly available data from][]{van_son_2024_14508864}, and a compilation of additional literature constraints including the BBH formation efficiency from \citet{Neijssel:2019} (a single BBH model, read from their Figure~1 using a plot digitizer), while colored markers indicate literature constraints at specific metallicities compiled in Table~\ref{tab:fc_Z_statements_bns}; arrows denote upper or lower limits and stars indicate approximate values.
    Columns show double compact object types (BBH, BHNS, BNS), while the left and right blocks correspond to systems formed with and without a CE phase, respectively.
    For \citet{vanSon:2024} BBH models, gray curves show fractions excluding chemically homogeneous evolution (CHE). Some \citet{Broekgaarden2022} models include a non-negligible ``other'' channel (CE not specified). Low-number statistics introduce noise near $\Zsun$ for BBH and BHNS models. 
     All points are converted to $\log_{10}(Z)$, and complementary fractions are inferred where needed assuming $f_{\rm with\,CE} + f_{\rm without\,CE} = 1$.     %
    The figure highlights the wide diversity of BBH and BHNS predictions and the consistent result that {BNS formation is dominated by CE channels}, with  $\lesssim 10\%$  forming without CE across models and metallicities. 
    See Appendix~\ref{sec:appendix-formation-channel-contributions-efficiency-metallicity} for details. 
    See  \href{https://floorbroekgaarden.github.io/Rates_of_Formation_Channels/interactive_figures_and_tables/formation_channel_rates_table.html}{interactive figure} and  \href{https://github.com/FloorBroekgaarden/Rates_of_Formation_Channels}{GitHub} for details and code.
    }
    \label{fig:formation-efficiency-results-detailed}
    \end{figure*}

Figure~\ref{fig:formation-efficiency-results} in the main text and Figure~\ref{fig:formation-efficiency-results-detailed} in this appendix show formation-channel fractions as a function of metallicity for \ac{BBH}, \ac{BHNS}, and \ac{BNS} mergers across multiple population-synthesis studies. Rather than intrinsic merger rates (which depend on the assumed cosmic star-formation history and metallicity evolution), these figures show formation efficiencies—the number of mergers formed per unit star-forming mass—as a function of progenitor metallicity. This representation removes the cosmological weighting and isolates the metallicity-dependent binary evolution physics.

\paragraph{Data sources.}
The formation efficiency curves shown are drawn from four sources:
\begin{itemize}
    \item \citet{Ioro:2023sevn}: formation efficiencies for \ac{BBH}, \ac{BHNS}, and \ac{BNS} mergers across a wide grid of physical assumptions, reproduced from the publicly available dataset \citep{iorio_2023_7794546} using the associated plotting code (\url{https://gitlab.com/iogiul/iorio22_plot/-/tree/v3/formation_channels}). The \citet{Ioro:2023sevn} models span variations in $\alpha_{\rm CE}$, natal kick prescriptions, remnant-mass prescriptions, and envelope binding energies, among others.

    \item \citet{Broekgaarden2022}: formation efficiencies for \ac{BBH}, \ac{BHNS}, and \ac{BNS} mergers across 20 binary evolution model variations, reproduced from the publicly available dataset \citep{Broekgaarden:Zenodo-DCO-formation-channels}. The displayed curves correspond to the same 20 primary model variations used in the main analysis.

    \item \citet{vanSon:2024}: formation efficiencies for \ac{BBH}, \ac{BHNS}, and \ac{BNS} mergers as a function of metallicity, reproduced from the publicly available dataset \citep{van_son_2024_14508864}. For \ac{BBH} systems, this study includes contributions from chemically homogeneous evolution (\ac{CHE}); Figure~\ref{fig:formation-efficiency-results-detailed} additionally shows curves with CHE excluded (gray) for direct comparison with the other studies.

    \item \citet{Neijssel:2019}: \ac{BBH}-only formation efficiency as a function of metallicity, digitized from their Figure~1 using WebPlotDigitizer \citep{rohatgi2020webplotdigitizer}, shown in the bottom row of both figures as a black curve.
\end{itemize}

Colored markers in the bottom row of both figures indicate literature constraints at individual metallicities, compiled in Table~\ref{tab:fc_Z_statements_bns} and drawn from studies that report formation-channel fractions at one or a few fixed metallicities \citep[e.g.,][]{Belczynski:2002, Dominik:2012, Andrews:2015, Chruslinska:2018, VignaGomez:2018, Mestichelli:2025, Xing:2024-BHNS-solarZ, Chattaraj:2026}. Arrows denote upper or lower limits and stars indicate approximate values. All metallicity values are converted to $\log_{10}(Z)$, and where only the with-CE or without-CE fraction is reported, the complementary fraction is inferred assuming $f_{\rm with\,CE} + f_{\rm without\,CE} = 1$.

\paragraph{Key trends.}
The figures reveal several consistent trends across models and metallicities. For \ac{BNS} mergers, all studies find that formation is overwhelmingly dominated by channels involving at least one \ac{CE} phase, with typically $\lesssim 10\%$ of systems forming without \ac{CE} across the full metallicity range explored. This agreement holds across different population-synthesis codes, physical assumptions, and metallicity regimes, and is further supported by the single-metallicity literature constraints compiled in Table~\ref{tab:fc_Z_statements_bns}.

For \ac{BBH} and \ac{BHNS} mergers, the formation-channel fractions show substantially more variation across models and metallicity. For \ac{BBH} systems in particular, the relative contribution of without-\ac{CE} channels generally increases toward the lowest metallicities ($\log_{10}(Z) \lesssim -2.5$), likely reflecting the reduced stellar wind mass loss and the resulting larger stellar radii and more extreme mass ratios at low metallicity. This trend is present in several of the \citet{Ioro:2023sevn} and \citet{Broekgaarden2022} model variations, as well as in the \citet{Neijssel:2019} results. For \ac{BHNS} systems, a similar but weaker trend is seen, with the without-\ac{CE} contribution modestly increasing at low metallicities in some models, while others show little dependence on metallicity.

Importantly, the large diversity in predicted \ac{BBH} and \ac{BHNS} formation-channel fractions across the different population-synthesis frameworks reinforces the conclusion from the main text: the balance between CE and without-CE channels is highly sensitive to the underlying physical assumptions, and this sensitivity persists across the full range of progenitor metallicities relevant for the observed gravitational-wave population.

\section{Appendix E: Observable features tables}
\label{sec:appendix-comparison-table}
To facilitate comparison across the literature, we summarize in Tables~\ref{tab:smt_statements}--\ref{tab:ce_statements} qualitative
statements from the literature about the observational imprints of the
SMT and CE formation channels on the gravitational-wave source
population. These entries are intentionally qualitative and study-level:
they are meant to highlight the dominant trends discussed in each work
rather than provide an exhaustive summary of every individual simulation
variant.

\section{Appendix F: Absolute merger rates parameter variations}
\label{sec:appendix-absolute-merger-rates-parameter-variations}

\begin{figure*}
    \centering
    \includegraphics[width=1\linewidth]{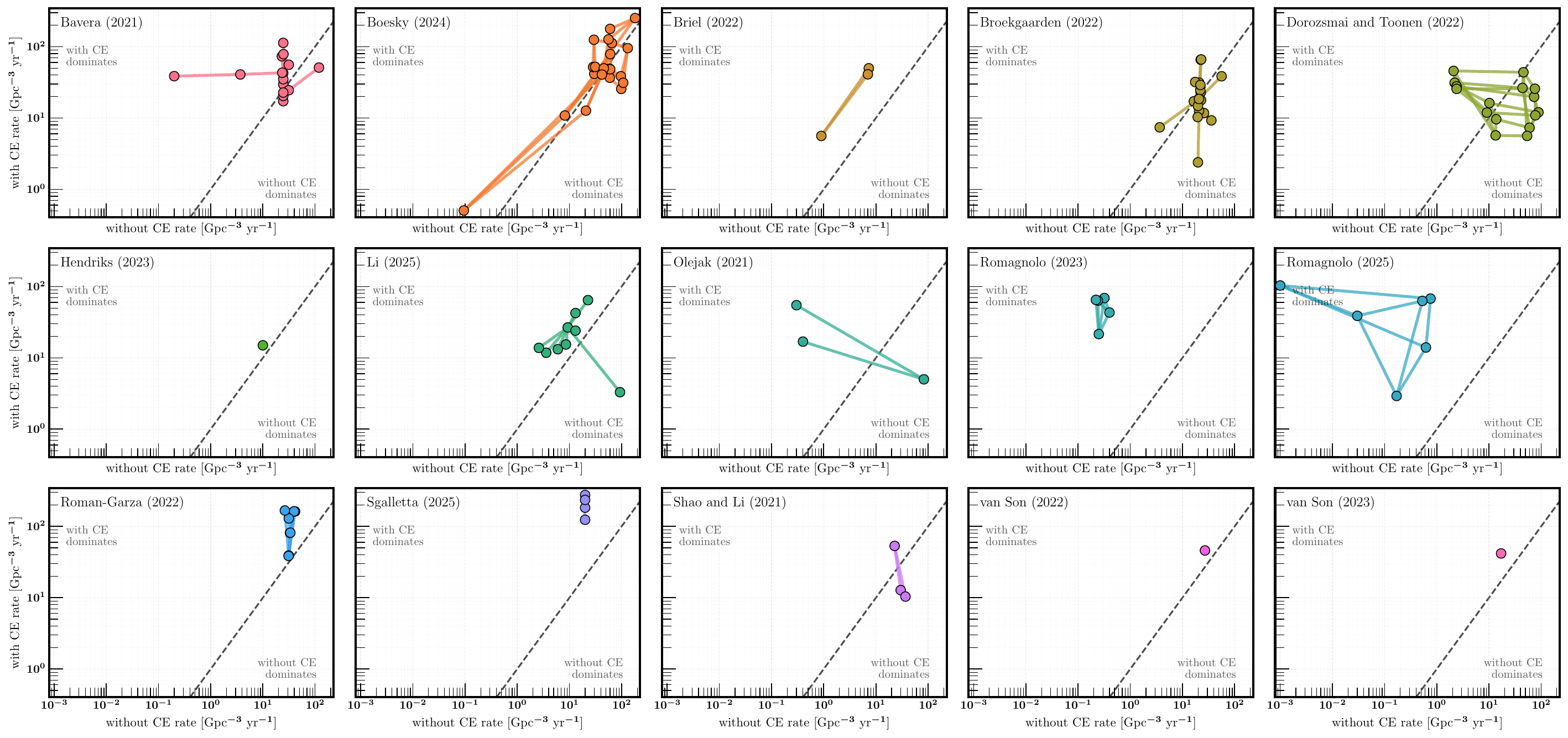}
    \caption{
    Without-\ac{CE} versus with-\ac{CE} intrinsic \ac{BBH} merger rates for the compiled \ac{BBH} population-synthesis simulations. 
    Scatter symbols, colors, and connecting lines are as in the bottom panel of Figure~\ref{fig-comparison-bbh-simulation-silo}. 
    The dashed diagonal line indicates equal contributions from with-\ac{CE} and without-\ac{CE} formation channels. Systems above the diagonal are dominated by with-\ac{CE} formation, whereas systems below the diagonal are dominated by without-\ac{CE} formation.
        See   \href{https://github.com/FloorBroekgaarden/Rates_of_Formation_Channels}{GitHub} for details and code.
    }
    \label{fig:BBH-with-CE-without-CE-absolute-rate-grid}
\end{figure*}

\begin{figure*}
    \centering
    \includegraphics[width=1\linewidth]{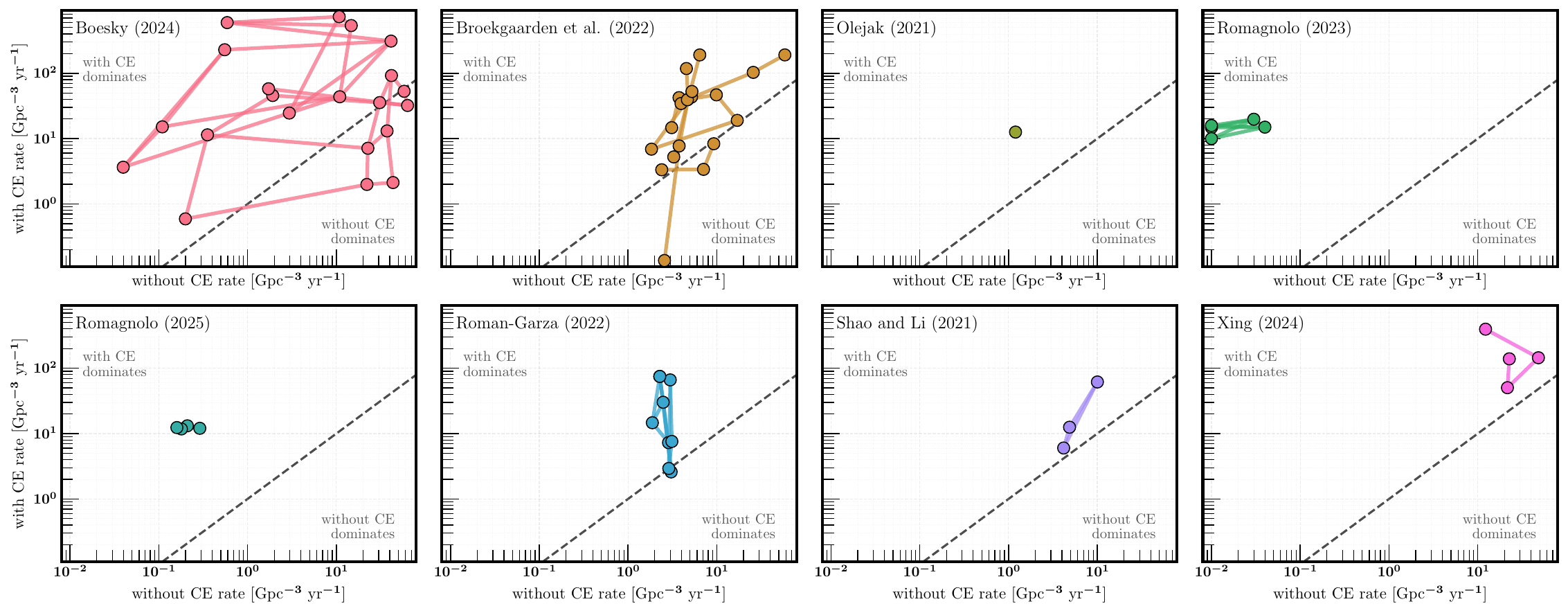}
    \caption{Same as \ref{fig:BBH-with-CE-without-CE-absolute-rate-grid} for BHNS mergers.
        See \href{https://github.com/FloorBroekgaarden/Rates_of_Formation_Channels}{GitHub} for details and code.}
    \label{fig:BHNS-with-CE-without-CE-absolute-rate-grid}
\end{figure*}

In Figure~\ref{fig:BBH-with-CE-without-CE-absolute-rate-grid} and \ref{fig:BHNS-with-CE-without-CE-absolute-rate-grid} we present the absolute intrinsic merger rates corresponding to the parameter variations discussed in Section~\ref{sec:results-Simulation-Variations-and-Parameter-Dependencies}. These figures complement the main-text analysis of relative formation-channel contributions (Figures~\ref{fig-comparison-bbh-simulation-silo} and~\ref{fig-comparison-bhns-simulation-silo}) by showing how variations in key binary-evolution assumptions affect both the total merger rates and the absolute rates associated with with-\ac{CE} and without-\ac{CE} pathways.

Importantly, an increase in the rate from the without-\ac{CE} formation channels does not necessarily correspond to an increase in total merger rate nor a decrease in the with-CE rate. This reflects the complex and often competing effects of binary-interaction physics on binary survival, orbital tightening, and merger timescales. In some cases, parameter variations can simultaneously increase the rate of without-\ac{CE} formation channel while also enhancing the with-CE rate overall production of merging systems. These figures therefore illustrate that interpreting changes in formation-channel fractions alone can be challenging without also considering the corresponding changes in the absolute merger-rate budget.

%%  ── TABLE 2 : SMT CHANNEL ─────────────────────────────────────────────

% \newlength{\SMTstatW}
% \setlength{\SMTstatW}{0.52\textwidth} 
\begin{table*}
\caption{Qualitative statements in the literature about the observational
imprint of the SMT channel. HM: high mass, LM: low mass, AM: angular momentum.}
\label{tab:smt_statements}
\centering
\small
\begin{tabular}{l l l} 
\hline
Study & Observable & Main statement about SMT channel \\
\hline
%--- Mass distribution ---
\citet{Li:2025}                      & Mass distribution & SMT produces somewhat higher-mass BBHs than CE \\
\citet{Neijssel:2019}                & Mass distribution & SMT contributes substantially to high-mass BBHs \\
\citet{Willcox:2025}                 & Mass distribution & SMT contributes to HM+HM BBHs; both CE and SMT contribute to LM+LM BBHs (chirp mass) \\
\citet{vanSon:2022}                  & Mass distribution & SMT contributes substantially to higher-mass BBHs \\
\citet{vanSon:2022-nopeaks}          & Mass distribution & SMT produces a low-mass cutoff ($\sim2.5$--$9\,\Msun$) and a peak around $\sim8$--$10\,\Msun$ set by MT stability \\
\citet{Hendriks:2023}                & Mass distribution & SMT preferentially produces high-mass BBHs \\
\citet{Belczynski:2022formationChannel} & Mass distribution & SMT produces substantially more massive BBHs than CE \\
\citet{Klencki:2026-smt}             & Mass distribution & SMT preferentially produces more massive BBHs \\
\citet{Xing:2024-BHNS-allZ}          & Mass distribution & SMT may be important for BHNS containing BHs $\gtrsim11\,\Msun$ \\
\hline
%--- Mass ratio ---
\citet{Briel:2026}   & Mass ratio & Without BH kicks, SMT BBH mergers peak at $q\sim0.8$--$1$, independent of metallicity \\
\citet{Briel:2026}   & Mass ratio & With BH kicks, SMT BBH mergers can have more unequal mass ratios \\
\citet{Banerjee:2024} & Mass ratio & SMT combined with the delayed remnant-mass model produces the most unequal-mass BBHs \\
\citet{Olejak:2024}  & Mass ratio & SMT can produce BBHs with $q\in[0.4,0.7]$ and relatively high $\chi_{\rm eff}$ \\
\citet{Zevin:2020}   & Mass ratio & Both CE and SMT channels can contribute to unequal-mass BBHs \\
\citet{Xing:2024-BHNS-solarZ} & Mass ratio \& period & Most SMT BHNS have $P\lesssim50\,\rm{d}$ and a wide spread of mass ratios (at \Zsun) \\
\citet{Zevin:2020}   & Mass ratio & At $Z\gtrsim\Zsun/30$ the SMT channel dominates formation of GW190814-like (unequal-$q$) BBHs \\
%--- New: chi_eff--q ---
\citet{Banerjee:2024} & Spin \& mass ratio & SMT with mass-ratio reversal produces a $\chi_{\rm eff}$--$q$  (anti-correlation disappears at high metallicity) \\
\hline
%--- Delay time ---
\citet{Klencki:2026-smt}   & Delay time & SMT produces long delay times ($\gtrsim1\,\Gyr$) due to a fundamental lower limit on post-SMT separation \\
\citet{vanSon:2022}        & Delay time & SMT preferentially produces long BBH delay times because orbital tightening is limited \\
\citet{vanSon:2022-nopeaks}& Delay time & SMT BBH/BHNS delay times depend on orbital tightening, can be long for inefficient AM loss  \\
\citet{Briel:2026}         & Delay time & SMT channel produces long delay times $\gtrsim0.2\,\Gyr$ \\
\citet{Olejak:2024}        & Delay time & SMT delay times depend sensitively on the degree of orbital tightening \\
\citet{Oh:2023}            & Delay time & SMT alone fails to shrink the orbit enough for asymmetric BBH/BHNS mergers (needs natal kicks) \\
\hline
%--- Spin ---
\citet{Klencki:2026-smt} & Spin & SMT produces low BH spins: minimum post-SMT separation prevents tidal spin-up of helium stars \\
\citet{Briel:2026}       & Spin & SMT BBH mergers have $\chi_{\rm eff}$ peaks at $0$ and $\sim0.15$ \\
\citet{Olejak:2024}      & Spin & SMT can produce low or moderate BH spins depending on AM transport and accretion efficiency \\
\citet{Xing:2024-BHNS-solarZ} & Spin & SMT BHNS can have large spin tilt angles ($\gtrsim\pi/2\,\rm{rad}$) and $\chi_{\rm eff}\in[-0.2,0.2]$ due to kicks (at \Zsun) \\
\hline
%--- Progenitors ---
\citet{Briel:2026} & SMT progenitors & SMT BBH mergers predominantly arise from case~A MT progenitors ($P_{\rm ZAMS}\lesssim10\,\rm{d}$) \\
\hline
\end{tabular}
\end{table*}

%%  ── TABLE 3 : CE CHANNEL ──────────────────────────────────────────────
\begin{table*}
\centering
\caption{Qualitative statements in the literature about the observational
imprint of the CE channel.}
\label{tab:ce_statements}
\small
\begin{tabular}{l l l} 
\hline
Study & Observable & Main statement \\
\hline
%--- Mass distribution ---
\citet{Li:2025}                      & Mass distribution & CE dominates the LM peak of the BBH mass distribution \\
\citet{Hendriks:2023}                & Mass distribution & CE struggles to produce the highest-mass BBHs; HM systems preferentially form through SMT \\
\citet{Belczynski:2022formationChannel} & Mass distribution & CE produces slightly LM BBHs than SMT \\
\citet{Willcox:2025}                 & Mass distribution & CE contributes to HM+LM BBHs, and both CE and SMT contribute to LM+LM \\
%--- new: vanSon+2022 mass/delay ---
\citet{vanSon:2022-nopeaks}          & Mass \& delay time & CE preferentially produces LM BBHs ($\lesssim30\,\Msun$) with short delay times ($\lesssim1\,\Gyr$) \\
\hline
%--- Mass ratio ---
\citet{Oh:2023}    & Mass ratio & The most asymmetric systems tend to undergo CE in the first MT phase \\
\citet{Zevin:2020} & Mass ratio & Both CE and SMT channels can contribute to unequal-mass BBHs \\
\citet{Olejak:2024}& Mass ratio \& spin & CE produces BBHs with near-equal masses ($q\sim1$) and a range of BH spins \\
\hline
%--- Spin ---
\citet{Bavera:2021}          & Spin & CE can produce spinning second-born BHs through tidal spin-up in post-CE orbits \\
\citet{Xing:2024-BHNS-solarZ}& Spin & CE BHNS have small spin tilt angles and small $\chi_{\rm eff}$ (peak at 0, extending $>0$) at \Zsun \\
\hline
%--- Period / delay time ---
\citet{Andrews:2015}  & Period   & All short-period BNS systems form via CE after SN1 \\
\citet{Neijssel:2019} & Delay time & CE preferentially produces shorter delay times than SMT \\
\citet{vanSon:2022}   & Delay time & CE contributes more strongly to rapidly merging BBHs than SMT \\
\citet{Andrews:2015}  & Delay time & BNS channels with case~BB MT after CE lead to shorter periods \\
\citet{Xing:2024-BHNS-solarZ} & Mass ratio \& period & Most CE BHNS have $P\gtrsim50\,\rm{d}$ and $q\gtrsim0.5$ at \Zsun \\
\citet{Andrews:2015}  & Eccentricity & BNS channels with electron-capture SN (low kicks) lead to lower eccentricities \\
\hline
\end{tabular}
\end{table*}

% %%  ── TABLE 4 : CHE CHANNEL ─────────────────────────────────────────────
% \begin{table*}
% \centering
% \caption{Qualitative statements in the literature about the observational
% imprint of the CHE channel.}
% \label{tab:che_statements}
% % \small
% % \setlength{\tabcolsep}{5pt}
% % \renewcommand{\arraystretch}{1.15}
% \begin{tabular}{l l l} 
% \hline
% Study & Observable & Main statement \\
% \hline
% \citet{Li:2025} & Mass distribution & CHE produces a mass peak around $\sim30\,\Msun$ \\
% \citet{Li:2025} & Mass distribution & CHE contributes to the high-mass part of the BBH population \\
%  & & \\
% \citet{Chattaraj:2026} & Donor structure & The donor's evolutionary state at CE onset determines whether systems merge within a Hubble time \\
% \citet{Chattaraj:2026} & Natal kicks      & Low SN kick velocities are required to reproduce the observed DNS population \\
% \hline
% \end{tabular}
% \end{table*}

\end{document}